\documentclass[twocolumn]{aastex631}

\usepackage{amsmath}

\defcitealias{Hutschenreuter2022}{H22}
\defcitealias{Edenhofer2024}{E24}
\defcitealias{Ordog2017}{O17}

\newcommand\change[1]{\textcolor{black}{{#1}}}

\usepackage{comment}

\newcommand{\radmsq}{\ensuremath{\textrm{ rad m}^{-2}}}

\begin{document}

\title{A three-dimensional model for the reversal in the local large-scale interstellar magnetic field }

\newcommand{\UBCO}{Department of Computer Science, Math, Physics, \& Statistics, University of British Columbia, Okanagan Campus, Kelowna, BC V1V 1V7, Canada}

\newcommand{\UCalgary}{Department of Physics and Astronomy, University of Calgary, 2500 University Drive NW, Calgary, Alberta, T2N 1N4, Canada}

\newcommand{\DRAO}{Dominion Radio Astrophysical Observatory, Herzberg Research Centre for Astronomy and Astrophysics, National Research Council Canada, PO Box 248, Penticton, BC, V2A 6J9, Canada}

\newcommand{\UTas}{School of Natural Sciences, University of Tasmania, Hobart, Tas 7000 Australia}

\newcommand{\INAF}{INAF-Istituto di Radioastronomia, Via Gobetti 101, 40129 Bologna, Italy}

\newcommand{\INAFOAA}{INAF – Osservatorio Astrofisico di Arcetri, Largo E. Fermi 5, 50125 Firenze, Italy}

\newcommand{\LPENS}{Laboratoire de Physique de l’École Normale Supérieure, ENS, Université PSL, CNRS, Sorbonne Université, Université de Paris, F-75005 Paris, France}

\newcommand{\RU}{Department of Astrophysics/IMAPP, Radboud University, PO Box 9010, 6500 GL Nijmegen, The Netherlands}

\newcommand{\CSIROBentley}{ATNF, CSIRO Space \& Astronomy, Bentley, WA, Australia}

\newcommand{\Caltech}{Caltech Owen's Valley Radio Observatory, Big Pine 93513, CA, USA}

\newcommand{\Onsala}{Department of Earth and Space Sciences, Chalmers University of Technology, Onsala Space Observatory, 439 92 Onsala, Sweden}

\newcommand{\Waterloo}{Department of Physics and Astronomy, University of Waterloo, 200 University Avenue West Waterloo, ON, N2L 3G1, Canada}

\newcommand{\USC}{Department of Physics \& Astronomy, University of South Carolina, Columbia, SC 29208, USA}

\newcommand{\Stanforda}{Department of Physics, Stanford University, Stanford, CA 94305, USA}

\newcommand{\Stanfordb}{Kavli Institute for Particle Astrophysics \& Cosmology, P.O. Box 2450, Stanford University, Stanford, CA 94305, USA}

\newcommand{\UWO}{Department of Physics \& Astronomy, University of Western Ontario, 1151 Richmond Street, London, ON, N6A 3K7, Canada}

\author[0000-0001-5181-6673]{Rebecca A. Booth}
\affiliation{\UCalgary}
\author[0000-0002-2465-8937]{Anna Ordog}
\affiliation{\UBCO}
\affiliation{\DRAO}
\affiliation{\UWO}

\author[0000-0003-4781-5701]{Jo-Anne Brown}
\affiliation{\UCalgary}

\author[0000-0003-1455-2546]{T. L. Landecker}
\affiliation{\DRAO}
\affiliation{\UCalgary}

\author[0000-0001-7301-5666]{Alex S. Hill}
\affiliation{\UBCO}
\affiliation{\DRAO}

\author[0000-0001-7722-8458]{Jennifer L. West}
\affiliation{\DRAO }

\author[0000-0002-2679-4609]{
Minjie Lei}
\affiliation{\Stanforda}
\affiliation{\Stanfordb}

\author[0000-0002-7633-3376]{S. E. Clark}
\affiliation{\Stanforda}
\affiliation{\Stanfordb}

\author[0000-0003-0932-3140]{Andrea Bracco}
\affiliation{\INAFOAA }
\affiliation{\LPENS }

\author[0000-0002-6300-7459]{John M. Dickey}
\affiliation{\UTas }

\author[0000-0002-3973-8403]{Ettore Carretti}
\affiliation{\INAF }

\begin{abstract}

We probe the three-dimensional geometry of the large-scale Galactic magnetic field within 1~kpc of the Sun using the Dominion Radio Astrophysical Observatory (DRAO) Global Magneto-Ionic Medium Survey (GMIMS) of the Northern Sky (DRAGONS). DRAGONS is a new full polarization survey of the Northern sky from 350 to 1030 MHz covering declinations $-20^\circ < \delta < 90^\circ$ and a component of GMIMS. The first moment of the Faraday depth spectra produced from DRAGONS above 500 MHz reveals large-angular-scale Faraday depth structures with signs that alternate only once in the Southern Galactic hemisphere and twice in the Northern hemisphere, patterns shared by other Faraday rotation datasets. DRAGONS is the first survey to achieve high Faraday depth resolution while maintaining sensitivity to broad Faraday depth structures, enabling the first use of Galactic longitude-Faraday depth plots. These plots reveal \change{Faraday-complex} structures across the sky, indicating a \change{slab-like} scenario in which emission and Faraday rotation are mixed. This complexity is overlaid on the same large-scale Faraday depth patterns that appear in the first moment map. We model these patterns as a magnetic reversal slicing through the disk on a diagonal and passing above the Sun in Galactic coordinates. We describe this reversal as a \change{plane} with a normal vector parallel to the line directed along $(\ell, b) = (168.5^\circ, -60^\circ)$ and estimate its distance to be between 0.25 and 0.55~kpc. Our results show that much of the observed Faraday sky may be dominated by the local magnetic field configuration.

 \end{abstract}


\section{Introduction} \label{sec:intro}

The Galactic magnetic field can be thought of as a large-scale magnetic field, which has coherence lengths on the order of 1~kpc, and small-scale magnetic fields with coherence lengths on the order of a few tens of parsecs \citep{Brown2001, Jaffe2010, Haverkorn2015, Beck2016}. Various models of the large-scale Galactic magnetic field describe the field as generally following the spiral arms, consistent with a logarithmic spiral, where the pitch angle decreases with increasing radius from the Galactic center \citep[e.g.,][]{Brown_2007, Sun2008, Jansson_2012, Han2017}. 

The `local' large-scale field (within 1~kpc of the Sun) is directed predominantly clockwise as viewed from the North Galactic pole, with a pitch angle of approximately 11.5$^\circ$ \citep{VanEck2011}. Models with at least one reversal between the Sun and the Sagittarius Arm, beyond which the field changes direction and spirals counterclockwise, explain the observed patterns in RMs of Galactic pulsars and extragalactic point sources \citep[e.g.,][]{SK80, Thomson80, Han1997, Brown_2007,Curtin2024}. In these observations, the reversal appears in the Galactic plane at longitude $\ell = 60^\circ$, where the line-of-sight \change{(LOS)} component of the field appears to abruptly switch direction. The physical location of this reversal is not well known, but is generally assumed to be around 0.5~kpc away in the direction of Galactic longitude $\ell = 0^\circ$ \citep[][]{Han1994, VanEck2011}, placing it in the local interstellar medium (ISM).

Using %diffuse polarization 
data from the Canadian Galactic Plane Survey, \citeauthor{Ordog2017} (\citeyear{Ordog2017}; hereafter \citetalias{Ordog2017}) showed that the position of the reversal in Galactic longitude has latitude dependence and follows a diagonal line. \citet{Ma2020} confirmed this observation. \citetalias{Ordog2017} proposed that a tilted, plane-like geometry for the reversal could explain their diagonal line, and a recent pulsar study by \citet{Curtin2024} supported this interpretation. Using pulsar distances to estimate the reversal position in different latitude ranges, they found the reversal is closer at positive Galactic latitudes, and extends farther away into the Sagittarius Arm at negative Galactic latitudes. This is what would be expected if the reversal were planar and tilted towards us in the Northern Galactic hemisphere. 

Much of our understanding of cosmic magnetism is based on observations of Faraday rotation, in which linearly polarized radiation undergoes a rotation of polarization angle when it propagates through a region containing a magnetic field and free electrons. The change in polarization angle, $\Delta \tau$, depends on wavelength squared, $\lambda^2$ (m$^2$), as well as the electron number density, $n_e$ (cm$^{-3}$), and the \change{LOS} magnetic field component, $B_\parallel$ ($\mu$G), along the path, $r$ (pc), as $\Delta \tau = \lambda^2  \phi$,  where
\begin{equation}
    \phi = 0.812 \int_\textrm{source}^\textrm{observer} n_e \, B_\parallel \, dr \textrm{ rad m}^{-2}. 
    \label{FD_def}
\end{equation}
Depending on the method of measurement, $\phi$
is referred to either as a Faraday depth \change{(FD)} or rotation measure (RM).

Earlier generations of Faraday rotation observations relied on polarization observations at only a small number of radio wavelengths. In these cases, the slope of a linear fit to a plot of $\tau$ versus $\lambda^2$ gives the RM of the particular LOS. Such a linear fit is only meaningful if there is a single dominant source of polarized emission along the path, Faraday rotating through a foreground screen, a scenario referred to as `Faraday simple'. Bright polarized point sources, such as the active galactic nuclei of external galaxies or pulsars, are excellent examples of this, and numerous RM catalogs have been published using such sources \citep[e.g.,][]{Manchester_2005, Brown_2007, Taylor2009, VanEck2021}. 

If the emission and Faraday rotation are mixed along the LOS, as in the case of diffuse Galactic synchrotron emission, a $\tau$ versus $\lambda^2$ plot will not always present as linear. In this case, the LOS ISM is `Faraday complex' as there are multiple $\phi$ values along the path. \change{An FD} spectrum is obtained by applying the Faraday synthesis technique \citep{Brentjens2005} to observations of polarized emission across a wide, and well-sampled, radio band. For a \change{Faraday-simple} LOS, the peak \change{FD} and RM are equivalent. For a \change{Faraday-complex} LOS, the \change{FD} spectrum gives the amount of polarized intensity emitted at each \change{FD}. 

While the output from Faraday synthesis is a three-dimensional \change{FD} cube, correlating \change{FD}s with physical distances is not straightforward, especially if there is a magnetic field reversal along the path. Despite this, distance is central to the concept of \change{FD} and, therefore, the three-dimensional nature of the \change{FD} spectra provides an opportunity to explore the geometry of the large-scale field. 

In this work, we adopt the \change{planar} geometry suggested by \citetalias{Ordog2017} and present a three-dimensional model to describe the magnetic field reversal within the local ISM. 
In \change{Sections 2 and 3}, we describe the data used in our analysis. In Section \change{4}, we identify the observational results that suggest that the \change{FD} patterns in our data are connected to the local large-scale magnetic field. Section \change{5} outlines the mathematical framework for our model of the three-dimensional structure of the reversal in the local ISM. In Section \change{6}, we fit our reversal model to the data, and in Section \change{7}, we discuss \change{the origin of the reversal}. Finally, Section \change{8} offers concluding remarks.

\section{Data} \label{sec:data}

We utilize data from the Global Magneto-Ionic Medium Survey (GMIMS) project \citep{Wolleben2010b, Wolleben2019, Wolleben2021,Sun2025}, which aims to map diffuse polarized emission across the entire sky over a wide frequency band using large single-dish radio telescopes. When all observations are complete, there will be six GMIMS surveys, with the Northern and Southern skies each divided into three frequency ranges: low, mid, and high. We use the Dominion Radio Astrophysical Observatory (DRAO) GMIMS of the Northern Sky (DRAGONS) survey \citep{ordog2025DRAGONS}, the low-frequency Northern GMIMS survey. We supplement the missing southern declinations in DRAGONS with data from the Southern Twenty centimeter All-sky Polarization Survey \citep[STAPS;][]{Sun2025, Raycheva2025}, the high-frequency Southern GMIMS survey, which observes 1324 to 1800 MHz.

\begin{table}[t!]
\centering
\begin{tabular}{ll}
\hline
\hline
\multicolumn{2}{c}{DRAGONS observation parameters} \\
\hline
Frequency range       &  500 to 1030 MHz \\ 
Telescope      &  DRAO-15  \\ 
Observation dates      &  June 2022 to January 2023 \\ 
Angular resolution     &  1.3$^\circ$ to 2.45$^\circ$,  \\ 
& all channels convolved to 2.45$^\circ$ \\
%Sensitivity of $Q$ and $U$         &  11\,mK \\ 
Frequency resolution          &  1 MHz  \\ 
Polarization            &  $I$, $Q$, $U$  \\ 
Calibrator                   &  Cygnus A \\ 
Sky coverage                   &  $\delta > -20^\circ$  \\
\hline
\hline
\multicolumn{2}{c}{DRAGONS Faraday synthesis parameters} \\
\hline
$\lambda^2$ range              &  0.08 $<\lambda^2 <$ 0.36 m$^2$     \\ 
$\phi$ resolution       & 14~\radmsq      \\ 
$\phi$ max-scale        & 37~\radmsq      \\ 
$\phi$ max detectable   &      1327~\radmsq \\ 
$\phi$ sampling         & 1~\radmsq      \\ 
\texttt{CLEAN} cut-off $\sigma$ range & 750 $<\phi <$ 1000~\radmsq\\
\change{Sensitivity in FD cube} & \change{5 mK~RMSF$^{-1}$}\\
%Moment 1 $\phi$ range   & $|\phi| <$ 100~\radmsq   \\
\hline
\end{tabular}
\caption{Parameters of DRAGONS data used in this paper.}
    \label{tab:survey_parameters}
\end{table}

\begin{figure}[t!]
  %\centering
  \resizebox{0.47\textwidth}{!}{%
    \begin{tabular}{l}
      \includegraphics{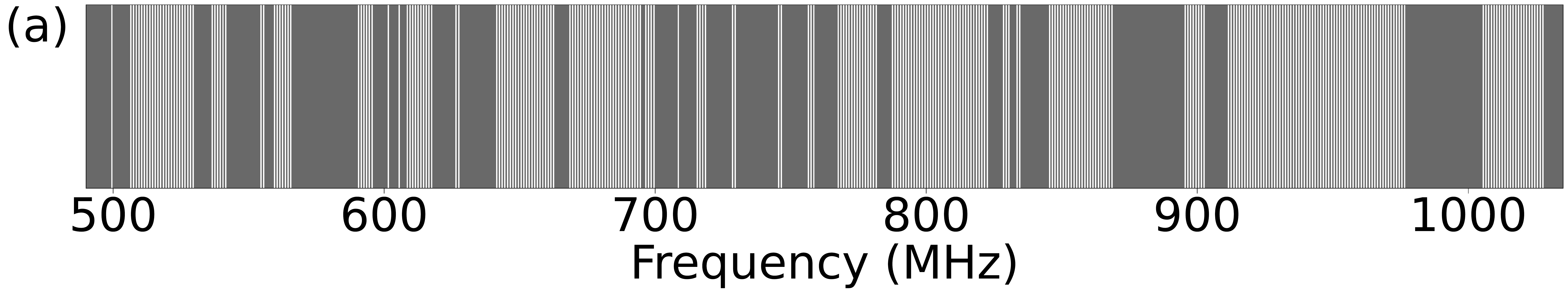} \\[50pt] % Another space
      \includegraphics{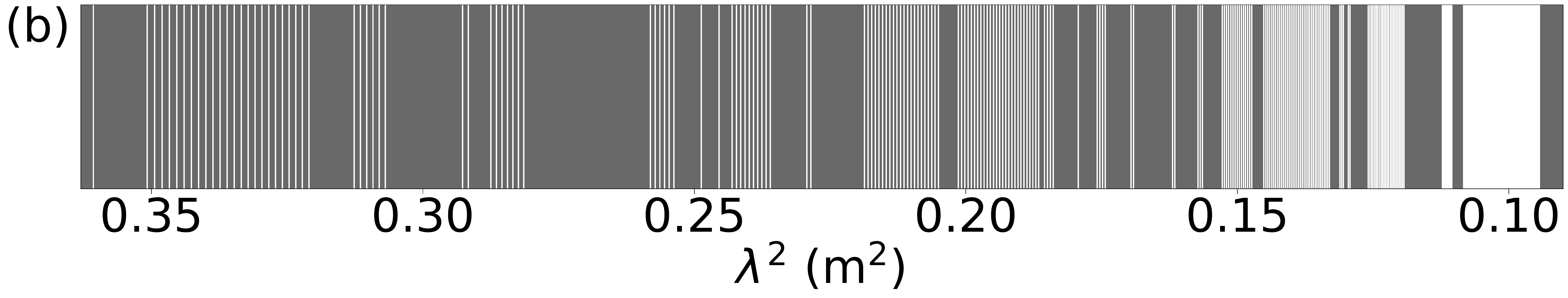} \\ [50pt]
      \includegraphics{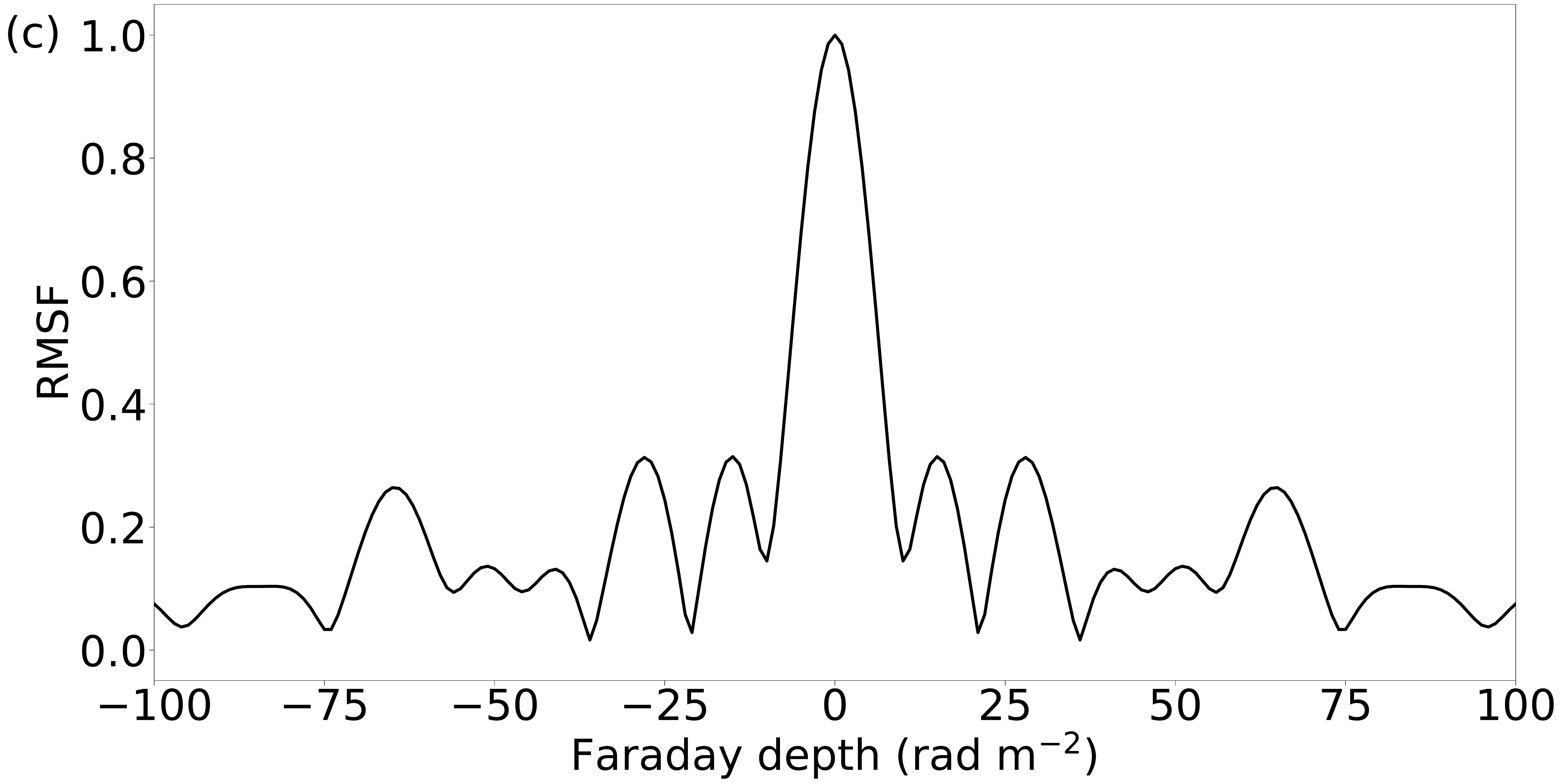} % Add 2pt vertical space
    \end{tabular}%
  }
  \caption{ Key DRAGONS parameters. (a) and (b) The DRAGONS frequency and $\lambda^2$ coverage respectively. White lines indicate the 302 frequency channels included in the survey. The dark gaps are due to channels omitted as a result of radio frequency interference. (c) The DRAGONS rotation measure spread function.% \textit{Bottom:} The DRAGONS $\lambda^2$ coverage.
  \label{fig:general}}
  \label{fig:rmsf}
\end{figure}

\subsection{DRAGONS}

The DRAGONS survey was observed between June 2022 and January 2023 using the DRAO-15 telescope, an offset paraboloid with an effective diameter of 15 m. Stokes $I$, $Q$, and $U$ observations cover frequencies 350 MHz to 1030 MHz, in declinations above $\delta= -20^\circ$ over the full range of right ascension. The observation strategy and data processing details are described in \citet{ordog2025DRAGONS}. The DRAGONS data used here span the top of the band, from 500 to 1030 MHz. Extending to 350 MHz would have yielded greater resolution in \change{FD}, at the expense of a broader beam, and we chose 500 MHz as the lowest frequency as a compromise. Our results are not sensitive to this choice.

To achieve consistent resolution across the DRAGONS survey, the maps were all convolved with a Gaussian kernel to a common angular resolution of 2.45$^\circ$, the beam size of the 500 MHz channel. We do not use DRAGONS data within $|b|<5^\circ$ as they are contaminated by the leakage of high total-power levels into Stokes \change{$Q$} and \change{$U$} near the Galactic plane.
%interaction of instrumental polarization with the high total-power levels near the Galactic plane. 

The details of 500 to 1030 MHz DRAGONS data are summarized in the top half of Table \ref{tab:survey_parameters}. The DRAGONS Stokes $I$, $Q$, and $U$ data cubes have 302 frequency channels, each 1 MHz wide, spread across the 500 MHz to 1030 MHz band. Figure \ref{fig:rmsf} (a) and (b) show the DRAGONS frequency coverage and corresponding wavelength squared ($\lambda^2$) respectively, with white bands representing channels included in the data cubes. 
Frequency channels with significant RFI contamination in the data were discarded, leaving 57\% of the band remaining. 

\subsubsection{The DRAGONS Faraday depth cube}

To apply Faraday synthesis to the DRAGONS data, we used the CIRADA \texttt{RM-Tools} package\footnote{\url{https://github.com/CIRADA-Tools/RM-Tools}} \citep{Purcell2020}. The 500 to 1030 MHz DRAGONS bandwidth provides $\lambda^2$ coverage from 0.08 to 0.36 m$^2$. Due to the finite $\lambda^2$ window, the observed \change{FD} spectrum is the convolution of the true \change{FD} spectrum with the rotation measure spread function (RMSF). The RMSF for DRAGONS is shown in Figure \ref{fig:rmsf}(c). The DRAGONS RMSF has a \change{FWHM} of $\delta \phi$ = 14~\radmsq, which determines the \change{FD} resolution of the survey. The $\lambda^2$ spacing limits the largest \change{FD} that can be detected, and the smallest $\lambda^2$ determines the maximum width of \change{FD} features that can be observed \citep{Brentjens2005,Dickey2019}. For DRAGONS, these are $\phi_{\textrm{max}}$ = 1327~rad~m$^{-2}$ and $\phi_{\textrm{max width}} = 37$~rad~m$^{-2}$ respectively. As $\phi_{\textrm{max width}}$ is significantly larger than ${\delta}{\phi}$, we can detect wide \change{FD} features; this is an exceptional feature of the DRAGONS data, one not achieved in earlier polarization surveys. 
 
The effects of the RMSF sidelobe contamination in the \change{FD} spectra can be reduced using \texttt{RM-CLEAN} \citep{heald2009}, which is also available in the \texttt{RM-Tools} package. In the first stage of Faraday synthesis, we produced `dirty' (contaminated by the RMSF sidelobes) \change{FD} spectra out to $\phi = \pm 1000$~\radmsq. Based on the \change{FD} values observed in other datasets \citep[e.g., GMIMS HBN;][]{Wolleben2021}, we did not expect \change{FD}s with such high magnitudes; however, the ends of the spectrum were used to assess the RMSF sidelobe contribution. Using the method developed by \citet{Raycheva2025}, we fit a Rayleigh distribution to the polarized intensity for  $|\phi| > 750$~\radmsq. Then, \texttt{RM-CLEAN} was applied to the dirty spectra and the \texttt{CLEAN} cut-off threshold for each LOS in the DRAGONS maps set to $3\sigma$, where $\sigma$ is the width of the fitted Rayleigh distribution. 

The DRAGONS \change{FD} cube that was prepared for this work has $\texttt{nside}=128$ in HEALPix format \citep{Gorski_2005}, with \change{an FD} sampling of 1~\radmsq. The properties of these data are summarized in the bottom half of Table \ref{tab:survey_parameters}.

\subsection{STAPS}

We use STAPS to fill missing information below the southern declination limit of DRAGONS (i.e., $\delta < -20^\circ$). STAPS was observed with the Parkes Murriyang 64-m telescope and covers the frequency range 1324 to 1800 MHz \citep{Sun2025}. STAPS and DRAGONS have no overlapping frequencies. The higher frequencies of STAPS lead to a poorer \change{FD} resolution of $\delta \phi = 140$~\radmsq, compared to DRAGONS, which has $\delta \phi = 14$~\radmsq. GMIMS Low-Band South \citep[LBS;][]{Wolleben2019}, which spans 300 to 400 MHz, was also an option to fill the missing DRAGONS declinations. However, despite being closer in frequency to the DRAGONS band used in this work, LBS is actually farther in $\lambda^2$ space than STAPS. 
For this work, we convolved the STAPS frequency cubes to 2.45$^\circ$ to give these data the same spatial resolution as DRAGONS. We then applied Faraday synthesis to the convolved STAPS cubes, following the processing steps described in \citet{Raycheva2025}.

\subsection{Supporting datasets}

To gain additional insight for our work, we utilized several existing datasets. In particular, we used the \citet{Edenhofer2024} three-dimensional dust map and the \citet{Hutschenreuter2022} RM map.

\subsubsection{Edenhofer et al. 2024 dust map}

Extinction measurements towards stars with parallax distances, determined using the European Space Agency's Gaia satellite \citep{GaiaDR1_2016}, have been employed to trace the three-dimensional interstellar dust distribution. These dust maps have emerged as a leading method to explore distances to objects in the local ISM. 
Several studies have found a correlation between Faraday rotation obervations from the LOw-Frequency ARray (LOFAR) Two-metre Sky Survey (LoTSS) and maps of interstellar dust \citep{Zaroubi2015, VanEck2017, Erceg2024b}, 
demonstrating the potential for using the three-dimensional dust maps to determine distances to Faraday rotation structures. 

Here we use the dust map of \citeauthor{Edenhofer2024} (\citeyear{Edenhofer2024}; hereafter \citetalias{Edenhofer2024}) to connect DRAGONS \change{FD} structures to distances. The \citetalias{Edenhofer2024} map is a statistical model of the interstellar dust distribution produced by applying Bayesian inference to the stellar extinction and Gaia distances of 54 million stars. The result is a three-dimensional map of dust extinction %, given in the units of \citet{Zhang_2023},
with \change{parsec-scale} resolution in logarithmically spaced distance bins from 69~pc to 1.25~kpc. 

\subsubsection{Hutschenreuter et al. 2022 rotation measure map}

We use the RM map of \citeauthor{Hutschenreuter2022} (\citeyear{Hutschenreuter2022}; hereafter \citetalias{Hutschenreuter2022}) as a complementary Faraday rotation dataset to test our magnetic field reversal model. The \citetalias{Hutschenreuter2022} map was produced using the RMs of 55190 extragalactic (EG) sources, compiled from 41 catalogs, and interpolated using Bayesian inference. The resulting map describes Faraday rotation across the entire sky along the full path through the Galaxy. 

\begin{figure*}[tb]
  \centering
  \resizebox{0.85\textwidth}{!}{%
    \begin{tabular}{c}
      \includegraphics{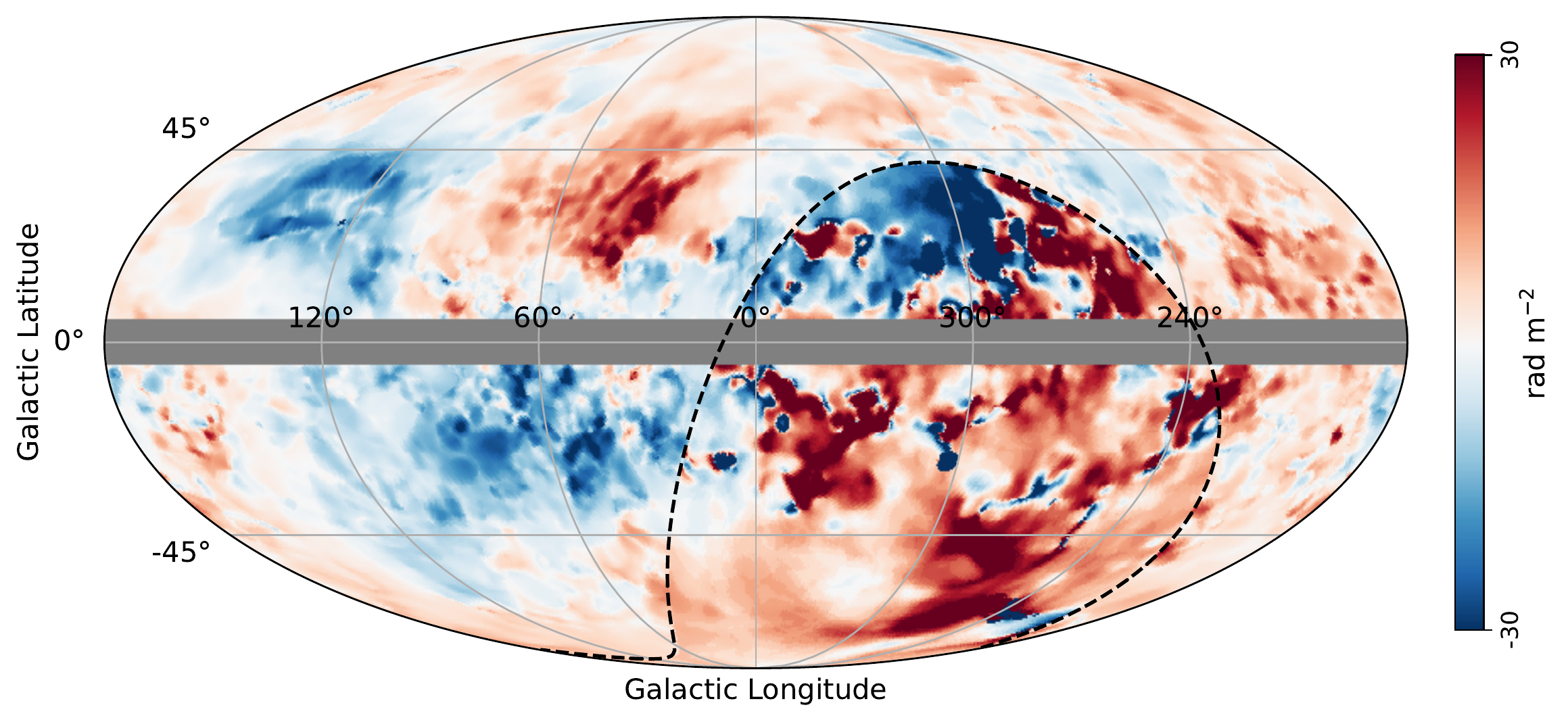} \\ 
    \end{tabular}%
  }
  \caption{DRAGONS and STAPS \change{FD} moment 1 combined. The STAPS data were convolved to the DRAGONS spatial resolution of 2.45$^\circ$. The dashed line indicates the boundary between the two datasets. We have masked out the Galactic plane, within $|b| < 5^\circ$, where instrumental effects make the data unreliable. %We note that the STAPS moment 1 values generally have higher magnitudes than DRAGONS. This is likely due to the farther STAPS polarization horizon that arises from its higher frequencies.
  }
\label{fig:STAPS_M1}
\end{figure*}

\begin{figure}[t!]
  \centering
  \resizebox{0.47\textwidth}{!}{%
      \includegraphics{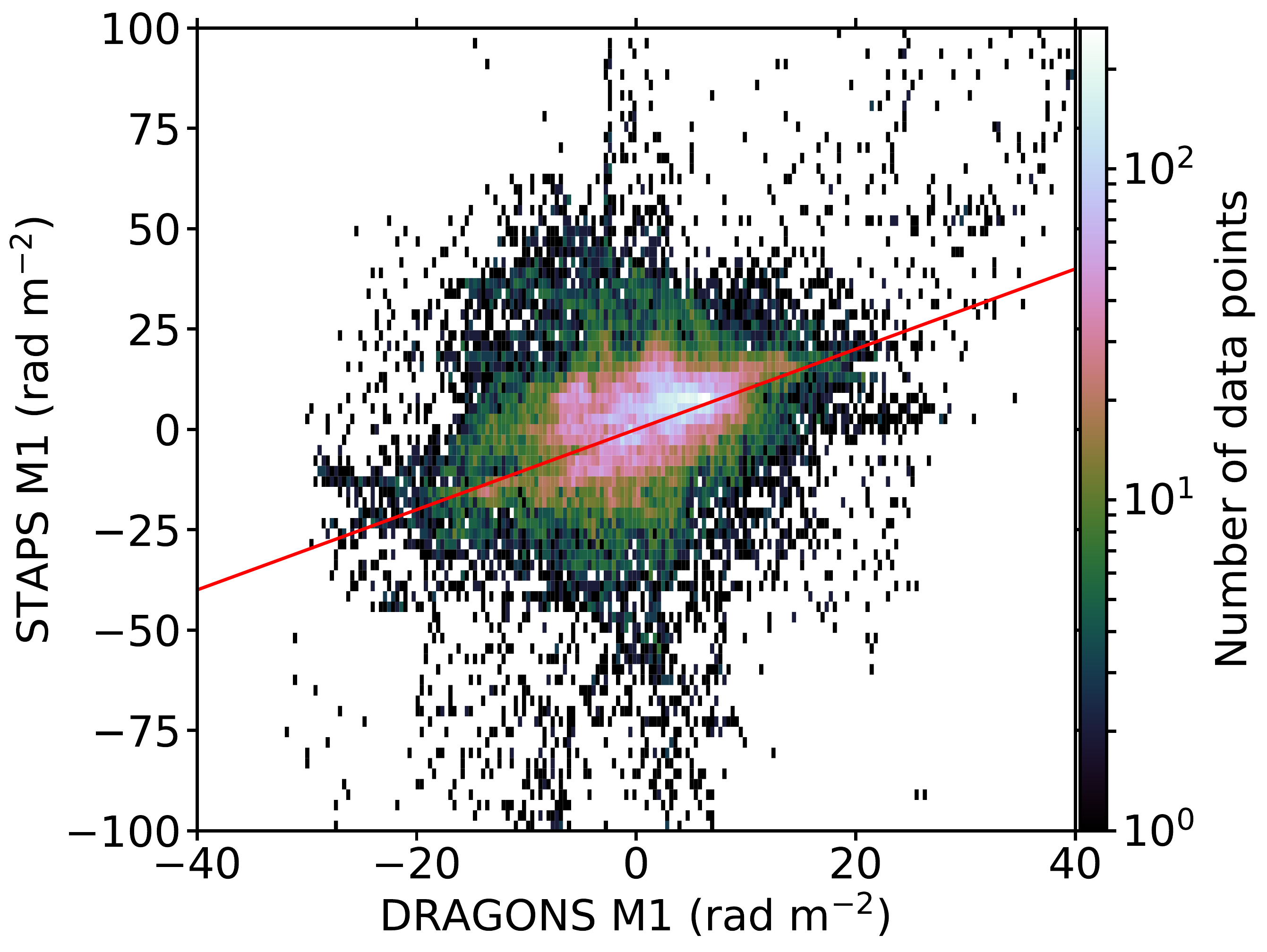} 
  }
  \caption{Comparison of STAPS M1 (vertical axis) to DRAGONS M1 (horizontal axis) in the region where the two surveys overlap. The red line is the 1:1 line where STAPS M1 $=$ DRAGONS M1.}
  \label{fig:ST_DR_compare}
\end{figure}

\section{The first moment of the DRAGONS and STAPS Faraday depth cubes}

Calculation of the first moment of the \change{FD} spectrum enables us to collapse the information contained in three-dimensional \change{FD} cubes into two-dimensional maps \citep{Dickey2019}. For each LOS, the first moment (M1) is calculated as,
\begin{equation}
    \textrm{M1} = \frac{\sum |\tilde{P_i}| \phi_i}{\sum{|\tilde{P_i}| }},
    \label{eq:M1}
\end{equation}
where $\tilde{P_i}$ is the complex polarized intensity of the $i$th \change{FD} channel, $\phi_i$. The combined DRAGONS and convolved STAPS M1 map is shown in Figure \ref{fig:STAPS_M1}. \change{FD} channels where polarized intensity is below a $6\sigma$ threshold were ignored in the summation, where $\sigma$ is the width of the fitted Rayleigh distribution in each pixel that was determined in the Faraday synthesis stage. For DRAGONS, we calculated the M1 summation in the range $|\phi| < 100$~rad~m$^{-2}$ as there is no significant polarised intensity in DRAGONS for \change{FD}s beyond this range. For STAPS, we used the full $\phi$ range, out to $|\phi| <$ 1000~\radmsq. 

 The dashed line in Figure \ref{fig:STAPS_M1} indicates the $\delta = -20^\circ$ boundary below which there are no DRAGONS observations and STAPS M1 was used. In the overlap region between the two datasets, $-20^\circ < \delta < 0^\circ$, only DRAGONS values are included, as DRAGONS is the primary dataset being considered in this work. A comparison of STAPS M1 to DRAGONS M1 in the overlap region is shown in Figure \ref{fig:ST_DR_compare}. In this region, approximately 30\% of STAPS and DRAGONS M1 values are within 10\% of each other. For the other 70\% of the data points, STAPS M1 tends to be considerably higher in magnitude than DRAGONS M1. This is consistent with visual inspection of Figure \ref{fig:STAPS_M1}, where the M1 values are generally darker across the region covered by STAPS. This discrepancy, where STAPS largely presents with higher \change{FD}s than DRAGONS, is likely due to depolarization effects. 

 \subsection{The DRAGONS polarization horizon}

The resolution of the single dish restricts our observations to diffuse synchrotron emission, and depolarization effects limit the distance probed to a finite polarization horizon \citep{Gaensler2001,Uyaniker2003,Hill2018}. Depolarization arises when a superposition of polarization angles results in lower observed polarized intensity than originally emitted. This may occur if synchrotron emission from different depths is emitted at different angles (geometric depolarization) or experiences varying amounts of Faraday rotation along the path, causing different angles to emerge at the point of observation (depth depolarization). In addition, there may be multiple \change{FD}s or emitted polarization angles distributed across the telescope beam (beam depolarization). We expect that depth and beam depolarization are dominant in DRAGONS, as Faraday rotation is enhanced at low frequencies, and the DRAGONS beam is broad.

\begin{figure*}[t!]
  \centering
  \resizebox{1\textwidth}{!}{%
  \begin{tabular}{cc}
      \includegraphics{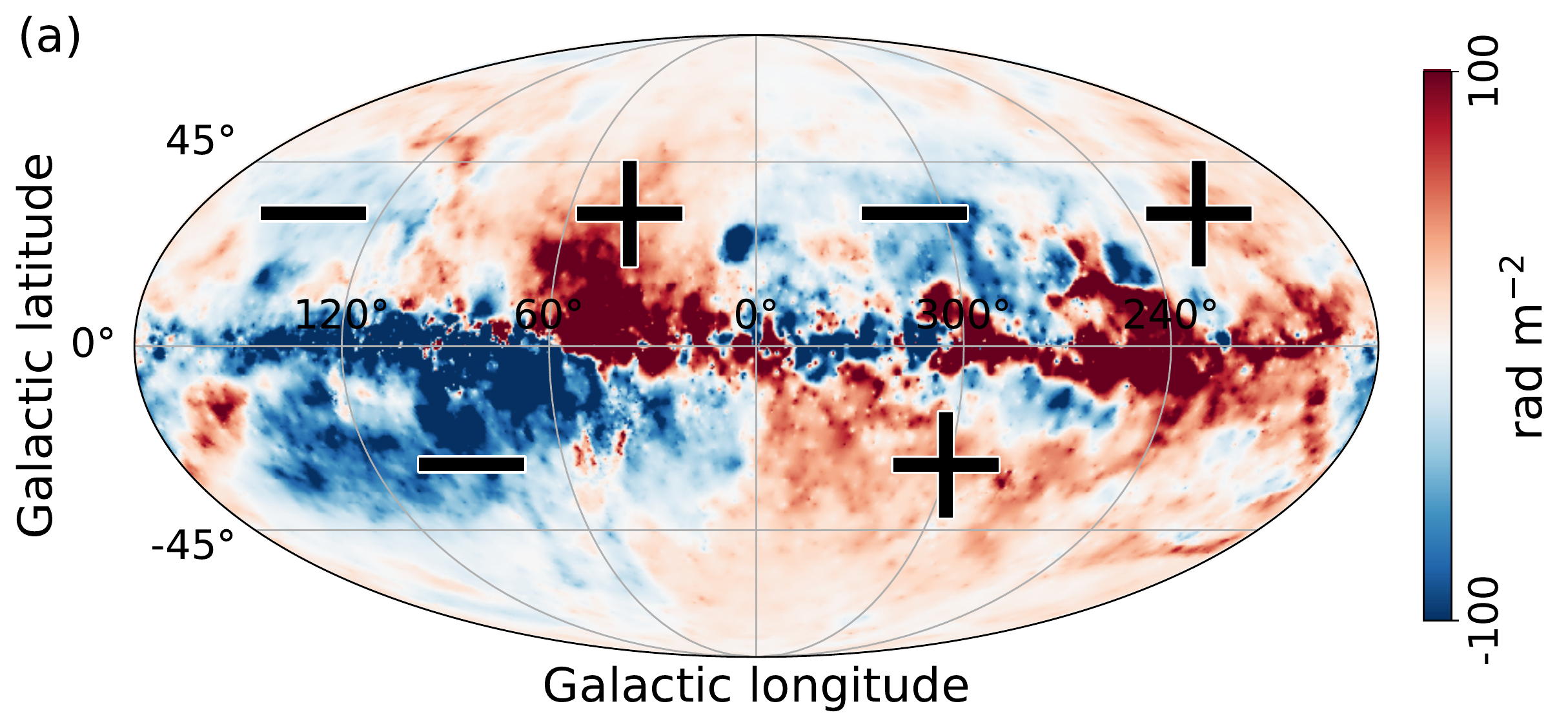} & 
      \includegraphics{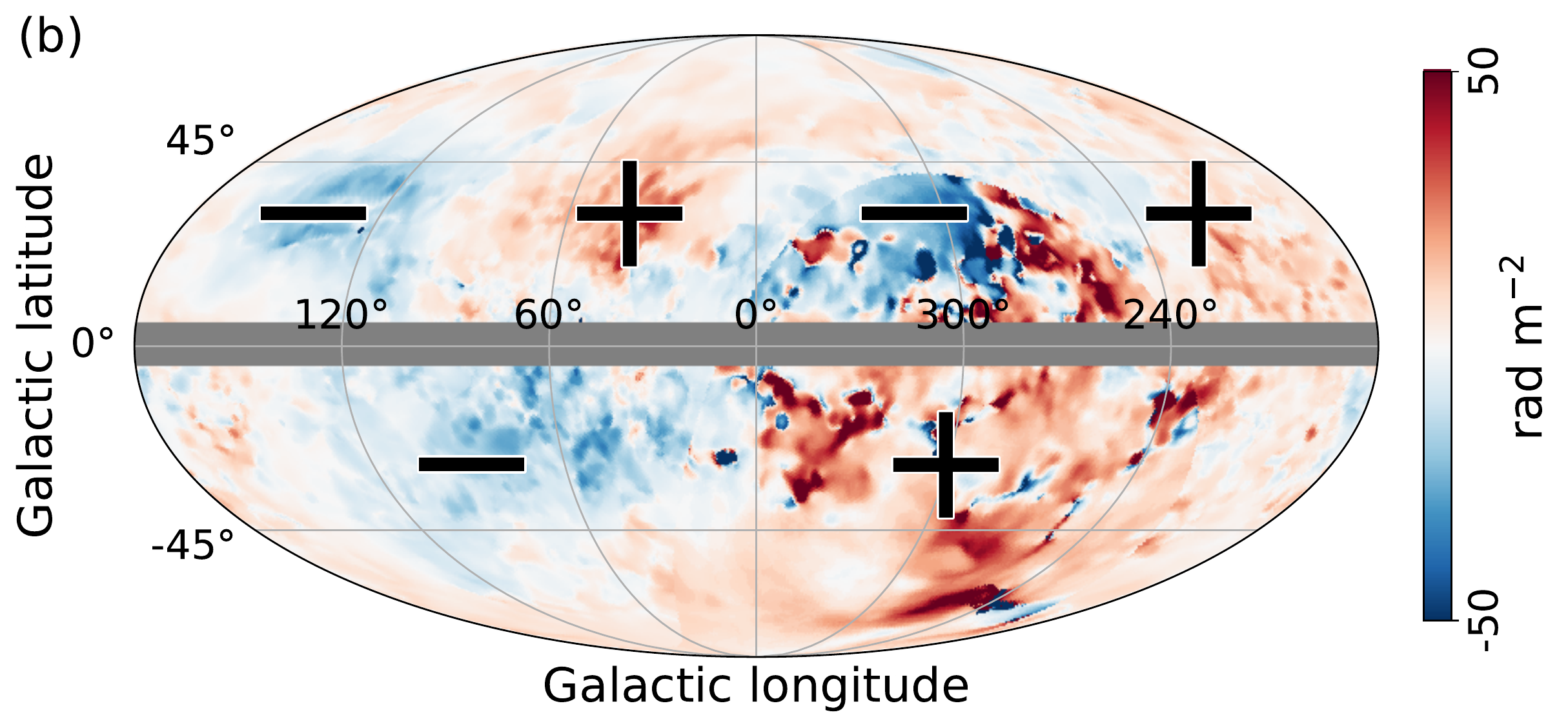} 
      \end{tabular}
  }
  \caption{(a) The peaks of the $\sin\ell$ and $\sin 2\ell$ patterns marked with $+$ and $-$ signs plotted on the \citetalias{Hutschenreuter2022} map. (b) The same $+$ and $-$ peaks marked on the combined DRAGONS and STAPS moment 1 map. 
\label{fig:SH}}
\end{figure*}

Through comparisons with pulsar RMs, \citet{Dickey2019} estimated a polarization horizon averaging around 700~pc for GMIMS High Band North \citep[HBN;][]{Wolleben2021}, which covers a frequency range from 1270 to 1750 MHz and has an angular resolution of 40$'$. In contrast, for GMIMS LBS, with frequencies spanning 300 to 480 MHz and an angular resolution of 80$'$, they estimated an average polarization horizon around 300~pc. The DRAGONS band lies between the HBN and LBS frequency ranges, and based on depth depolarization, we would expect DRAGONS to be sensitive to emission originating at distances between the HBN and LBS limits, around 500~pc. With its coarser angular resolution, beam depolarization may play a significant role in DRAGONS, and therefore, the average polarization horizon may be even closer. Given this expected nearby polarization horizon, DRAGONS is well-suited to explore the local magnetic environment.
 
Depth depolarization is reduced at the shorter wavelengths of STAPS, and the polarization horizon for that dataset is likely at a greater distance. However, even if STAPS `sees' farther than DRAGONS, the emission recorded in the STAPS data has passed through the same volume that DRAGONS is sensitive to. The Faraday rotation information contained in STAPS is not completely independent of DRAGONS, and therefore, the datasets will still complement each other.

\section{Analysis of Observations} \label{sec:results}

In this section, we first identify the large angular scale patterns present across the DRAGONS and STAPS M1 maps. Then, we compare the morphology of structures in DRAGONS M1 to regions in the \citetalias{Edenhofer2024} dust map in order to estimate distances to the observed \change{Faraday-rotation} patterns. 

\begin{figure*}[tb]
  %\centering
  \resizebox{0.98\textwidth}{!}{
      \includegraphics{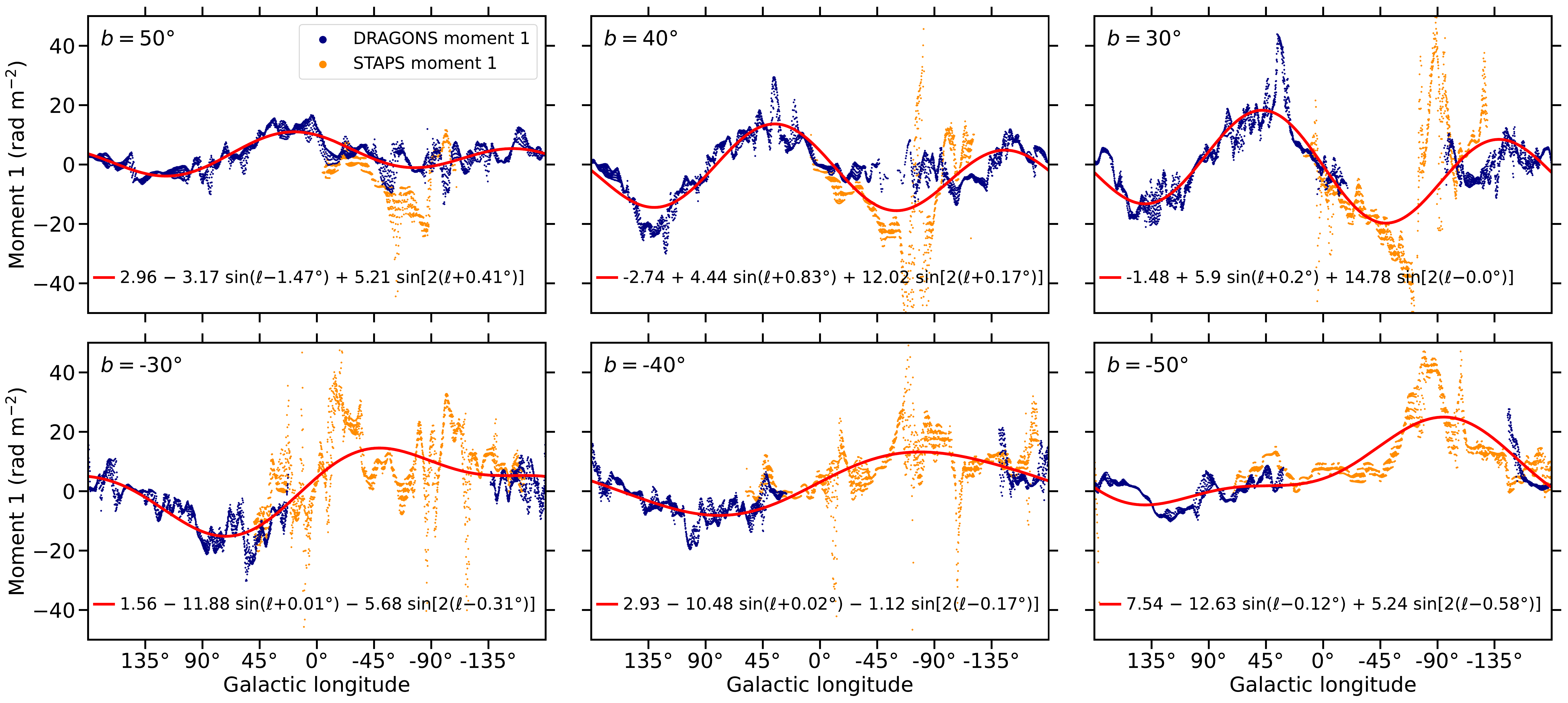}
  }
  \caption{DRAGONS (blue) and STAPS (orange) moment 1 values plotted along lines of constant latitude. Sinusoids fitted to the combined DRAGONS and STAPS moment 1 map (Figure \ref{fig:STAPS_M1}) along the corresponding latitudes are plotted in red.}
  \label{fig:sin2l_lat}
\end{figure*}

\subsection{The Southern $\sin\ell$ and Northern $\sin 2\ell$ patterns}
\label{sin2l}

Many studies of different Faraday rotation datasets have found a common pattern on large angular scales across the sky \citep[e.g.,][]{Han1997, Dickey2022}. In mid latitudes in the Southern Galactic hemisphere ($b < 0^\circ$), the \change{FD}s change signs only once as a function of longitude, appearing as ($-$ $+$) when centered at $\ell =0^\circ$, while at mid latitudes in the Northern Galactic hemisphere ($b > 0^\circ$), the sign changes twice, appearing as ($-$ $+$ $-$ $+$). This is shown in Figure \ref{fig:SH}, which displays these patterns marked on the \citetalias{Hutschenreuter2022} \change{map} and the combined DRAGONS and STAPS M1 maps. 

\citet{Dickey2022} fitted a Fourier sine series,
\begin{equation}
    \phi = C_0 + C_1\sin(\ell + \phi_1) + C_2\sin(2[\ell + \phi_2]),
    \label{Fourier}
\end{equation}
in Galactic longitude, $\ell$, to GMIMS HBN M1 and the \citetalias{Hutschenreuter2022} map along lines of constant Galactic latitude. They showed that the $\sin\ell$ term dominates the Southern ($-$ $+$) pattern, while the Northern ($-$ $+$ $-$ $+$) pattern is better described by the $\sin2\ell$ term.

Examining Figure \ref{fig:SH} by eye, the $\sin\ell$ and $\sin2\ell$ patterns appear to also be present in the DRAGONS and STAPS combined M1 map. While the magnitudes of the \change{FD}s are smaller in DRAGONS and STAPS than \citetalias{Hutschenreuter2022}, the positions of the large positive and negative regions appear to peak at roughly the same coordinates. 

To mathematically verify this observation, we followed the methods of \citet{Dickey2022} and used \texttt{scipy.optimize.curve\_fit} \citep[hereafter \texttt{curve\_fit};][]{Virtanen2020} to fit Equation \ref{Fourier} to the DRAGONS and STAPS combined M1 map as a function of Galactic longitude, within 3$^\circ$ wide latitude bins. The fitting results for three Northern latitude ranges, with bins centered at $b=50^\circ$, $40^\circ$, and $30^\circ$, and three Southern latitude ranges, centered at $b=-30^\circ$, $-40^\circ$, and $-50^\circ$, are shown in Figure \ref{fig:sin2l_lat}. The higher magnitude STAPS M1 values do tend to skew the data in the fourth quadrant, creating uncertainty in the fitting results. However, omitting STAPS and just fitting to DRAGONS led to poor convergence, as \texttt{curve\_fit} was free to assume any shape across the wide gap where there is no DRAGONS data. 

From the \texttt{curve\_fit} results, we find that the $\sin2\ell$ coefficient, $C_2$, of the fitted sinusoids is approximately twice as large as the $\sin\ell$ coefficient, $C_1$, in the Northern latitudes. In the Southern latitudes, $C_1$ is consistently more than twice $C_2$. This is confirmation that the Northern $\sin2\ell$ and Southern $\sin\ell$ patterns are present in both DRAGONS and STAPS.

It is notable that the $\sin\ell$ and $\sin2\ell$ patterns persist across datasets sensitive to different distance ranges. \citet{Dickey2022} identified them in GMIMS HBN and \citetalias{Hutschenreuter2022}, and we find them in DRAGONS and STAPS. The most significant difference in how the patterns present in these datasets is the magnitude of the \change{FD}s, suggesting integration along different path lengths through the same large-scale field configuration. As the persistence of this trend across data with different polarization horizons implies a coherence in distance, we consider this evidence that the $\sin\ell$ and $\sin2\ell$ patterns may result from Faraday rotation through the large-scale Galactic magnetic field. We will explore this possibility in Section \ref{sec:reversalplane} with our planar reversal model.

\subsection{The distance to the $\sin2\ell$ pattern in DRAGONS}
\label{sec:FDregions}

In several magnetic field models, the $\sin\ell$ pattern has been related to the large-scale disk field, and the $\sin2\ell$ pattern to a more distant field in the Galactic halo, transitioning around a height of $z\approx0.8$~kpc above the Galactic mid-plane \citep[e.g.,][]{Sun2008, Pshirkov_2011, Unger2024, Korochkin2025}. Others have argued that it is possible for the magnetic field in the shell of the Local Bubble alone to produce such sinusoidal patterns \citep[e.g.,][]{Maconi2025}. Dust polarisation has been shown to correlate with Faraday rotation structures \citep{Zaroubi2015}, and \citet{Halal2024} found that the magnetic field in dust beyond the Local Bubble contributes to the measured dust polarization fraction on large scales. This suggests that we can expect fields beyond the Local Bubble to contribute to the large-scale \change{FD} patterns we observe across the sky. Here, we present evidence that at least some of the magnetic field configuration that produces the $\sin2\ell$ pattern in DRAGONS is located beyond the Local Bubble, but still within the local ISM.

\begin{figure*}[t!]
  \centering
  \resizebox{0.85\textwidth}{!}{%
      \includegraphics{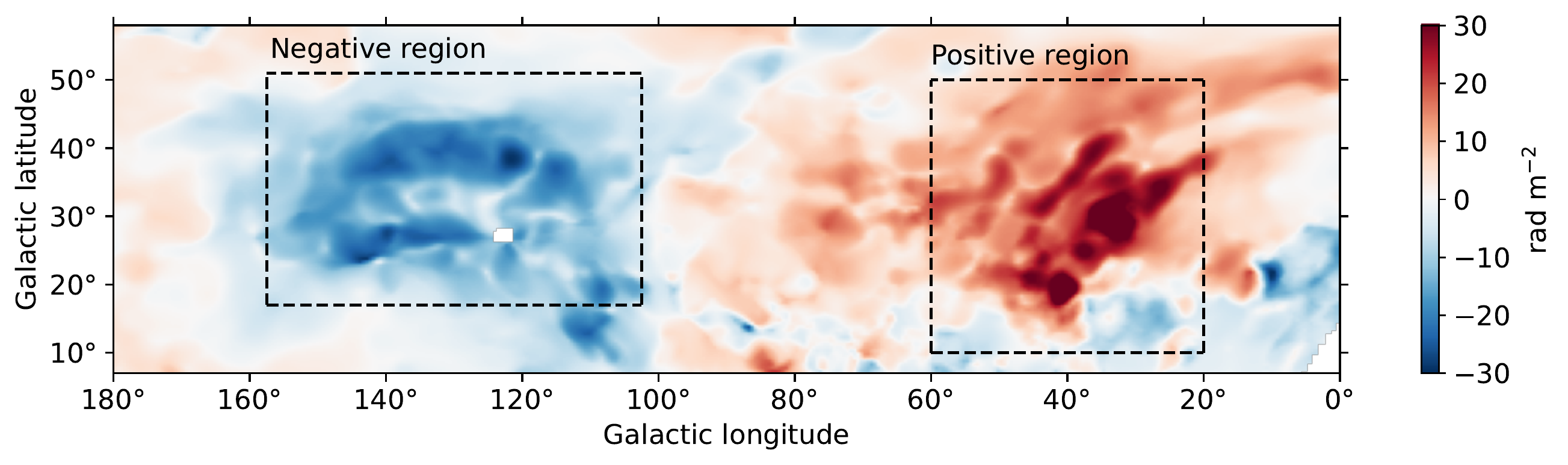}
  }
  \caption{A cut-out of DRAGONS M1 displaying the two regions with the highest \change{FD}s in the survey.}
  \label{fig:dust_compare}
\end{figure*}

There are two regions in DRAGONS M1 that stand out as having the largest negative (centered at $\ell = 130^\circ,\, b = 35^\circ$) and positive (centered at $\ell = 40^\circ,\, b = 30^\circ$) \change{FD}s in the survey. They are positioned at adjacent peaks in the $\sin2\ell$ pattern, and we show them in a cut-out of the DRAGONS M1 map in Figure \ref{fig:dust_compare}. As Faraday rotation occurs along the entire LOS, out to the polarization horizon, it is not always possible to associate individual distances with \change{FD} observations. However, the negative and positive regions identified here have distinct morphologies with clear boundaries. This suggests that magnetic field or electron density enhancements from discrete structures in the ISM may dominate the Faraday rotation along these lines of sight. By identifying the associated ISM structures, we can estimate a distance for these regions.

To identify ISM structures that may be related to the negative and positive M1 regions, we utilized the \citetalias{Edenhofer2024} dust map and applied the Histogram of Oriented Gradients (HOG) method, using the tools available in the AstroHOG library\footnote{\href{https://github.com/solerjuan/astroHOG}{\change{https://github.com/solerjuan/astroHOG}}} \citep{Soler2019a}. AstroHOG compares two images by computing the two-dimensional spatial gradient of each and performing a statistical analysis of the distribution of gradient angle differences across the images, quantifying their morphological agreement.

\begin{figure*}[t!]
  \centering
  \resizebox{1\textwidth}{!}{%
  \begin{tabular}{c}
      \includegraphics{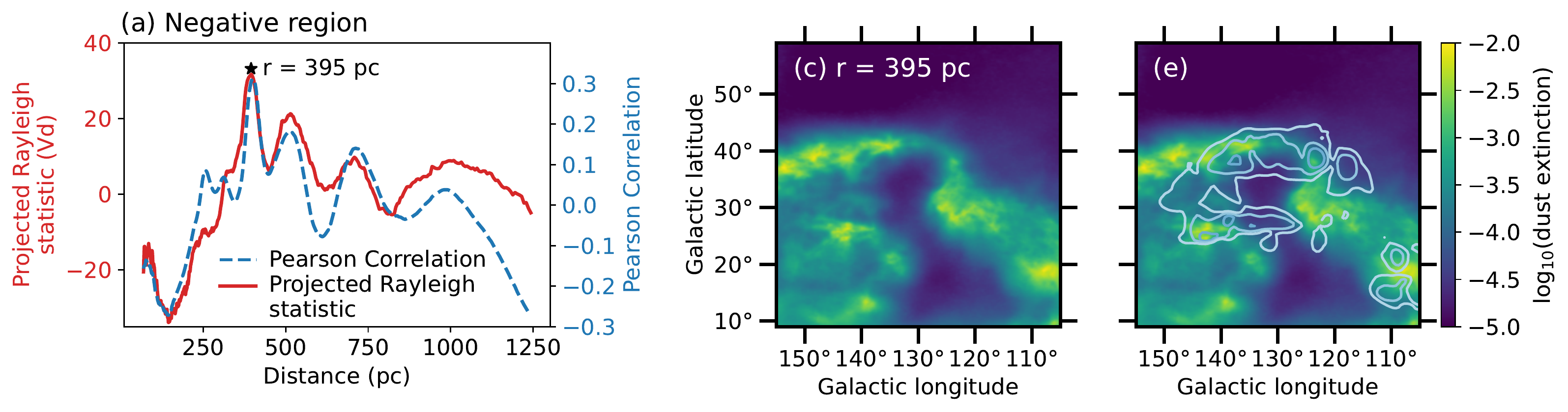}\\
      \includegraphics{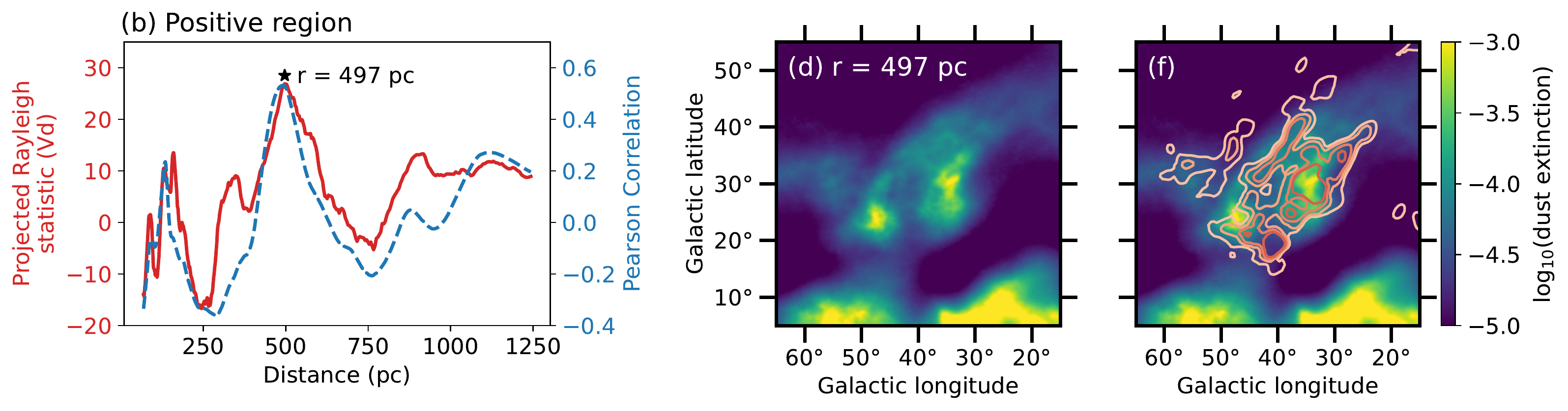}
\end{tabular}
  }
  \caption{The results of comparing DRAGONS M1 for the two high-Faraday-depth regions to dust at each distance in the \citetalias{Edenhofer2024} map using the AstroHOG \citep{Soler2019a}. In (a) and (b) selected statistics as a function of distance are plotted, with the projected Rayleigh statistic ($V_d$) in red and the Pearson correlation in blue. For both regions, $V_d$ and the Pearson correlation coefficient peak at a preferred distance marked with a star, and we include the dust map for that distance, with the same map plotted twice in the middle, (c) and (d), and right, (e) and (f), columns. The dust images in (e) and (f) have the DRAGONS M1 contours overlaid, with contours at $|$M1$|$ = 15, 20, 25, and 30~\radmsq.}
  \label{fig:HOG_result}
\end{figure*}

We rotated the DRAGONS M1 maps to obtain two-dimensional images with 0.5$^\circ$ wide pixels centered on each region of interest to prevent spurious correlations due to projection effects. For the negative region, we centered the images at $\ell, b$ = 130$^\circ$, 34$^\circ$, and cropped them to 55$^\circ$ by 30$^\circ$. For the positive region, we centered at $\ell, b$ = 40$^\circ$, 30$^\circ$, and cropped to 40$^\circ$ by 40$^\circ$. We then sampled the \citetalias{Edenhofer2024} map out to 1.25~kpc, using the same logarithmically spaced distance bins as \citetalias{Edenhofer2024}, and the same Galactic coordinate grids as the DRAGONS M1 images. We used AstroHOG to compare DRAGONS M1 to the \citetalias{Edenhofer2024} map for each distance, using a 2.5$^\circ$ Gaussian kernel. Figures \ref{fig:HOG_result} (a) and (b) display the projected Rayleigh statistic for the gradient vector direction, $V_d$ \citep{Jow2018, Soler_2025}, as a function of distance for the two regions. The $V_d$ statistic is positive when the vector angles in the two images are preferentially more parallel and negative when they are more perpendicular. We also show the Pearson correlation coefficient as a function of distance, which assesses the linear correlation of pixel brightness. 

In Figure \ref{fig:HOG_result}, for both the negative M1 region (a) and the positive M1 region (b), the $V_d$ and Pearson correlation peak together. For the negative region, they have a common peak at $d = 395$~pc, and for the positive, they peak at $d = 497$~pc. This demonstrates strong morphological agreement between DRAGONS M1 and the dust at these distances, as not only do the pixel brightnesses correlate (Pearson coefficient), the orientations of the shapes in the images also agree ($V_d$). We show the \citetalias{Edenhofer2024} dust images\footnote{We plot $\log_{10}$ of dust extinction, which is recorded by \citetalias{Edenhofer2024} in the units of \citet{Zhang_2023}, and can be translated to extinction at different wavelengths using their extinction curve.} for these distances in Figures \ref{fig:HOG_result} (c) and (d). The dust feature in (c), corresponding to the negative \change{FD} region, is the North Celestial Pole Loop (NCPL), a well-studied dust and \ion{H}{1} cavity \citep{Heiles1989, Meyerdierks1991,Marchal2023}. The region of strongly positive \change{FD}s was identified by \citet{Wolleben2010a}, who associated it with an \ion{H}{1} bubble. 

On Figures \ref{fig:HOG_result} (e) and (f), we include the DRAGONS M1 contours for $|$M1$|$ = 15, 20, 25, and 30~rad~m$^{-2}$ overlaid on the dust images. From this, we are able to see a strong morphological match by eye between the regions in M1 and dust. In both cases, it is notable that some of the M1 structures with the largest \change{FD}s match with the dark gaps adjacent to regions of dust (e.g., the arch of positive \change{FD} centered at $\ell = 30^\circ$ and $b=35^\circ$). This is expected, as the dust extinction and Faraday rotation from the same region may not have exact spatial coincidence, but may be instead layered, as different regions of the cloud are exposed differently to sources of ionization. It has been shown that Faraday rotation is particularly sensitive to the ionized and partially ionized ISM phases in ways that other tracers may not be able to detect \citep[e.g.,][]{Uyaniker2003, Mohammed2024}. The two regions identified here are examples of how the interconnectedness of the different ISM phases makes multi-ISM tracer analysis important for capturing a complete picture. 

In this work, we are interested in determining if a reversal in the local large-scale field can produce the observed $\sin2\ell$ pattern in the Northern sky. The two M1 regions that we have identified here contribute to this pattern, and by associating their Faraday rotation with dust structures, we have estimated their distances to be 395~pc and 497~pc for the negative and positive regions, respectively. At a latitude of $b=30^\circ$, these distances represent heights of approximately 250~pc above the Galactic mid-plane, confirming that the associated Faraday rotation occurs within the local disk ISM. From this, we conclude that at least some of the DRAGONS portion of the $\sin2\ell$ pattern emerges from the local disk.

Some uncertainty does remain as to whether the field direction traced by the regions is connected to the large-scale field. The observed \change{FD} enhancements have likely emerged due to conditions local to the regions themselves. The NCPL, for example, has been shown to be host to an unusually strong magnetic field \citep{Heiles1989}, likely resulting from the compression of the ambient magnetic field as the object expanded. However, we argue that their \change{FD}s are still informed by the large-scale field in which they are embedded. First, while the two regions have higher \change{FD} magnitudes than their surroundings, they share the same sign as the surroundings, suggesting that the ambient LOS field direction was preserved during formation. Second, the \change{FD} sign is uniform across each region. If the field lines had been significantly redirected or warped by small-scale perturbations, such as an expanding shock or a dense cloud moving along the LOS, we would expect to see the sign change across the regions as the magnetic field lines that were carried away return to their initial orientation farther along in the ISM. Though we cannot completely eliminate the possibility that the $\sin2\ell$ pattern is generated by randomly oriented fields due to small-scale perturbations, in the modeling that follows, we demonstrate that a reversal in the local large-scale field is able to reproduce this observed Faraday rotation pattern. 

\section{A model for the reversal in the local large-scale field} \label{sec:reversalplane} % methods

The key features of the $\sin\ell$ and $\sin2\ell$ patterns are the changes in \change{FD} sign on large angular scales over the sky. %, once from negative to positive in the South, and twice in the North. 
\change{An FD} sign change results when the LOS component of the magnetic field changes direction relative to the observer. This may transpire in one of two ways:
\begin{enumerate}
    \item A geometric sign change occurs when the physical direction of the magnetic field remains the same, but our perspective of it changes as we scan around in Galactic longitude.
    \item A magnetic shear across which the magnetic field physically switches direction.
\end{enumerate}

The Southern $\sin\ell$ pattern can be explained by the first case, in which the large-scale field spirals clockwise about the Galactic center as viewed from above the North Galactic pole. The LOS component of such a field can be modeled as\footnote{\change{Note that in the coordinate system where the observer is at the origin, as depicted in Figure \ref{fig:reversal_diagrams}, the LOS component of the magnetic field is positive when directed away from the observer. The convention of integrating from source to observer, as in Equation \ref{FD_def}, reverses the FD sign such that FD is positive when the LOS field is directed towards the observer.}}
\begin{equation}
    B_\parallel = |B| \,\bigg\{ \change{-} a \, \cos b \, \cos \beta \, \sin(\ell + p - \alpha) + \sin b \, \sin\beta \bigg\},
    \label{eq:Bparallel}
\end{equation}
where $|B|$ is the total field strength, $a$ is $+1$ ($-1$) in the case of a clockwise %\footnote{All directions indicated are as viewed from the North Galactic pole.}
(counterclockwise) field, and $p$ is the field pitch angle. We take the Galactocentric azimuthal angle, $\alpha$, to be $0^\circ$ in the direction away from the Sun and increasing in the direction of increasing Galactic longitude. To calculate $\alpha$, we use a Galactocentric distance $r_{GC}$ of 8.15~kpc \citep{Reid2019}. The parameter $\beta$ describes the vertical tilt angle between the horizontal Galactic plane and the field vector.

The $B_\parallel$ described by Equation \ref{eq:Bparallel} follows a simple $\sin\ell$ relationship, shifting horizontally along the longitude axis with $p$ and $\alpha$, and with a midline that increases or decreases with latitude, $b$, and the vertical tilt, $\beta$. Thus, the $\sin\ell$ pattern emerges from geometry alone, as we observe the same field from different directions.

The geometric case cannot produce two sign changes around $360^\circ$ of Galactic longitude. Therefore, a field reversal is needed to reproduce the $\sin2\ell$ pattern in the Northern sky. In the case of the large-scale field, such a reversal has been known to exist for decades, having been first identified in pulsar RMs \citep{Thomson80} and EG source RMs \citep{SK80}. More recently, the three-dimensional and large-scale nature of the reversal was confirmed with a larger set of pulsar RMs by \citet{Curtin2024}. These observations showed that a magnetic shear exists somewhere between the Local Arm and the Sagittarius Arm. We now investigate how this reversal might be extended to higher latitudes within the local disk ISM.

\begin{figure}[t!]
  \centering
  \resizebox{0.47\textwidth}{!}{%
      \includegraphics{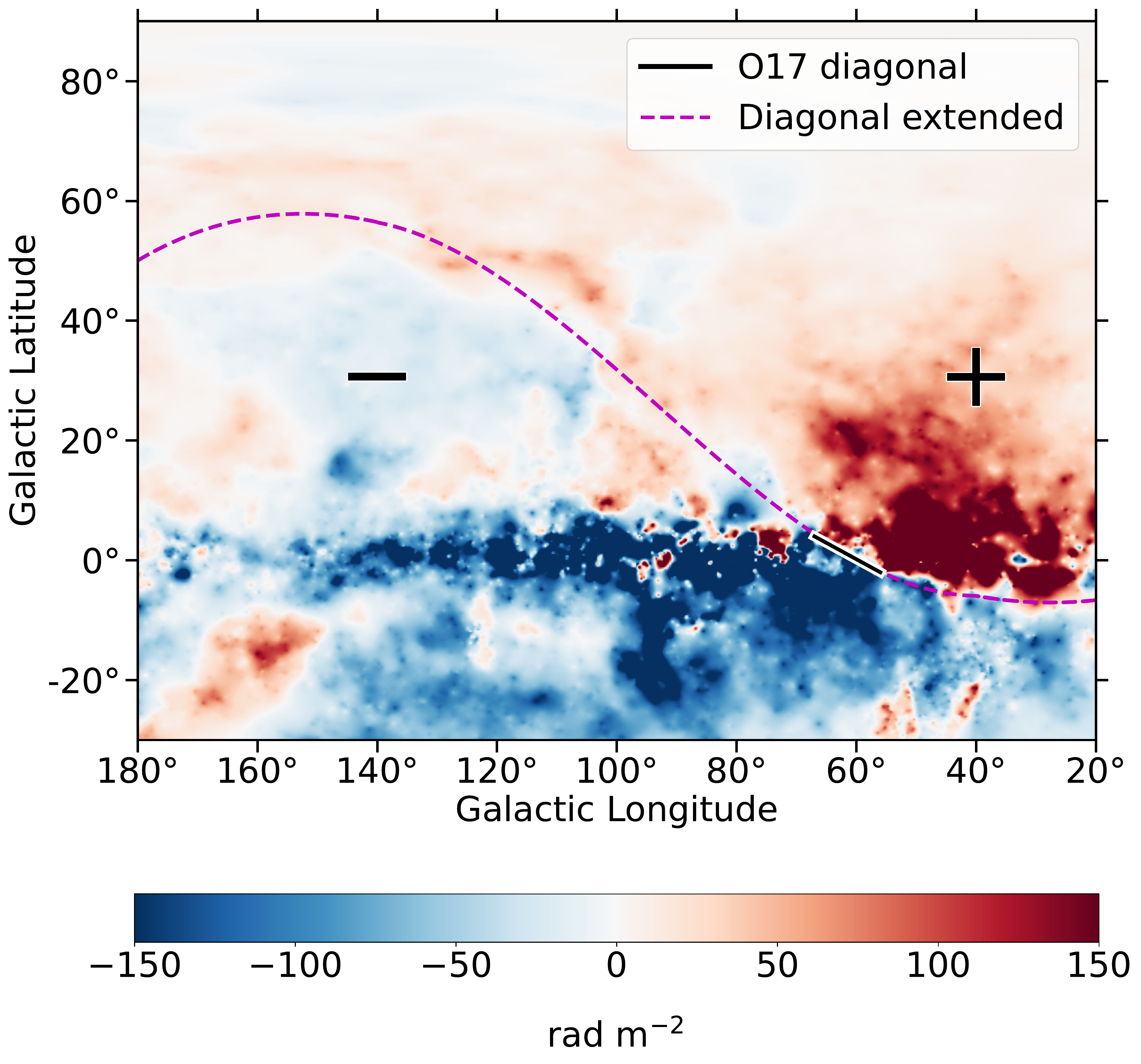} 
  }
  \caption{The \citetalias{Ordog2017} diagonal (black) plotted on the \citetalias{Hutschenreuter2022} map. The dashed magenta line demonstrates how a curved extension of the diagonal separates adjacent peaks of the $\sin2\ell$ pattern, marked with $-$ and $+$ signs.}
  \label{fig:ordog_diag}
\end{figure}

\begin{figure*}[tb]
  \centering
    
      \resizebox{0.85\textwidth}{!}{\includegraphics{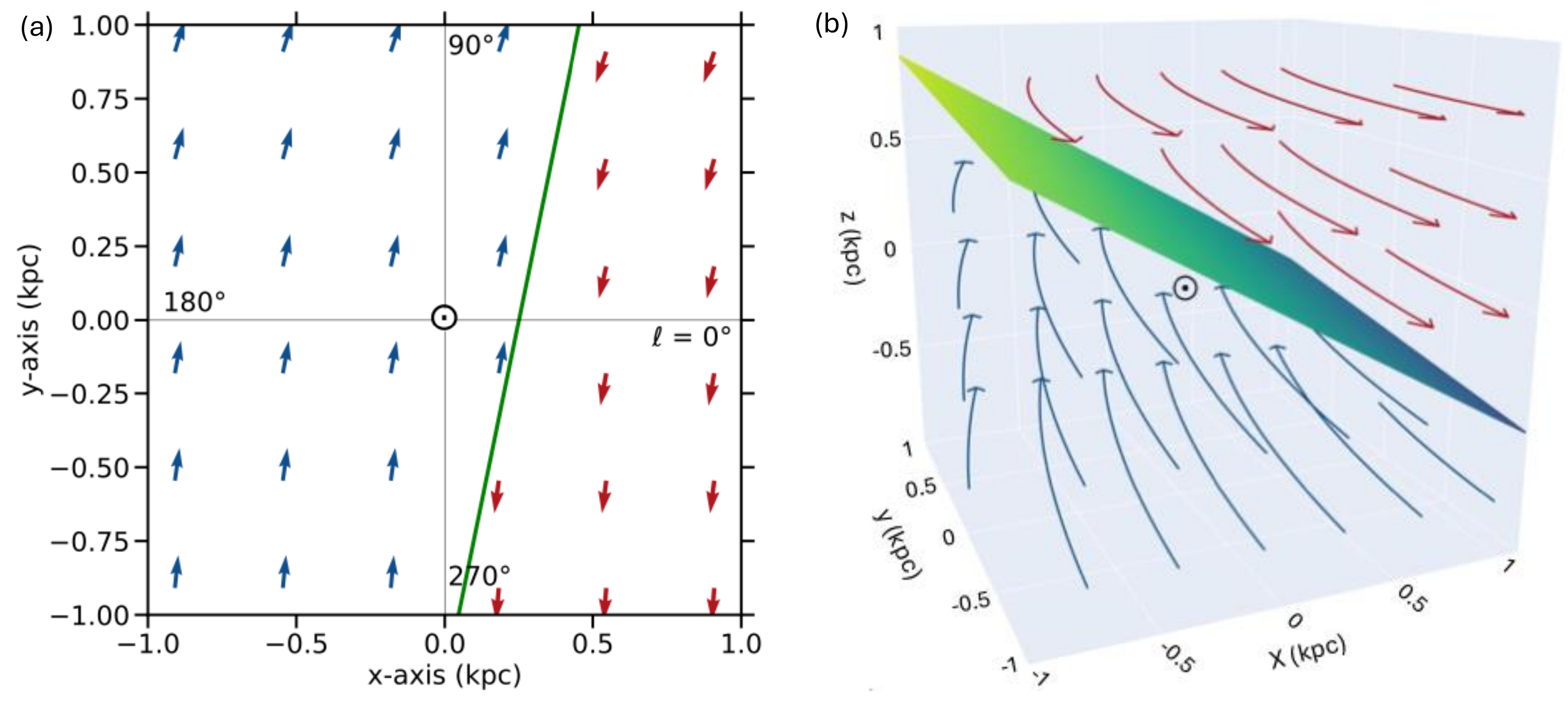}} 
  \caption{The \change{planar reversal} model. (a) A cross-section along the \change{$x$-$y$} plane with the positive \change{$x$-axis} towards the Galactic center, located at $(8.15, 0, 0)$~kpc. The green line indicates the orientation of the reversal, with a clockwise field (blue arrows) on the near side, and counterclockwise field (red arrows) on the far side. (b) The reversal model in three-dimensional space. The reversal is depicted as a diagonal plane separating a region with a clockwise field (blue arrows) below and a counterclockwise field (red arrows) above. The blue to yellow plane colors reflect the $z$ values to enhance three-dimensional visualization. The position of the Sun is shown by $\odot$. 
\label{fig:reversal_diagrams}}
\end{figure*}

\vspace{5 em}

\subsection{The \change{planar reversal} model}

 Our model for the reversal is motivated by the work of \citetalias{Ordog2017}. Using RMs of EG sources as well as Galactic extended emission from the Canadian Galactic Plane Survey, they identified that the reversal follows a diagonal line in Galactic coordinates, extending from $(\ell, b) = (67^\circ, 4^\circ)$ to $(56^\circ, -2^\circ)$. We plot the \citetalias{Ordog2017} diagonal as a black line overlaid on the \citetalias{Hutschenreuter2022} map in Figure \ref{fig:ordog_diag}. \citet{Ma2020} extended this investigation into the longitude range $20^\circ$ to $52^\circ$ with the addition of 194 new EG source RMs. They suggested that there is an odd-parity disk field in the Sagittarius Arm, switching direction from counterclockwise above to clockwise below the \citetalias{Ordog2017} diagonal line.

 \citetalias{Ordog2017} suggested a planar geometry for the reversal would be consistent with their diagonal. This idea was further supported by \citet{Curtin2024}, who analyzed pulsar RMs and found that the reversal is more nearby in the Northern Galactic hemisphere than in the South, where it extends back into the Sagittarius Arm. They describe a potential reversal plane\change{,} tilted towards the Galactic center at negative latitudes\change{,} to account for this observed configuration. In the local ISM, projection effects are significant, and a plane in physical space would curve when observed in Galactic coordinates. If the \citetalias{Ordog2017} line were extended up to higher latitudes and allowed to curve, as illustrated with the dashed magenta line in Figure \ref{fig:ordog_diag}, the reversal boundary would separate the adjacent peaks of the $\sin2\ell$ pattern, marked with $-$ and $+$ signs. This motivates our study of a planar reversal as an explanation for the Northern $\sin2\ell$ pattern.

 We model the reversal as a \change{plane}\footnote{\change{A Python library for our planar reversal model is\\ included at \href{https://github.com/beckybooth/Local_reversal}{https://github.com/beckybooth/Local\_reversal}.}}, as shown in Figure \ref{fig:reversal_diagrams}. The reversal plane is angled towards the outer Galaxy in the Northern Galactic hemisphere, and towards the Galactic center in the South. Above the plane, the magnetic field is directed counterclockwise, and below, the magnetic field is clockwise. The Sun is positioned below the reversal, in the clockwise region.
 
 We acknowledge that the true nature of the reversal cannot be an infinite flat plane as described here. For example, it is most likely that the reversal curves parallel to the spiral arms. However, near the Sun, this curvature may be considered relatively `flat', as the Galactocentric azimuthal angle varies by less than 10\% within distances inside 1~kpc. Our objective is to test if a plane can successfully approximate the reversal geometry when confined to the narrow `local' region where DRAGONS is most sensitive. 

\begin{figure}[t!]
  \centering
  \resizebox{0.47\textwidth}{!}{%
      \includegraphics{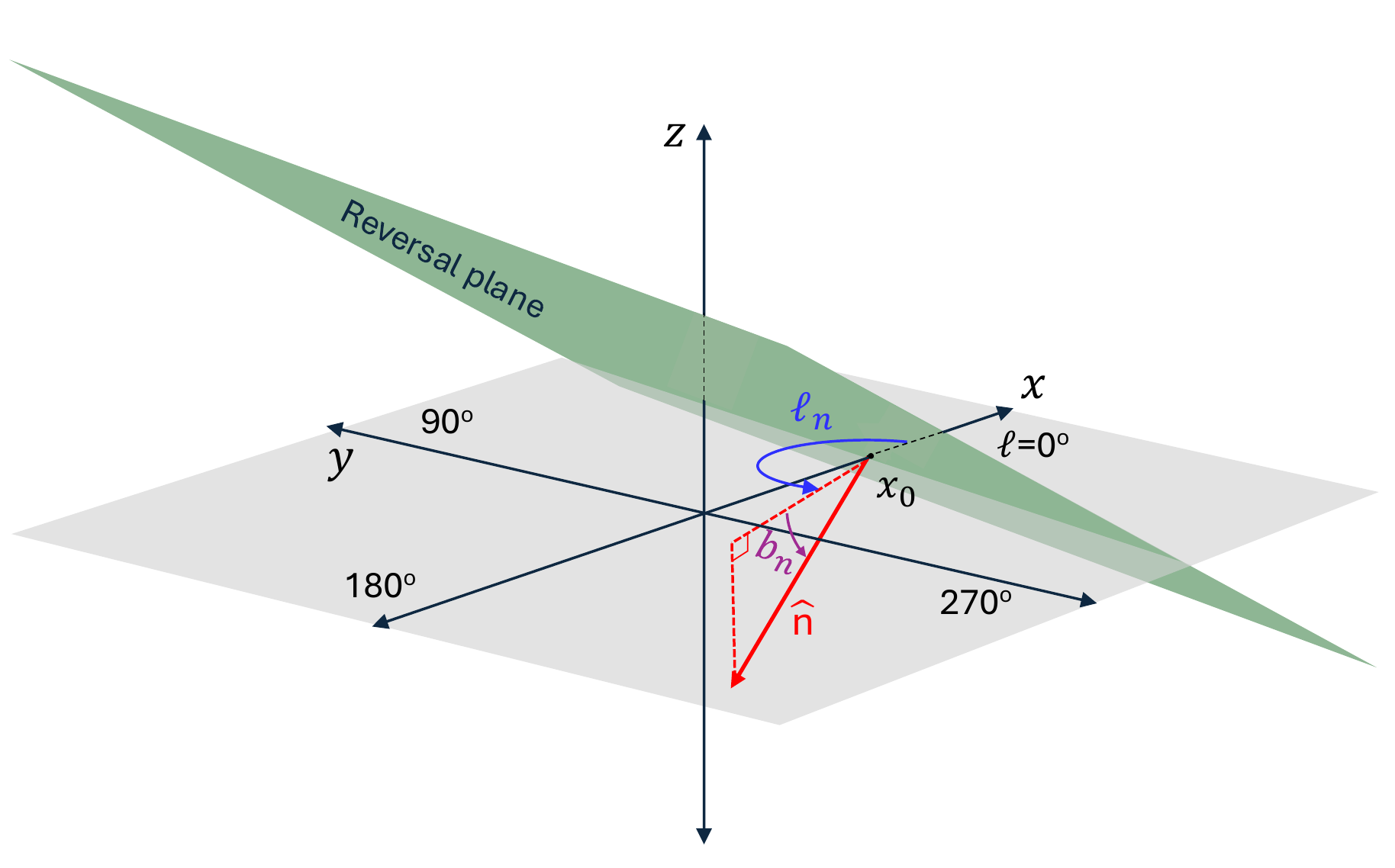} 
  }
  \caption{The geometry of the reversal plane model. The Sun is at the origin, and the positive \change{$x$-axis} points in the direction of the Galactic center. The reversal plane intersects the positive \change{$x$-axis} at $x_0$. The tilt of the plane is defined by the normal line, $\hat{n}$, which is rotated horizontally by angle $\ell_n$ and vertically by angle $b_n$.}
  \label{fig:geometry}
\end{figure}

The equation describing the reversal plane is
\begin{equation}
    A\,(x-x_0) + B\,y + C\,z = 0,
\end{equation}
where $x_0$ is the point where the plane intersects the positive $x$-axis, in the direction of the Galactic center. The coefficients $A$, $B$, and $C$, give the normal vector of the plane, $\hat{n} = A\,\hat{x} + B\,\hat{y} + C\,\hat{z}$, which is directed parallel to the line along Galactic longitude $\ell_n$ and latitude $b_n$ such that
\begin{equation} 
\begin{split}
A &  = \cos b_n \, \cos \ell_n,\\
B &  = \cos b_n \, \sin \ell_n,\\
C &  = \sin b_n. 
\end{split}
\label{eq:normal_coefs}
\end{equation}
The parameters that describe the geometry of the reversal plane, the $x$-intercept, $x_0$, and the two tilt angles, $\ell_n$ and $b_n$, are depicted in Figure \ref{fig:geometry}.

For any LOS, $(\ell, b)$, the distance to the reversal plane is
\begin{equation}
    R_{p} = \frac{A \, x_0}{A \cos b \cos\ell + B \cos b \sin \ell + C \sin b}.
    \label{eq:rintersect}
\end{equation}
If $R_p < 0$, the LOS passes below the reversal plane and we consider a clockwise large-scale field only. In this case, we calculate the simulated \change{FD} for a LOS with no reversal, $\phi_{\textrm{no-rev}}$, as
\begin{equation}
    \phi_{\textrm{no-rev}}(r) = 0.812  \int_{r}^{0} n_e\, B_{\parallel CW} \, dr^\prime,
    \label{eq:phi_sim_1}
\end{equation}
where $r$ is the path length, and $B_{\parallel CW}$ is given by Equation \ref{eq:Bparallel} with $a=+1$. We also use Equation \ref{eq:phi_sim_1} when the path ends before the reversal, i.e., when $r < R_p$. 

In the case where a LOS does cross the reversal, we have a clockwise field contribution below, with LOS component $B_{\parallel CW}$, and a counterclockwise field contribution above, with LOS component $B_{\parallel CCW}$. For the counterclockwise field, $B_{\parallel CCW}$ is given by Equation \ref{eq:Bparallel} with $a=-1$. We denote the vertical tilt angle below the reversal as $\beta_{CW}$ and above as $\beta_{CCW}$, allowing for the possibility that the vertical field component may behave differently above and below the reversal. To compute the simulated \change{FD} for a LOS that crosses the reversal, $\phi_{\textrm{rev}}$, we integrate 
\begin{equation}
    \phi_{\textrm{rev}}(r) = 0.812 \bigg\{ \int_{r}^{R_p} n_e\, B_{\parallel CCW} \, dr^\prime +\int_{R_p}^0 n_e\, B_{\parallel CW} \, dr^\prime\bigg\}.
    \label{eq:phi_sim_2}
\end{equation}

If models describing the varying electron density and magnetic field strength along the path are included, simulated \change{FD}s may be obtained through Equations \ref{eq:phi_sim_1} and \ref{eq:phi_sim_2} by numerical integration. This option remains available for future work. Here, our objective is to demonstrate that a planar reversal can explain the observed large-scale \change{Faraday-rotation} patterns in the Northern sky. To do so, we fit the model to our data, testing many different combinations of model parameters. As numerical integration can be a computationally intensive process, for the purpose of fitting, we apply two approximations to obtain simplified analytical expressions for the simulated \change{FD}s. First, we assume a smooth ISM, with constant electron density and magnetic field strength. These we combine into a single parameter $n_e\,|B|$, which serves to scale the magnitude of the simulated \change{FD}s. We are aware that this approximation does not represent the complexity of the ISM. However, we would require knowledge of the true magnetic field variations and electron density distribution for each LOS to accurately reproduce the complexity we see in the DRAGONS \change{FD} spectra. Without this knowledge, a smooth ISM is the simplest approach to demonstrating that our model is able to predict the observed \change{Faraday-rotation} patterns. 

In addition to assuming uniform $n_e\,|B|$, we approximate the Galactocentric azimuthal angle as $\alpha \approx 180^\circ$. Using this, we integrate Equation \ref{eq:phi_sim_1} and obtain an approximate expression for the \change{FD} for LOS with no reversal,
\begin{equation}
    \phi_{\textrm{no-rev}}(r) = \epsilon\, \zeta \,r.
    \label{phi_approx_1}
\end{equation}
For a LOS that crosses the reversal, we integrate Equation \ref{eq:phi_sim_2} and have
\begin{equation}
    \phi_{\textrm{rev}}(r) = \epsilon\big[\zeta\,R_p + \eta\,(r-R_p)\big].
    \label{phi_approx_2}
\end{equation}
In Equations \ref{phi_approx_1} and \ref{phi_approx_2}, for any given LOS toward $\ell$, $b$, the constants $\epsilon$, $\zeta$, and $\eta$ are,
\begin{equation} 
\begin{split}
\epsilon &  = - 0.812 \,n_e \, |B|,\\
\zeta &  =\cos(b)\, \cos(\beta_{CW}) \,\sin(\ell + p) + \sin(b)\sin(\beta_{CW}), \textrm{ and}\\
\eta &  =-\cos(b)\, \cos(\beta_{CCW}) \,\sin(\ell + p) + \sin(b)\sin(\beta_{CCW}). 
\label{phi_constants}
\end{split}
\end{equation}
Maintaining the assumption of uniform $n_e\,|B|$, these expressions remain valid within $r<1.5$~kpc. In this distance range, the percent difference between the approximate simulated $\phi$ values and the results of numerically integrating Equations \ref{eq:phi_sim_1} and \ref{eq:phi_sim_2} through the full range of azimuthal angles remains, on average across the sky, within 5\%.

\subsection{Simulating emission: screens and slabs}

Equations \ref{phi_approx_1} and \ref{phi_approx_2} describe \change{FD}s independent of emission and can be utilized in different emission scenarios. The two simplest models for polarized emission are a Faraday screen and a slab. For a screen, all emission is behind the Faraday rotating medium. In this case, the \change{FD} spectrum is simple, with a single peak, and M1 is equal to the \change{FD} of that peak. In the case of a screen, for a LOS that does not cross the reversal, the simulated M1 is
\begin{equation}
    \textrm{M1}_{\textrm{screen}_{\textrm{no-rev}}} = \phi_{\textrm{no-rev}}(R) = \epsilon\, \zeta \,R,
    \label{M1_screen_nr}
\end{equation}
and for a LOS that does cross the reversal, the simulated M1 is
\begin{equation}
    \textrm{M1}_{\textrm{screen}_{\textrm{rev}}} =\phi_{\textrm{rev}}(R) =  \epsilon\big[\zeta\,R_p + \eta\,(R-R_p)\big].
    \label{M1_screen_r}
\end{equation}
Here, we use $R$ to denote the total observed path length. Throughout our exploration of the model, the overall path length, $R$, remains a free parameter that can be adjusted to fit observations. 

For an idealized slab, emission and Faraday rotation are uniformly mixed, and the observed \change{FD} spectrum is complex. If there is no reversal along the LOS, M1 is half of the \change{FD} of the far end of the slab \citep{Burn1966}. If there is a reversal, M1 could be any fraction of the farthest \change{FD}, depending on at what depth along the slab the magnetic field reverses.

An analytical expression for M1 of a slab can be obtained by integrating Equations \ref{phi_approx_1} and \ref{phi_approx_2} through all distances to emission along the LOS. For a slab, we assume uniform polarized intensity at all distances. In this case, the simulated M1 can be determined from
    \begin{equation}
        \textrm{M1}_{\textrm{slab}} = \frac{\int_0^R \phi(r) dr}{\int{_0^R dr}}.
        \label{general_slab_M1}
    \end{equation}
    If the LOS does not cross the reversal within the path length, R, we integrate Equation \ref{phi_approx_1}, and the simulated M1 is
\begin{equation}
    \textrm{M1}_{{\textrm{slab}}_{\textrm{no-rev}}} = \frac{\int_0^R \phi_{\textrm{no-rev}}(r) dr}{\int{_0^R dr}} = \frac{1}{2} \, \epsilon \, \zeta \, R.
    \label{M1_slab_nr}
\end{equation}
    If the LOS does cross the reversal, in the numerator of Equation \ref{general_slab_M1} we integrate Equation \ref{phi_approx_1} through distances up to the reversal, $R_p$, and add this to Equation \ref{phi_approx_2} integrated from $R_p$ to the total path length, $R$. This gives the simulated M1 for a slab that passes through the reversal to be
    \begin{equation}
    \begin{split}
    \textrm{M1}_{\textrm{slab}_{\textrm{rev}}} & = \frac{\int_0^{R_p} \phi_{\textrm{no-rev}}(r) dr + \int_{R_p}^{R} \phi_{\textrm{rev}}(r) dr}{\int{_0^R dr}}\\&=
        \frac{\epsilon \,\big[\zeta\, (R_pR - \frac{1}{2}R_p^2) + \eta\, (\frac{1}{2}R^2 + \frac{1}{2}R_p^2 - R_pR)\big]}{R}. 
    \end{split}
    \label{M1_slab_r}
    \end{equation}

We will fit Equations \ref{M1_screen_nr} and \ref{M1_screen_r}, the simulated M1 for a screen, and Equations \ref{M1_slab_nr} and \ref{M1_slab_r}, the simulated M1 for a slab, to DRAGONS M1 in Section \ref{sec:fit_rev}.

\subsection{Reversal model parameters}
\label{sec:parameters}

Our \change{planar reversal} model has eight parameters. We list these in Table \ref{tab:FS_parameters}. Complex models with large numbers of parameters are prone to overfitting. If a model with fewer parameters can agree with the data, the result is more meaningful and the model is more trustworthy. Thus, we sought to reduce the number of free parameters before fitting our reversal model to the DRAGONS data. 
To do this, we set the values for the magnetic field pitch angle, $p$, and the parameters that describe the geometry of the \change{reversal plane}, $\ell_n$, $b_n$, and restrict the plane $x$-intercept distance, $x_0$. 

Literature values for the pitch angle of the large-scale \change{field vary} between 0$^\circ$ and 55$^\circ$  \citep{Haverkorn2015}. As the field is thought to follow the spiral arms \citep{Beck2016}, we use a pitch angle of $p=11.5^\circ$, a value that is used in \citet{VanEck2011} and is consistent with the pitch of the Local Arm reported by \citet{Reid2019}. 

\begin{table}[t]
\begin{tabular}{lll}
\hline
Parameter & symbol & value \\
\hline
Electron density and & $n_e \, |B|$ & free parameter \\
magnetic field strength & & \\
Total path length & $R$ & $<$ 1~kpc \\
\hline
Magnetic field: & &\\
$\circ$ Pitch angle & $p$ & 11.5$^\circ$\\
$\circ$ Tilt below reversal & $\beta_{CW}$ & free parameter \\
$\circ$ Tilt above reversal & $\beta_{CCW}$ & free parameter \\
\hline
Reversal plane: & & \\
$\circ$ Horizontal tilt & $\ell_n$ & 168.5$^\circ$ \\
$\circ$ Vertical tilt & $b_n$ & -60$^\circ$ \\
$\circ$ $x$-axis intercept & $x_0$ & 0.25 to 0.55~kpc \\
\hline
\end{tabular}
\caption{Reversal model parameters}
    \label{tab:FS_parameters}
\end{table}

The horizontal tilt angle of the plane, $\ell_n$, can be chosen by considering two requirements. First, in order to maintain $\nabla \cdot \vec{B} = 0$, the field lines should not intersect the reversal. This requires that the dot product between the plane normal vector and the magnetic field vanishes. Second, we require that the reversal plane should pass between the Local and Sagittarius Arms, without intersecting either arm. The first requirement leaves $\ell_n$ relatively flexible in the range $160^\circ$ to 175$^\circ$, assuming the vertical tilt angle of field lines remains moderate, within $\beta = \pm 30^\circ$, and $p = 11.5^\circ$. To meet the second requirement, we assume the reversal plane runs parallel to the Local Arm in the $x$-$y$ plane, and set $\ell_n = 168.5^\circ$. This ensures that the plane normal is perpendicular to the pitch angle of the Local Arm, as we look towards the Local Arm along the $x$-axis, and meets the first requirement if $\beta$ remains small near the reversal plane.

For the vertical tilt of the reversal, $b_n$, we fit the plane to the \citetalias{Ordog2017} diagonal. This RM = 0~rad~m$^{-2}$ line has a slope in Galactic coordinates of $db/d\ell = 0.55$ and passes through $(\ell, b) = (60^\circ, 0^\circ)$. The line is not visible in DRAGONS M1, either due to obscuration by instrumental effects in the Galactic plane, or because the DRAGONS polarization horizon does not reach far enough for M1 to reach zero in this direction. The observations used by \citetalias{Ordog2017} include the RMs of EG sources, which describe Faraday rotation through the entire path through the Galaxy, and thus probe an entirely different distance than DRAGONS. However, the reversal is a local feature, positioned between the Sun and the Sagittarius Arm, consistent with pulsar observations \citep{Curtin2024}. Since we are modeling a planar reversal, the slope of the reversal observed by \citetalias{Ordog2017} would be the same as we observe in DRAGONS. By fitting the slope of the plane to the \citetalias{Ordog2017} diagonal, the planar approximation likely best describes the true reversal geometry in the direction of the diagonal, i.e., $(\ell, b) = (60^\circ, 0^\circ)$. If the reversal has any curvature, over longer path lengths the slope would start to deviate from what we have modeled here in directions away from the diagonal. For the short path lengths we use in this work, we are assuming minimal curvature.

Assuming the observations used by \citetalias{Ordog2017} have uniform path length, the field reversal must be at the same distance everywhere along the diagonal line, with the total path length evenly split on either side of the reversal. Taking Galactic latitude, $b$, to be a function of Galactic longitude, $\ell$, and differentiating Equation \ref{eq:rintersect} with respect to $\ell$, lines of constant distance to the reversal plane, at $b=0^\circ$, have slope
\begin{equation}
    \frac{db}{d\ell}\bigg|_
    {\substack{ b=0^\circ \\ {R_p = \text{constant}}
    }} = \frac{\sin(\ell - \ell_n)}{\tan(b_n)}.
\end{equation}
At $(\ell, b) = (60^\circ, 0^\circ)$, for all reasonable choices of $\ell_n$, a slope of $db/d\ell = 0.55$ can be achieved if the vertical tilt of the reversal plane is $b_n \approx -60^\circ$.

We can identify the possible distance range to the reversal based on the dust correlations that we identified in section \ref{sec:FDregions}. With $\ell_n = 168.5^\circ$ and $b_n = -60^\circ$, the reversal would need to intercept the positive \change{$x$-axis} between $x_0 = 0.25$ and $0.55$~kpc in order for it to pass between the NCPL and the \citet{Wolleben2010a} bubble. %The conclusions about this distance range do not change meaningfully as $\ell_n$ varies through $160^\circ$ to 175$^\circ$. 

\begin{figure}[t]
  \centering
  \resizebox{0.47\textwidth}{!}{%
      \includegraphics{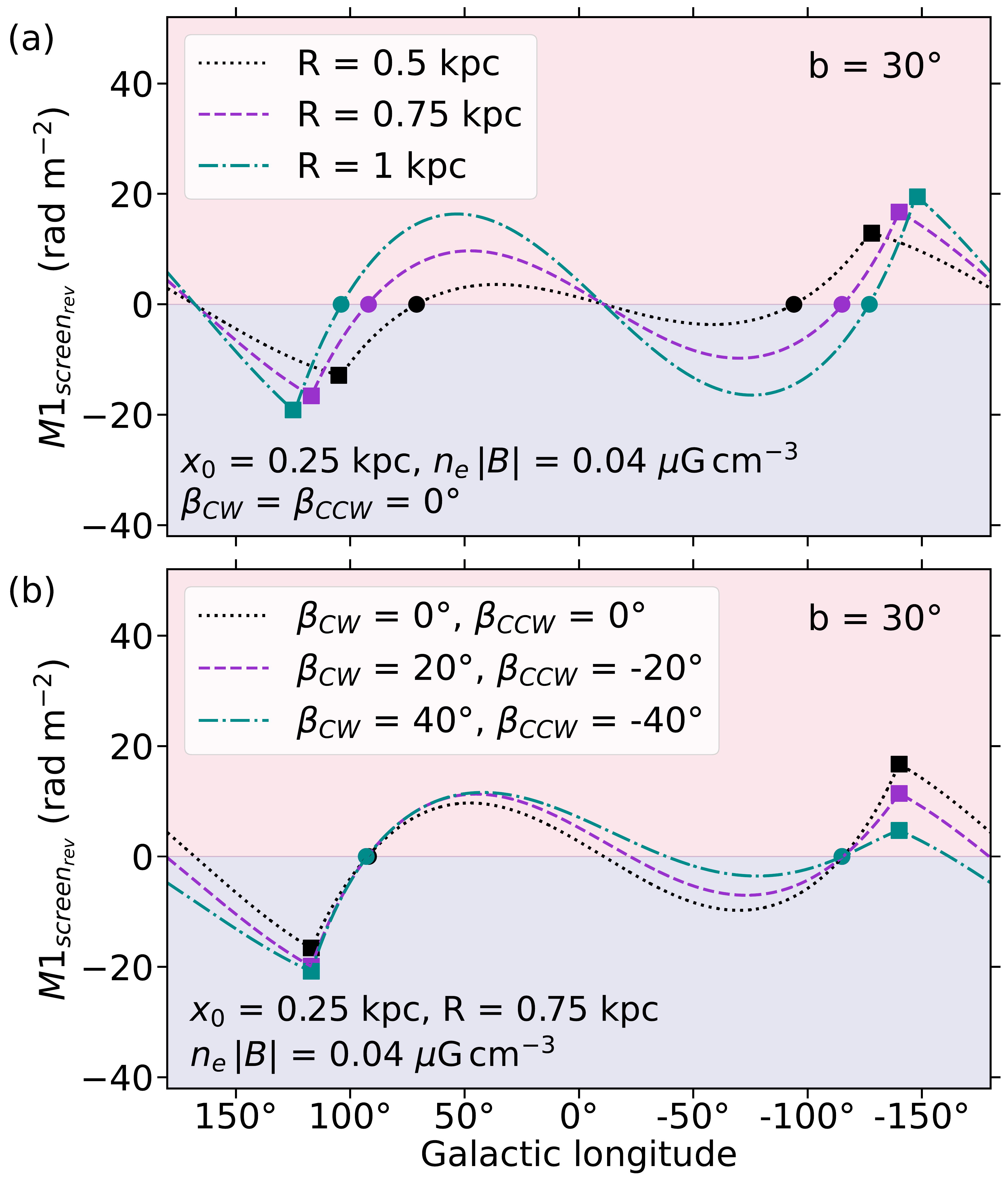}  }
  \caption{A demonstration of how changing the \change{parameters of the planar reversal model} adjusts the simulated $\textrm{M1}_{\textrm{screen}_\textrm{rev}}$ profiles for Galactic latitude, $b = 30^\circ$. Red and blue shading are included for quick visual identification of the positive and negative peaks. (a) Varying the path length, $R$, shifts the longitude of the zeros (circles) and peaks (squares) horizontally. (b) Varying the vertical tilt angles, $\beta_{CW}$ and $\beta_{CCW}$, shifts the profiles vertically.
  \label{fig:examples}}
\end{figure}

\section{Fitting the reversal model to DRAGONS moment 1} % Analysis
\label{sec:fit_rev}

In this section, we will show that our reversal model is able to reproduce the large-scale \change{FD} patterns across DRAGONS and identify the best-fit parameter set, informed by DRAGONS data. First, we demonstrate that our model predicts the Northern $\sin 2\ell$ pattern. Figure \ref{fig:examples} shows M1 profiles at a latitude of $b = 30^\circ$ for the screen model. We note that varying the free parameters modifies the simulated M1 profiles, but the $\sin 2\ell$ pattern persists. In Figure \ref{fig:examples} (a), we vary the path length, $R$, and hold the other parameters constant. In Figure \ref{fig:examples} (b), we vary the tilt angles so that in each case the magnitudes of $\beta_{CW}$ and $\beta_{CCW}$ are equal, but the signs are opposite, flipping the tilt angle across the reversal. For the purposes of this demonstration, we use a constant $n_e\,|B| = 0.04 \,\mu$G$\,$cm$^{-3}$ everywhere, based on estimates that $|B| \approx$ 2~$\mu$G for the large-scale field \citep{Sun2008} and the space-averaged $\langle n_e \rangle \approx$ 0.02~cm$^{-3}$ within 1~kpc of the Sun \citep{Yao2017}.

In \change{Figure} \ref{fig:examples} (a), we observe how increasing path length shifts the position of the zeros and peaks of the simulated M1 profile. There are two zeros that shift with changing $R$ (circles). For the \change{Faraday-screen} scenario, these are at the points in the profiles where the total path is roughly evenly split on each side of the reversal, i.e., when $R/R_p \approx 2$. A longer path length moves this point farther along the reversal. The peaks in the profile (squares) are at the points where $R/R_p\approx 1$, and the reversal just enters the \change{LOS}. Figure \ref{fig:examples} (b) shows how changing tilt angles shifts the profiles up or down, but does not significantly change their shapes. 

The M1 profiles plotted in Figure \ref{fig:examples} are not unique to the parameter sets used. Any parameter set with the same ratio, $R/x_0$, and product, $n_e\,|B|\,R$, as used in the plots, would produce approximately the same profiles. %(there could be small variations due to the slowly changing Galactocentric azimuthal angle). T
This is because $R$ and $n_e|B|$ together modulate the amplitude of the peaks, and since $x_0$ is proportional to $R_p$ (Equation \ref{eq:rintersect}), $R/x_0$ determines the position of the shifting zeros.  

We emphasize that all reasonable variations of the parameters are able to produce the observed Northern sky pattern, with two negative M1 peaks and two positive. This demonstrates the robustness of the model, as fine-tuning of the parameters is not required to obtain agreement with the observed large-scale pattern. In Section \ref{sin2l}, we identified this pattern as having a $\sin2\ell$ relationship, which is characterized by four evenly spaced peaks. Here, we have shown that the \change{planar reversal} model predicts the same two sign reversals with longitude as a $\sin2\ell$ function, but their spacing is also an important quantity to measure. Next, we seek to fit the parameters such that the amplitude and the spacing of the simulated M1 peaks agree with DRAGONS. 

 \begin{table*}[t]
\begin{tabular}{cccccccc}
\hline
Latitude & Emission model & $R/x_0$ & $n_e\,|B|\, R$ (cm$^{-3}\,$$\mu$G$\,$kpc )& $\beta_{CW}$ ($^\circ$)& $\beta_{CCW}$ ($^\circ$) & Pearson correlation\\
\hline
\hline
All-sky & screen & 3.0  & 0.02 & 18  & -23 & 0.60\\
& slab & 5.1 & 0.03 & 18  & -23  & 0.60\\
\hline

-50$^\circ$ & screen & - & 0.01 & 26 & - & 0.53 \\
 & slab & - & 0.02 & 26 & - & 0.53 \\
-40$^\circ$ & screen & - & 0.02 & 15 & - & 0.79 \\
 & slab & - & 0.03 & 15 & - & 0.79 \\
-30$^\circ$ & screen & - & 0.02 & -6 & - & 0.74 \\
 & slab & - & 0.03 & -6 & - & 0.74 \\
30$^\circ$ & screen & 4.0 & 0.03 & 23 & -20 & 0.78 \\
 & slab & 6.7 & 0.07 & 21 & -18 & 0.79 \\
40$^\circ$ & screen & 2.1 & 0.04 & 17 & -12 & 0.86 \\
 & slab & 3.5 & 0.11 & 15 & -10 & 0.85 \\
50$^\circ$ & screen & 1.7 & 0.03 & 4 & -21 & 0.69 \\
 & slab & 2.8 & 0.09 & 7 & -20 & 0.71 \\
\hline
\end{tabular}
\caption{Best-fit parameters for the reversal plane simulated M1 to DRAGONS M1. There is a 10\% \change{systematic} uncertainty in these fitting results.}
    \label{tab:BF_parameters}
\end{table*}

\subsection{All-sky best-fit parameters}
\label{all_sky_fit}

\begin{figure*}[t]
  \centering
      \resizebox{0.49\textwidth}{!}{\includegraphics{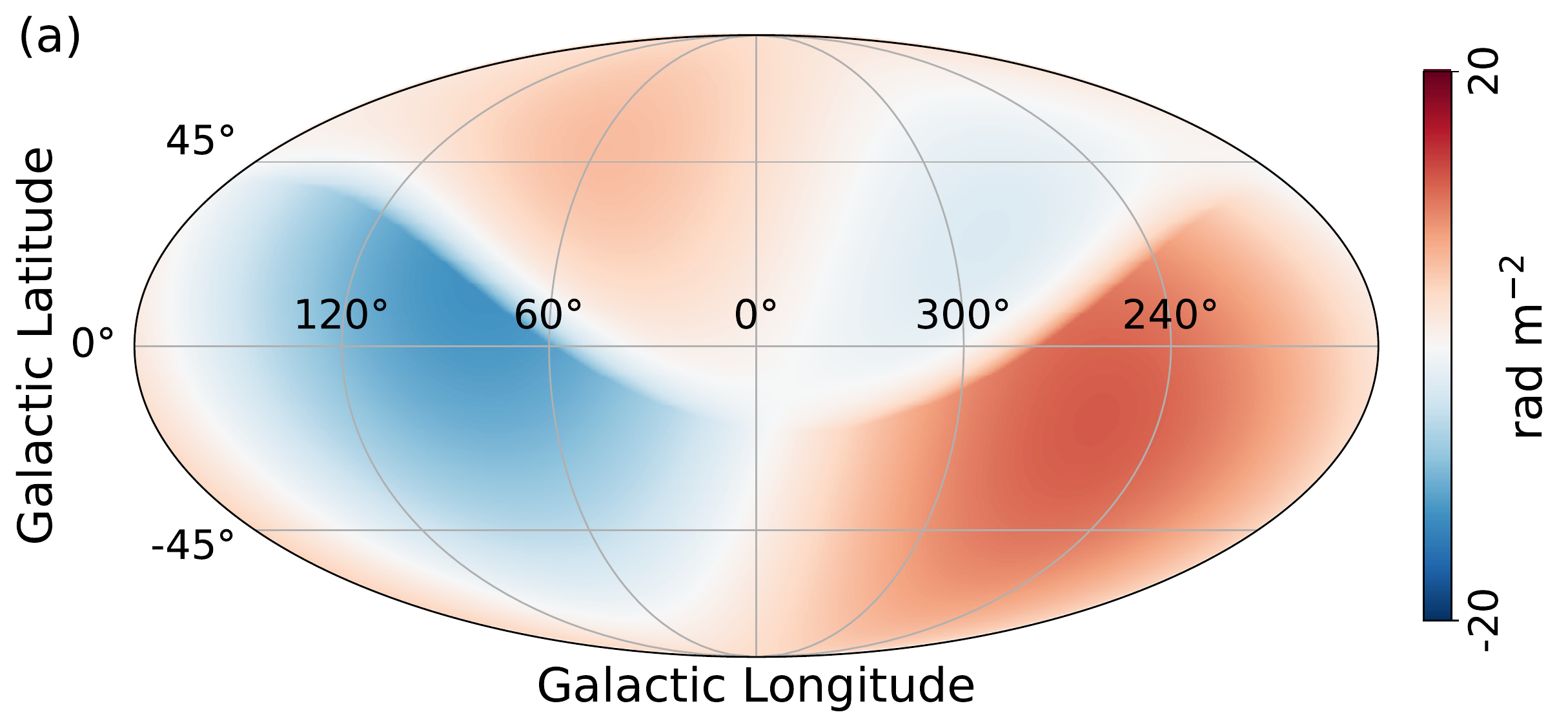}}
      \resizebox{0.49\textwidth}{!}
{\includegraphics{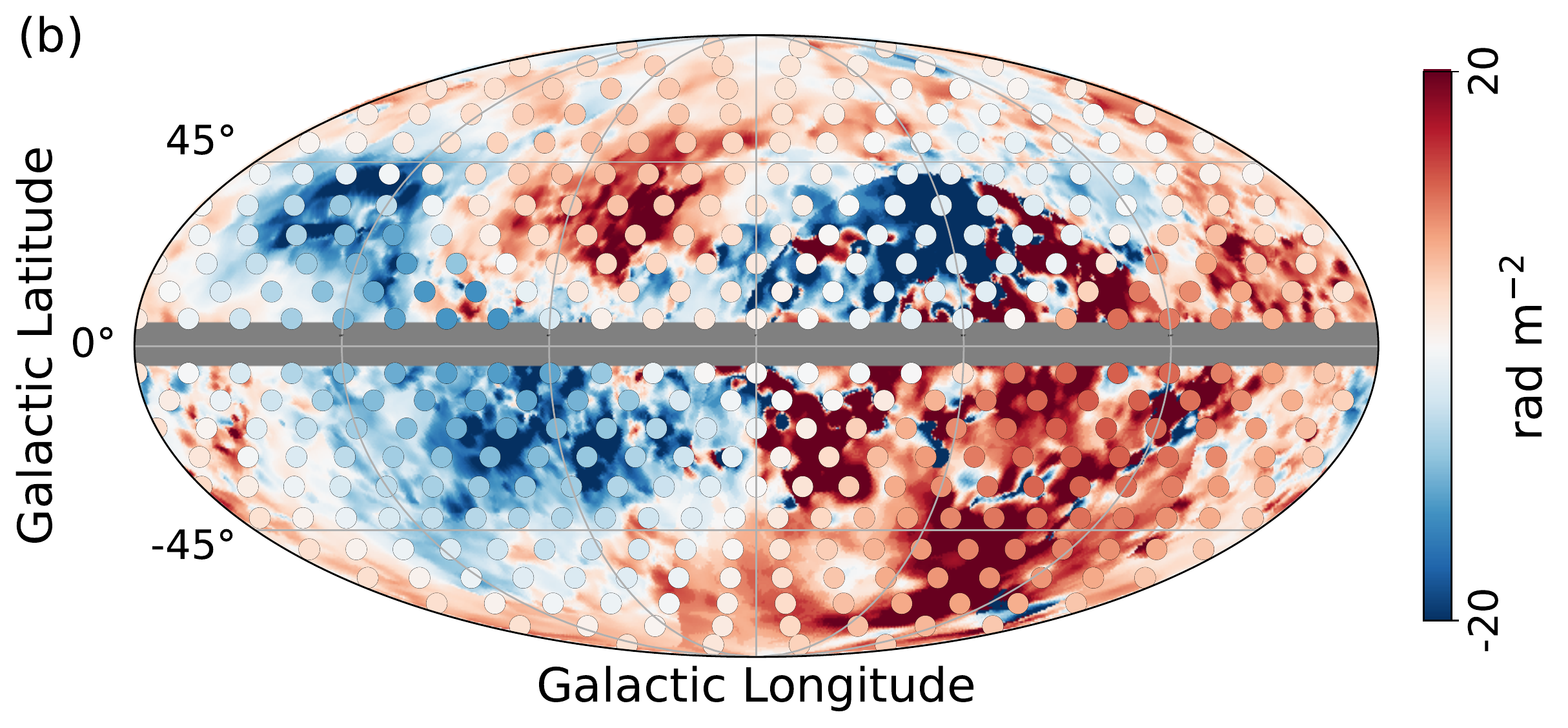}}\\

  \caption{The best-fit simulated M1 map, based on the all-sky best-fit parameters listed in Table \ref{tab:BF_parameters}. (a) The simulated M1 map for the reversal model. As the screen and slab emission models produce nearly identical all-sky maps, this map represents the results produced by both models. (b) The modeled M1 values plotted on the DRAGONS and STAPS combined M1 map.
\label{fig:phi_sim_map}}
\end{figure*}

We used \texttt{curve\_fit} to determine the best-fit parameters that minimize the residuals between the reversal model and DRAGONS M1 data. The goal of this fitting is to demonstrate that the planar reversal geometry that we have modeled here is able to produce M1 maps that are consistent with observations. We fit the reversal model for both kinds of emission scenarios, using the simulated M1 equations for a Faraday screen (Equations \ref{M1_screen_nr} and \ref{M1_screen_r}) and a slab (Equations \ref{M1_slab_nr} and \ref{M1_slab_r}). Inputs to \texttt{curve\_fit} were the Galactic coordinates and corresponding M1 values of the DRAGONS M1 HEALPix array. The best-fit model parameters, $R/x_0$, $n_e\,|B|\, R$, $\beta_{CW}$, and $\beta_{CCW}$, were determined to minimize the residuals between the M1 data and the models, $\textrm{M1}_{\textrm{screen}}$ and $\textrm{M1}_{\textrm{slab}}$. 

For simplicity, we initially fit one set of parameters for the entire DRAGONS M1 map. We present these fitting results in the top section of Table \ref{tab:BF_parameters}. There is a $10\%$ \change{systematic} error in the fitting results, determined by taking twice the square root of the diagonal values of the covariance matrix returned by \texttt{curve\_fit}. The maps produced from the all-sky fitting are indistinguishable by eye for both emission models. Quantitatively, they differ on average by 4\%, which is within the uncertainty introduced by using $\alpha \approx 180^\circ$. We present only the version produced using the best-fit screen parameters in Figure \ref{fig:phi_sim_map}, which includes the simulated M1 map for the model (a), and the model plotted on the combined DRAGONS and STAPS M1 map (b). Though STAPS M1 was not used for fitting, we include it to demonstrate that the best-fit model also agrees well with the \change{FD} patterns in STAPS. %systemic

In the last column of Table \ref{tab:BF_parameters}, we include the Pearson correlation coefficient, which describes the pixel by pixel\footnote{These comparisons are done using \textit{nside }= 128 HEALPix maps to ensure that the pixels represent roughly equal sky areas.} linear correlation, comparing the model to DRAGONS M1. For the all-sky fitting, we find a Pearson correlation of 0.6 in both the screen and slab scenarios. In addition, the best-fit all-sky map has 74\% agreement in sign with the DRAGONS M1 by sky area, 70\% in the North, and 80\% in the South. Both of these indicate good agreement on large scales between the model and the data. 

Given a reversal distance of $x_0 = 0.25$ to 0.55~kpc, the best-fit $R/x_0$ values predict path lengths ranging from $R=0.76$ to 1.7~kpc for a screen, and $R=1.3$ to 2.8~kpc for a slab. Aside from the lowest value of $R=0.76$~kpc, these are beyond what we expect for the DRAGONS polarization horizon. In Section \ref{sec:FDregions}, we showed that the Northern $\sin2\ell$ pattern is dominated by nearby Faraday rotation, originating between 400 and 500~pc. This discrepancy is likely due to the simplifications we have made to demonstrate the model. While the uniform $n_e\, |B|$ configuration that we modeled requires farther distances to produce profiles that agree with the data, a non-uniform ISM would be able to generate the same profiles with shorter path lengths.

\begin{figure*}[tb]
  \centering
  \resizebox{0.98\textwidth}{!}{%
      \includegraphics{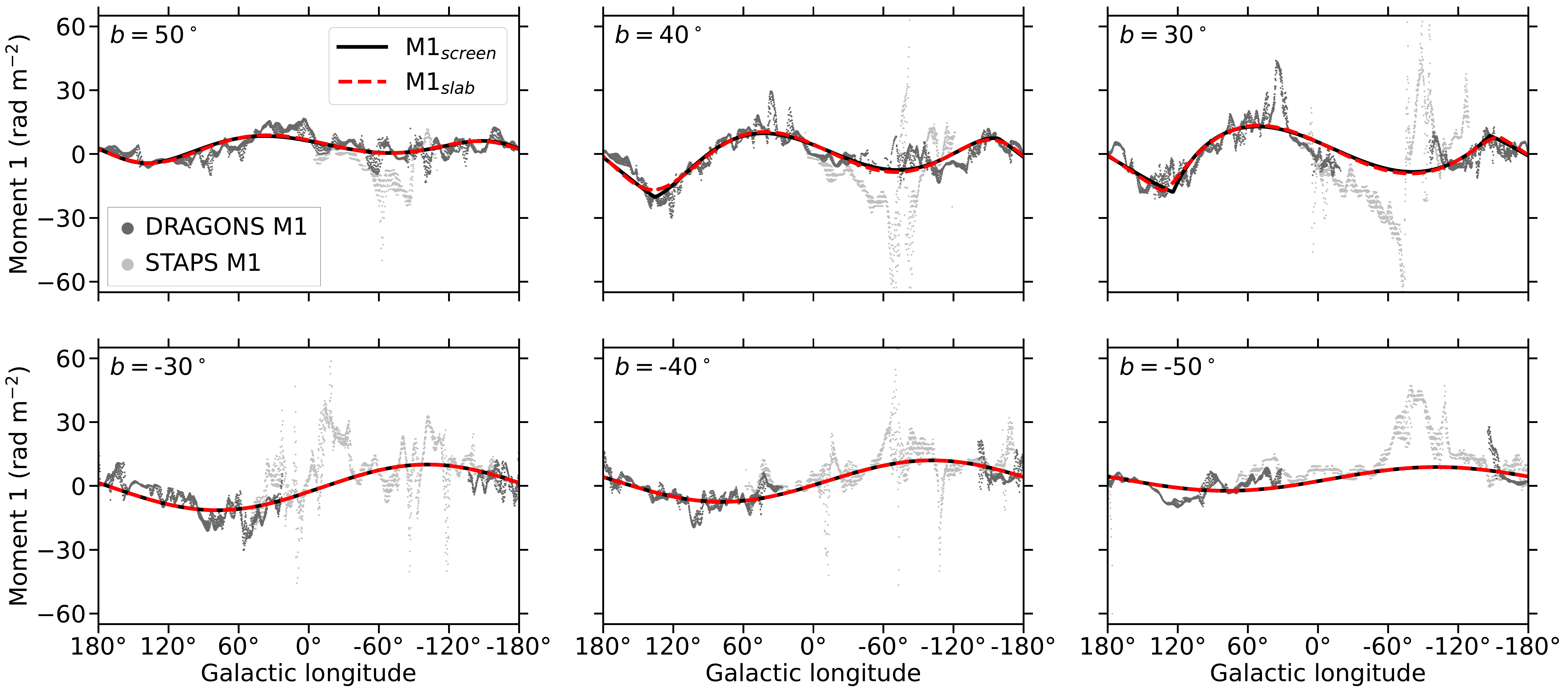} 
  }
  \caption{The best-fit modeled M1 profiles fit to to DRAGONS M1 within 3$^\circ$ wide bins. The central latitude of the bin is indicated. In each plot, the best-fit screen (black solid line) and slab (red dashed line) models are nearly identical. The models were fit to DRAGONS M1 (dark grey); STAPS M1 values (light grey) are shown for reference only.
  \label{fig:fit_to_M1}}
\end{figure*}

\subsection{Fitting within latitude bins}
\label{vary_with_latitude}
% tilted plane reversal
It is unlikely that a single set of parameters can describe the DRAGONS observations across the full sky. In particular, we would expect a varying path length, as the polarization horizon can change with direction. Here, we will explore how the parameters vary with latitude. We fit the reversal model for both screen and slab emission to DRAGONS M1 within $3^\circ$ wide latitude bins, centered on Northern mid-latitudes, $b=30^\circ$, $40^\circ$, and $50^\circ$, and Southern mid-latitudes, $b=-50^\circ$, $-40^\circ$, and $-30^\circ$. For each Northern latitude, we found best-fit values for the ratio $R/x_0$, the product $n_e\,|B|\,R$, and the two tilt angles, $\beta_{CW}$ and $\beta_{CCW}$. In the South, only $n_e\,|B|\,R$ and $\beta_{CW}$ were required, as the \change{reversal plane} does not extend into the Southern mid-latitudes that we examined. The fitting results are summarized in the bottom section of Table \ref{tab:BF_parameters}. In all cases, the Pearson correlation is positive, and for most latitudes, the agreement between the model and data is strong, with $\textrm{correlation}>0.7$. This demonstrates that the reversal model is able to produce simulated M1 values that agree very well with observations.

Figure \ref{fig:fit_to_M1} shows the best-fit M1 profiles for both the screen and slab models plotted as a function of longitude on the DRAGONS and STAPS M1 data for each latitude bin. The screen and slab profiles are nearly the same.
In both cases, the modeled M1 values are able to reproduce the number of peaks in the observed data, and also the distance between the peaks and the position of the zeros. 

\begin{figure*}[tb]
  \centering
  \resizebox{0.98\textwidth}{!}{%
      \includegraphics{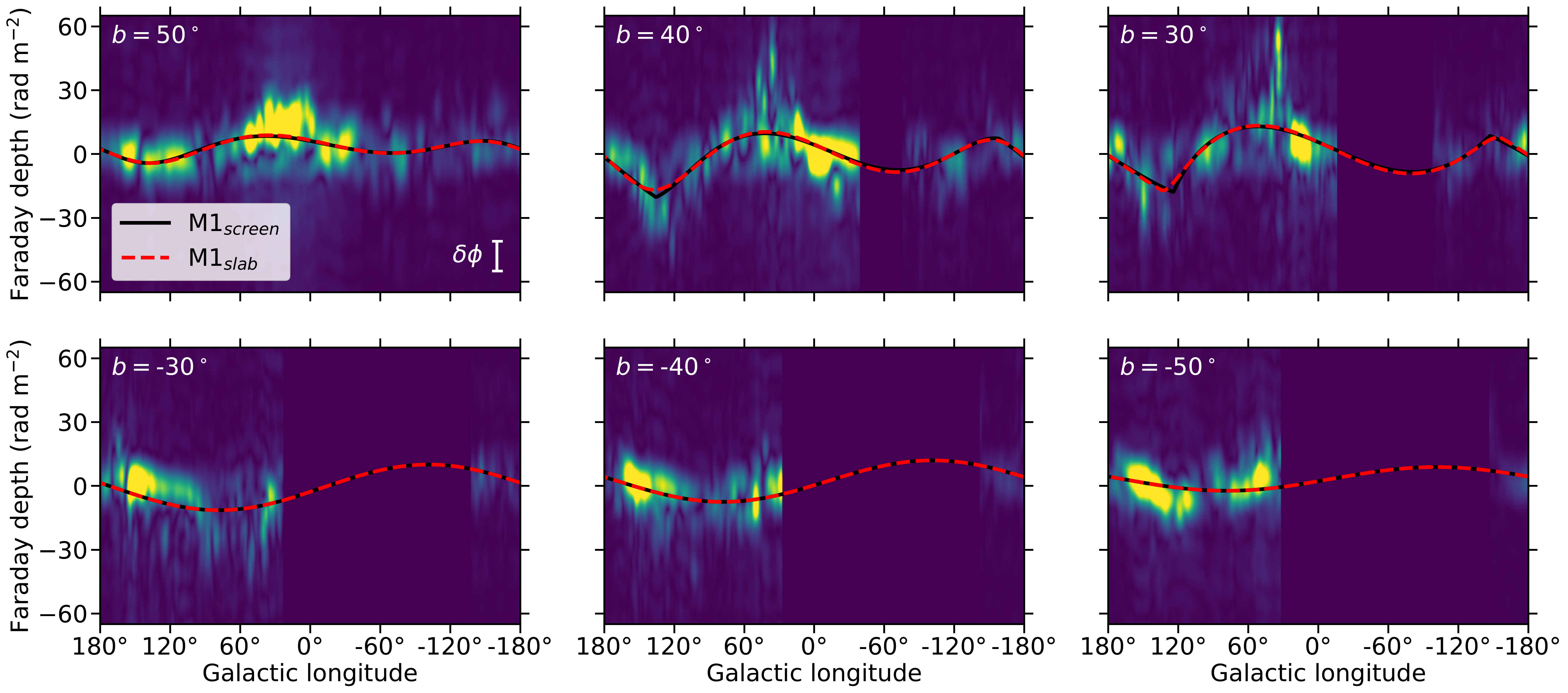} 
  }
  \caption{The \change{best-fit} models for each latitude range from Figure \ref{fig:fit_to_M1} plotted on the cross-section through the DRAGONS \change{FD} cube corresponding to that latitude.
  \label{fig:phi-ell-model}}
\end{figure*}

Figure \ref{fig:phi-ell-model} shows the best-fit M1 profiles plotted on cross sections of the DRAGONS \change{FD} cube, cut along the same latitudes. For these plots, the horizontal axis is Galactic longitude and the vertical axis is \change{FD}. We refer to them as longitude-\change{FD}, or $\ell$-$\phi$ plots, an analog to the $\ell$-$v$ plots commonly used in spectroscopy. The $\ell$-$\phi$ plots provide a visual representation of how the Faraday complexity in the DRAGONS spectra varies with longitude. They show that the smaller-scale \change{Faraday-complex} structures are overlaid on top of the larger-scale coherent pattern predicted by the model and traced by M1 in the data.

\begin{figure*}[t]
  \centering
      \resizebox{0.49\textwidth}{!}{\includegraphics{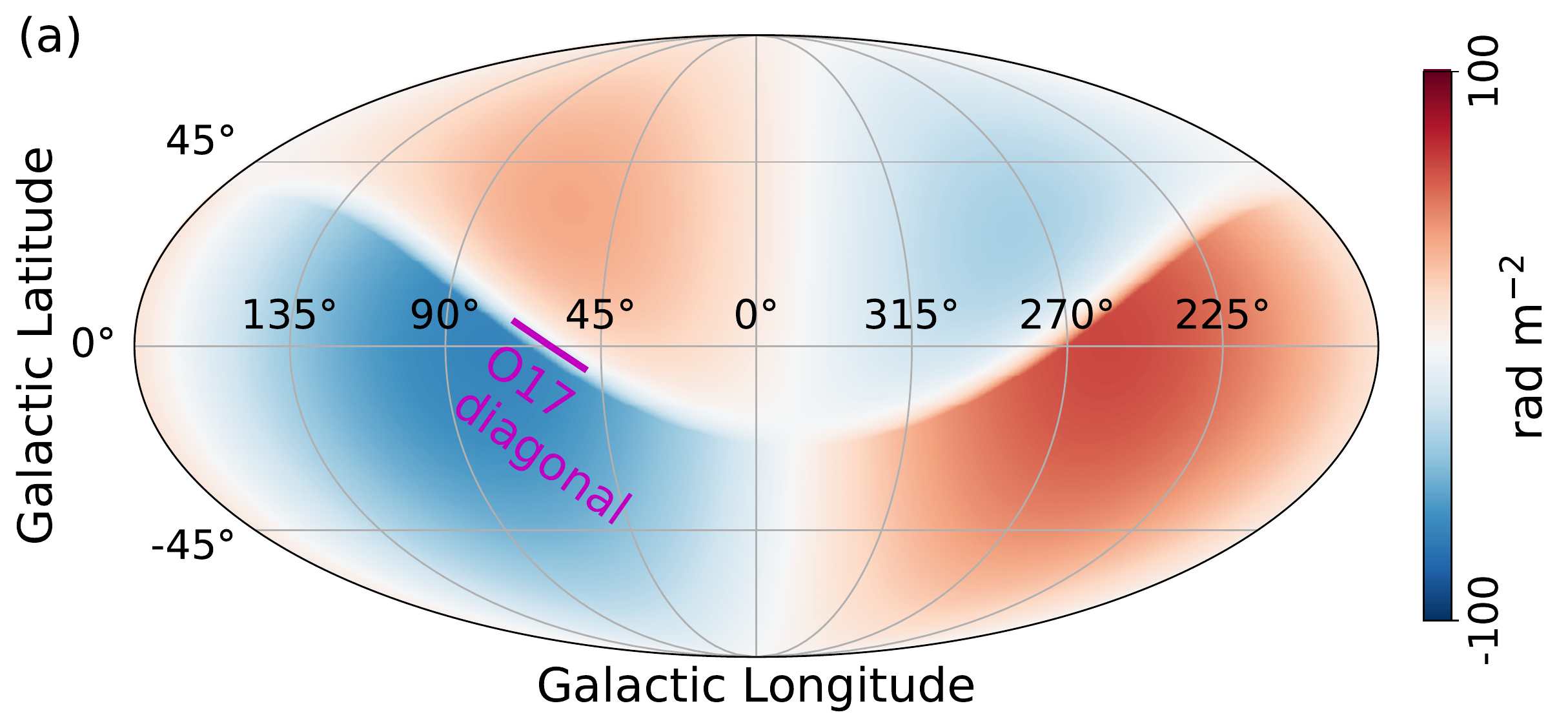}}
      \resizebox{0.49\textwidth}{!}
{\includegraphics{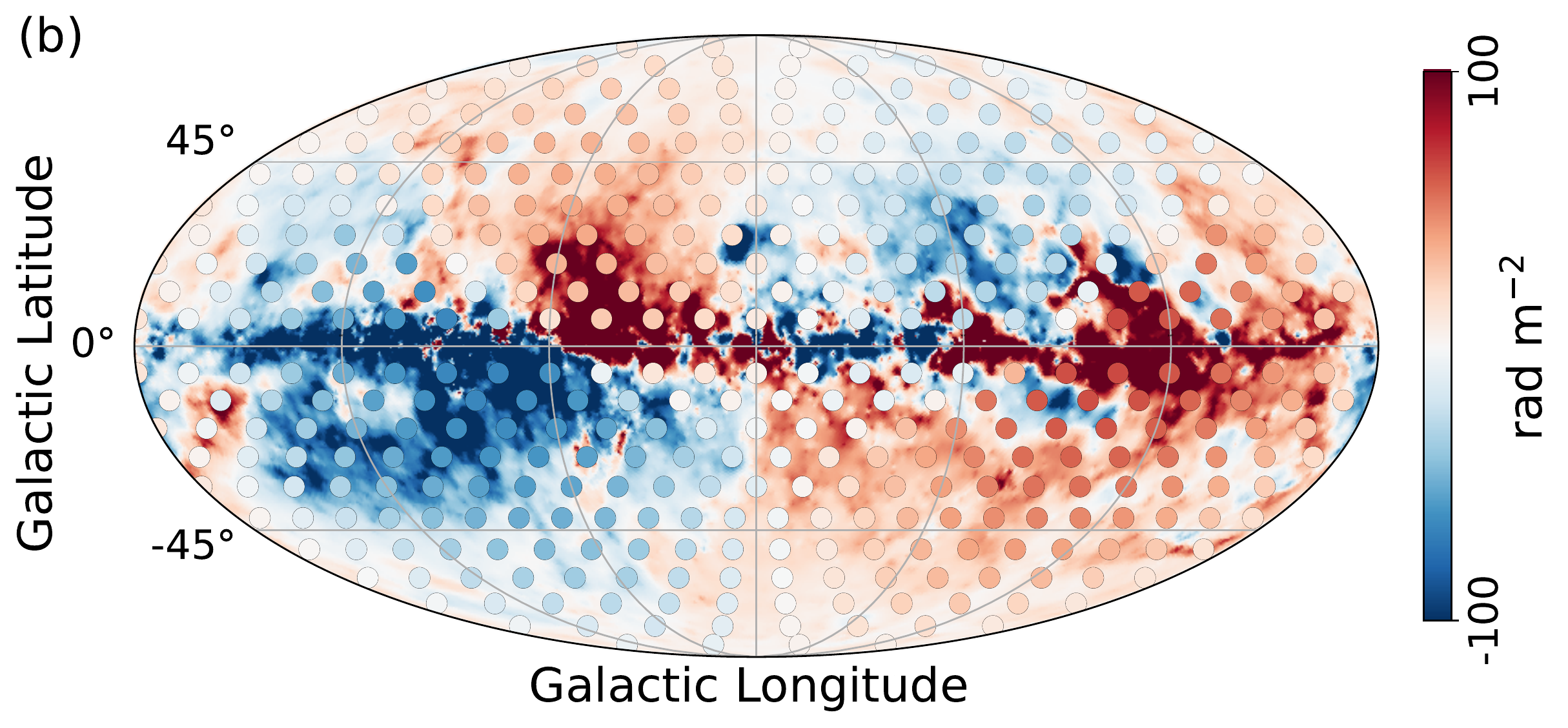}}\\

  \caption{The best-fit simulated RM map for the \citetalias{Hutschenreuter2022} map. The best-fit parameters are $R/x_0 = 4.5$, $n_e\,|B|\,R =0.07$~cm$^{-3}\,$$\mu$G$\,$kpc, $\beta_{CW} = 0^\circ$, $\beta_{CCW} = -7^\circ$. (a) The simulated RM map with the \citetalias{Ordog2017} diagonal indicated in magenta. (b) The simulated RM values plotted on top of the \citetalias{Hutschenreuter2022} map.
\label{fig:h22_model}}
\end{figure*}

The $\ell$-$\phi$ plots also highlight the exceptional nature of the DRAGONS \change{FD} data, as we can readily identify emission structures broader than the DRAGONS \change{FD} resolution, $\delta\phi$. Broad \change{FD} features in the spectra are an indication of ``slab-like'' emission, in which Faraday rotation and emission are mixed, though not necessarily uniformly as in an idealized slab. DRAGONS is the first \change{FD} data set covering large portions of the sky with $\phi_{\textrm{max width}}$ significantly greater than $\delta \phi$, enabling the detection of slab-like emission. \citet{Dickey2022} showed that the Southern sky could be described as slab-like, as the Southern M1 values observed in GMIMS HBN were approximately one half of the RMs in the \citetalias{Hutschenreuter2022} map. They could confirm no such relationship in the North, as is expected when there is a field reversal along the LOS. Using DRAGONS, we can now directly observe the slab-like emission in the Northern sky through the $\ell$-$\phi$ plots in Figure \ref{fig:phi-ell-model}. 

\subsection{Fitting the reversal model to the \citetalias{Hutschenreuter2022} map}
\label{Fit_to_H22}

Our simulated M1 map in Figure \ref{fig:phi_sim_map} does not reproduce the \citetalias{Ordog2017} diagonal. Instead, the M1 = 0~rad~m$^{-2}$ line in Figure \ref{fig:phi_sim_map} passes through the Galactic mid-plane at $(\ell, b) = (50^\circ, 0^\circ)$. However, as we showed in Figure \ref{fig:examples}, the zeros of the model change position depending on the path length. As the path length is increased, we expect the 0~rad~m$^{-2}$ line to shift higher in longitude. As the \citetalias{Ordog2017} diagonal was identified in EG source RMs that were later incorporated into the \citetalias{Hutschenreuter2022} map, we fit the reversal plane model to the \citetalias{Hutschenreuter2022} map to show how the model can be adapted to agree with the \citetalias{Ordog2017} diagonal.

We used the same fitting procedure as described in Section \ref{all_sky_fit}. As the \citetalias{Hutschenreuter2022} map was constructed with EG RMs, only the screen model was required, described by Equations \ref{phi_approx_1} and \ref{phi_approx_2}. We fit the model to data in latitudes $|b| > 15^\circ$, to omit the likely more distant and complex structures in the Galactic plane. The resulting best-fit parameters are $R/x_0 = 4.5$, $n_e\,|B|\,R =0.07$~cm$^{-3}\,$$\mu$G$\,$kpc, $\beta_{CW} = 0^\circ$, $\beta_{CCW} = -7^\circ$. The simulated RM map for these parameters is shown in Figure \ref{fig:h22_model} (a). With the farther path length, the RM=0~rad~m$^{-2}$ line now matches the \citetalias{Ordog2017} diagonal. Figure \ref{fig:h22_model} (b) shows the model values plotted on the \citetalias{Hutschenreuter2022} map for a visual comparison. The Pearson correlation comparing the model to the data is 0.46 for the whole sky, and 0.57 for data in latitudes $|b| > 15^\circ$. There is a 74\% sign agreement between the map and the model, indicating agreement in the LOS field direction on \change{large-angular-scales} predicted by the model.

\subsection{The vertical tilt angle}

We find the vertical tilt angle of the field, $\beta$, remains robust against the choice of emission model. In our fitting, we found $\beta$ to be generally positive below the reversal and negative above, with $\beta_{CW} = 18^\circ$ and $\beta_{CCW} = -23^\circ$ in the DRAGONS all-sky results. In both cases, this represents a field with a vertical component directed towards the reversal. It is of note that these results are consistent with the magnetic field tilt angles determined using Planck dust polarization observations. In the South, the Planck observations suggest a positive tilt angle of $24^\circ\pm 5^\circ$ \citep{PlanckCollaborationIntXLIV2016}, and in the North, they suggest a negative tilt angle of $\sim - 18^\circ$ \citep{Adak2020}.

\section{\change{The origin of the reversal}} \label{sec:discussion}

%Identifying a physical origin for the reversal is beyond the scope of this paper, and we restrict our discussion to some general remarks. 
As the reversal is part of the large-scale field, any origin theory should consider the Galactic-scale mechanisms that generate the field as a whole. There are several mechanisms that could lead to the reversal geometry we have identified here. For example, dynamo models, in which kinetic energies of turbulence, convection, and differential rotation shape, amplify, and maintain the Galactic magnetic field, have been shown to predict reversals \citep[e.g.,][]{Chamandy2013}. \citet{Dickey2022} presented a number of possible combinations of dipolar and quadrupolar dynamo modes, based on the dynamo models of \citet{Henriksen2017} and \citet{Henriksen2018}, demonstrating that these combinations can lead to field reversals. It may be possible in future work to identify a combination of dynamo modes that predict our planar reversal geometry.

Another model that gives rise to reversals was described by \citet{Dobbs2016}. They utilized magnetohydrodynamics (MHD) simulations to demonstrate that the interaction between the field and spiral arm density shocks can result in reversals. This would explain the spatial correspondence between the reversal and the Local Arm, and indicates that there should be a reversal associated with each arm. Some observations suggest that it is possible for many such reversals to exist \citep{Han2018}, while others find no evidence of additional reversals outside the Solar circle \citep{VanEck2011}. Our results agree with \citet{VanEck2011} in that even if there are many reversals, only one local reversal is necessary to reproduce a significant portion of the observed Faraday sky. 

\citet{Pakmor2018} analyzed the magnetic fields of simulated galaxies generated using MHD simulations as part of the Auriga project \citep{Grand2017}. The magnetic fields of the simulated galaxies evolved due to an early turbulent dynamo, followed by a later dynamo phase driven by differential rotation \citep{Pakmor2017}. \citet{Pakmor2018} produced Faraday rotation maps from different positions within the simulated galaxies. We note in particular that their Figure 8 shows a galactic cross-section with a reversal in the azimuthal magnetic field component that appears to follow a \change{planar} orientation, much like we have described here. Moreover, the resulting simulated Faraday rotation maps for a position below that reversal \citep[Figure 9 of][]{Pakmor2018} exhibit the same Northern $\sin2\ell$-like pattern as observed in our own Galaxy. 

By integrating Faraday rotation through different distances in the simulated galaxies, \citet{Pakmor2018} demonstrated that the local environment of the observer strongly dominates the resulting map, regardless of the path length chosen. This is particularly true at mid to high latitudes, where the scale height of the Faraday rotating medium may be relatively close. Our results support this conclusion. In Section \ref{Fit_to_H22}, we were able to reproduce the large-scale \change{Faraday-rotation} patterns in the \citetalias{Hutschenreuter2022} map using only our local reversal model and a relatively short path length.

\section{Summary}

We have utilized the \change{FD} data from the DRAGONS survey \citep{ordog2025DRAGONS}, in the frequency range 500 to 1030 MHz, to test a model for the reversal in the local, large-scale Galactic magnetic field. Here, we define `local' as within a radius of $r = 1$~kpc of the Sun. Our model describes the reversal as a plane positioned between the Sun and the Sagittarius Arm, tilted towards us in the Galactic Northern hemisphere, such that the plane passes above us in Galactic coordinates. Below the reversal plane, the large-scale field is directed clockwise, as viewed from the North Galactic pole, and above the reversal, the field is directed counterclockwise. Our key findings are summarized as follows:
 \begin{itemize}
     \item  In Section \ref{sec:FDregions}, we used the \citetalias{Edenhofer2024} dust map to identify the distance to Faraday rotation structures in the local ISM. By doing this, we have demonstrated the potential for using three-dimensional dust maps to determine Faraday rotation distances. From the identified Faraday rotation structures, we concluded that at least some of the characteristic Northern sky ($-$, $+$, $-$, $+$), or $\sin2\ell$, pattern results from Faraday rotation in the local ISM. In Section \ref{sec:parameters}, we used the identified distances to locate the reversal between 0.25 and 0.55~kpc in the direction of the Galactic center. 
     \item The key feature of our reversal model is the \change{planar} geometry. In Section \ref{sec:fit_rev}, we demonstrated that our planar reversal is capable of reproducing the positive and negative peaks characteristic of the \citet{Dickey2022} Northern $\sin2\ell$ pattern in both the screen and slab emission scenarios, across various combinations of model parameters. This shows that the $\sin2\ell$ pattern can be explained simply by a single local field reversal, with no additional magnetic field components required. The planar geometry also explains why no signature of a reversal has been observed in the Southern sky, as these lines of sight pass below the reversal. 
     \item In Figure \ref{fig:phi-ell-model}, we presented $\ell-\phi$ plots as an effective tool for visualizing \change{FD} spectra. The subset of DRAGONS used here has \change{FD} resolution of $\delta \phi = 14$~\radmsq, while maintaining sensitivity to broader \change{FD} features as wide as 37~\radmsq. This was instrumental in making $\ell-\phi$ plots possible for the first time. From these plots, we directly observe broadened \change{FD} spectra indicating the presence of slab-like emission, in which emission and Faraday rotation are mixed, across a significant portion of the sky. The $\ell-\phi$ plots also demonstrate how the small-scale Faraday rotation features, appearing as Faraday complexity in the spectra, are overlaid on top of the large-scale patterns predicted by our reversal model.
     
     \item By fitting our model to the \citetalias{Hutschenreuter2022} map in Section \ref{Fit_to_H22}, we showed that the \change{planar reversal model} can be adapted for data that probe farther distances. As our reversal model only describes a local feature, this supports the idea that Faraday rotation information may be dominated by the local magnetic field configuration, even if the observed emission originates from a greater distance.
 \end{itemize}

While the details of our modeling should be considered in the context of a preliminary model, intended to generate discussion and lay the framework for future investigations, our \change{planar reversal model} describes a simple geometry that reproduces much of the large-angular-scale patterns seen in Faraday rotation. We have demonstrated that a full galaxy model is not necessary to produce the observed \change{Faraday-rotation} structure; this structure can be reproduced by a single planar reversal integrated through the local volume.

\vspace{1 em}

\section*{Acknowledgments}

This paper relies on observations obtained using a telescope at the Dominion Radio Astrophysical Observatory, which is located on the traditional, ancestral, and unceded territory of the syilx people. We benefit enormously from the stewardship of the land by the syilx Okanagan Nation and the radio frequency interference environment protection work by the syilx Okanagan Nation and NRC. DRAO is a national facility operated by the National Research Council Canada.

\change{We thank Marijke Haverkorn, Roland Kothes, Mehrnoosh Tahani, and Alec Thomson for their input during the development of this project. We thank DRAO staff, especially Charl Baard, David Del Rizzo, Andrew Gray, Richard Hellyer, Dustin Lagoy, Rob Messing, and Benoit Robert for their work in enabling the DRAGONS survey.}

This work benefited from discussions during the Interstellar Institute’s program ``II7’’ at the Paris-Saclay University’s Institut Pascal in July 2025 and ``Structure and polarization in the interstellar medium: A Conference in Honor of Prof. John Dickey'', a hybrid meeting hosted jointly at Stanford University and at the Australia Telescope National Facility in February 2025. We acknowledge support from the National Science Foundation (NSF Award No. 2502957), from the Kavli Institute for Particle Astrophysics and Cosmology, from the Commonwealth Scientific and Industrial Research Organisation, and from the Australian Research Council. 

R.A.B. was supported by the Natural Sciences and Engineering Research Council of Canada (NSERC) Vanier scholarship and the University of Calgary Izaak Walton Killam Doctoral Scholarship. A.O. was partly supported by the Dunlap Institute at the University of Toronto. The Dunlap Institute is funded through an endowment established by the David Dunlap family and the University of Toronto. This research has been supported by NSERC Discovery Grants (PIs: J.C.B., T.L.L., and A.S.H).

\change{\facilities{DRAO:DRAO-15}}
\change{\software{AstroHOG \citep{Soler2019a}; Astropy \citep{Astropy_2022}; healpy \citep{Zonca2019,Gorski_2005}; Matplotlib \citep{Hunter_2007}; NumPy \cite{harris2020array}; RM-tools \citep{Purcell2020}; }}

\newpage 

\bibliography{reversal}{}

@preamble{ "\newcommand{\noop}[1]{}" }

@string{june = {June}}

@article{Beck2016,
 author = {Rainer Beck},
 doi = {10.1007/s00159-015-0084-4},
 issn = {0935-4956},
 issue = {1},
 journal = {The A\&AR},
 month = {12},
 pages = {4},
 title = {Magnetic fields in spiral galaxies},
 url = {http://link.springer.com/10.1007/s00159-015-0084-4},
 volume = {24},
 year = {2016}
}

@article{Brentjens2005,
 author = {M. A. Brentjens and A. G. de Bruyn},
 doi = {10.1051/0004-6361:20052990},
 issn = {0004-6361},
 issue = {3},
 journal = {A\&A},
 month = {10},
 pages = {1217-1228},
 title = {Faraday rotation measure synthesis},
 volume = {441},
 year = {2005}
}

@article{Burn1966,
 author = {B. J. Burn},
 doi = {10.1093/mnras/133.1.67},
 issn = {0035-8711},
 issue = {1},
 journal = {MNRAS},
 month = {7},
 pages = {67-83},
 title = {On the Depolarization of Discrete Radio Sources by Faraday Dispersion},
 volume = {133},
 year = {1966}
}

@article{Curtin2024,
 author = {Alice P. Curtin and Joel M. Weisberg and Joanna M. Rankin},
 doi = {10.3847/1538-4357/ad7b15},
 issn = {0004-637X},
 issue = {2},
 journal = {ApJ},
 month = {11},
 pages = {217},
 title = {Determining the Magnetic Field in the Galactic Plane from New Arecibo Pulsar Faraday Rotation Measurements},
 url = {https://iopscience.iop.org/article/10.3847/1538-4357/ad7b15},
 volume = {975},
 year = {2024}
}

@article{Dickey2019,
 author = {John M. Dickey and T. L. Landecker and Alec J. M. Thomson and M. Wolleben and X. Sun and E. Carretti and K. Douglas and A. Fletcher and B. M. Gaensler and A. Gray and M. Haverkorn and A. S. Hill and S. A. Mao and N. M. McClure-Griffiths},
 doi = {10.3847/1538-4357/aaf85f},
 issn = {0004-637X},
 issue = {1},
 journal = {ApJ},
 month = {1},
 pages = {106},
 title = {The Galactic Magneto-ionic Medium Survey: Moments of the Faraday Spectra},
 url = {https://iopscience.iop.org/article/10.3847/1538-4357/aaf85f},
 volume = {871},
 year = {2019}
}

@article{Dickey2022,
 author = {John M. Dickey and Jennifer West and Alec J. M. Thomson and T. L. Landecker and A. Bracco and E. Carretti and J. L. Han and A. S. Hill and Y. K. Ma and S. A. Mao and A. Ordog and Jo-Anne C. Brown and K. A. Douglas and A. Erceg and V. Jelić and R. Kothes and M. Wolleben},
 doi = {10.3847/1538-4357/ac94ce},
 issn = {0004-637X},
 issue = {1},
 journal = {ApJ},
 month = {11},
 pages = {75},
 title = {Structure in the Magnetic Field of the Milky Way Disk and Halo Traced by Faraday Rotation},
 url = {https://iopscience.iop.org/article/10.3847/1538-4357/ac94ce},
 volume = {940},
 year = {2022}
}

@article{Edenhofer2024,
 author = {Gordian Edenhofer and Catherine Zucker and Philipp Frank and Andrew K. Saydjari and Joshua S. Speagle and Douglas Finkbeiner and Torsten A. Enßlin},
 doi = {10.1051/0004-6361/202347628},
 issn = {0004-6361},
 journal = {A\&A},
 month = {5},
 pages = {A82},
 title = {A parsec-scale Galactic 3D dust map out to 1.25 kpc from the Sun},
 url = {https://www.aanda.org/10.1051/0004-6361/202347628},
 volume = {685},
 year = {2024}
}

@INPROCEEDINGS{Haverkorn2015,
       author = {{Haverkorn}, Marijke},
        title = "{Magnetic Fields in the Milky Way}",
    booktitle = {Magnetic Fields in Diffuse Media},
         year = 2015,
       editor = {{Lazarian}, Alexander and {de Gouveia Dal Pino}, Elisabete M. and {Melioli}, Claudio},
       series = {Astrophysics and Space Science Library},
       volume = {407},
        month = jan,
        pages = {483},
          doi = {10.1007/978-3-662-44625-6_17},
archivePrefix = {arXiv},
       eprint = {1406.0283},
 primaryClass = {astro-ph.GA},
       adsurl = {https://ui.adsabs.harvard.edu/abs/2015ASSL..407..483H}
}

@article{Heiles1989,
 author = {Carl Heiles},
 doi = {10.1086/167051},
 issn = {0004-637X},
 journal = {ApJ},
 month = {1},
 pages = {808},
 title = {Magnetic fields, pressures, and thermally unstable gas in prominent H I shells},
 volume = {336},
 year = {1989}
}

@article{Hutschenreuter2022,
 author = {S. Hutschenreuter and C. S. Anderson and S. Betti and G. C. Bower and J.-A. Brown and M. Brüggen and E. Carretti and T. Clarke and A. Clegg and A. Costa and S. Croft and C. Van Eck and B. M. Gaensler and F. de Gasperin and M. Haverkorn and G. Heald and C. L. H. Hull and M. Inoue and M. Johnston-Hollitt and J. Kaczmarek and C. Law and Y. K. Ma and D. MacMahon and S. A. Mao and C. Riseley and S. Roy and R. Shanahan and T. Shimwell and J. Stil and C. Sobey and S. P. O’Sullivan and C. Tasse and V. Vacca and T. Vernstrom and P. K. G. Williams and M. Wright and T. A. Enßlin},
 doi = {10.1051/0004-6361/202140486},
 issn = {0004-6361},
 journal = {A\&A},
 month = {1},
 pages = {A43},
 title = {The Galactic Faraday rotation sky 2020},
 url = {https://www.aanda.org/10.1051/0004-6361/202140486},
 volume = {657},
 year = {2022}
}

@article{Jaffe2010,
 adsurl = {https://ui.adsabs.harvard.edu/abs/2010MNRAS.401.1013J},
 archiveprefix = {arXiv},
 author = {{Jaffe}, T.~R. and {Leahy}, J.~P. and {Banday}, A.~J. and {Leach}, S.~M. and {Lowe}, S.~R. and {Wilkinson}, A.},
 doi = {10.1111/j.1365-2966.2009.15745.x},
 eprint = {0907.3994},
 journal = {\mnras},
 month = {January},
 number = {2},
 pages = {1013-1028},
 primaryclass = {astro-ph.GA},
 title = {{Modelling the Galactic magnetic field on the plane in two dimensions}},
 volume = {401},
 year = {2010}
}

@article{Korochkin2025,
 author = {Alexander Korochkin and Dmitri Semikoz and Peter Tinyakov},
 doi = {10.1051/0004-6361/202451440},
 issn = {0004-6361},
 journal = {A\&A},
 month = {1},
 pages = {A284},
 title = {The coherent magnetic field of the Milky Way halo, the Local Bubble, and the Fan region},
 url = {https://www.aanda.org/10.1051/0004-6361/202451440},
 volume = {693},
 year = {2025}
}

@article{Ma2020,
 author = {Y K Ma and S A Mao and A Ordog and J C Brown},
 doi = {10.1093/mnras/staa2105},
 issn = {0035-8711},
 issue = {3},
 journal = {MNRAS},
 month = {9},
 pages = {3097-3117},
 title = {The complex large-scale magnetic fields in the first Galactic quadrant as revealed by the Faraday depth profile disparity},
 url = {https://academic.oup.com/mnras/article/497/3/3097/5873678},
 volume = {497},
 year = {2020}
}

@article{Marchal2023,
 author = {Antoine Marchal and Peter G. Martin},
 doi = {10.3847/1538-4357/aca4d2},
 issn = {0004-637X},
 issue = {2},
 journal = {ApJ},
 month = {1},
 pages = {70},
 title = {On the Origin of the North Celestial Pole Loop},
 url = {https://iopscience.iop.org/article/10.3847/1538-4357/aca4d2},
 volume = {942},
 year = {2023}
}

@article{Mohammed2024,
 author = {Nasser Mohammed and Anna Ordog and Rebecca A. Booth and Andrea Bracco and Jo-Anne C. Brown and Ettore Carretti and John M. Dickey and Simon Foreman and Mark Halpern and Marijke Haverkorn and Alex S. Hill and Gary Hinshaw and Joseph W. Kania and Roland Kothes and T. L. Landecker and Joshua MacEachern and Kiyoshi W. Masui and Aimee Menard and Ryan R. Ransom and Wolfgang Reich and Patricia Reich and J. Richard Shaw and Seth R. Siegel and Mehrnoosh Tahani and Alec J. M. Thomson and Tristan Pinsonneault-Marotte and Haochen Wang and Jennifer L. West and Maik Wolleben and Dallas Wulf},
 doi = {10.3847/1538-4357/ad5099},
 issn = {0004-637X},
 issue = {1},
 journal = {ApJ},
 month = {8},
 pages = {100},
 title = {Faraday Tomography with CHIME: The “Tadpole” Feature G137+7},
 url = {https://iopscience.iop.org/article/10.3847/1538-4357/ad5099},
 volume = {971},
 year = {2024}
}

@article{Ordog2017,
 author = {A. Ordog and J. C. Brown and R. Kothes and T. L. Landecker},
 issn = {23318422},
 journal = {A\&A},
 pages = {A15},
 title = {Three-dimensional structure of the magnetic field in the disk of the milky way},
 volume = {603},
 year = {2017},
alias = {O17}
}

@article{PlanckCollaborationIntXLIV2016,
 author = {{Planck Collaboration Int. XLIV}},
 doi = {10.1051/0004-6361/201628636},
 issn = {0004-6361},
 journal = {A\&A},
 month = {12},
 pages = {A105},
 title = {<i>Planck</i> intermediate results},
 url = {http://www.aanda.org/10.1051/0004-6361/201628636},
 volume = {596},
 year = {2016}
}

@article{Raycheva2025,
 author = {N. Raycheva and M. Haverkorn and S. Ideguchi and J. M. Stil and X. Sun and J. L. Han and E. Carretti and X. Y. Gao and A. Bracco and S. E. Clark and J. M. Dickey and B. M. Gaensler and A. S. Hill and T. L. Landecker and A. Ordog and A. Seta and M. Tahani and M. Wolleben},
 doi = {10.1051/0004-6361/202449556},
 issn = {0004-6361},
 journal = {A\&A},
 month = {3},
 pages = {A101},
 title = {Faraday moments of the Southern Twenty-centimeter All-sky Polarization Survey (STAPS)},
 url = {https://www.aanda.org/10.1051/0004-6361/202449556},
 volume = {695},
 year = {2025}
}

@article{Reid2019,
 author = {M. J. Reid and K. M. Menten and A. Brunthaler and X. W. Zheng and T. M. Dame and Y. Xu and J. Li and N. Sakai and Y. Wu and K. Immer and B. Zhang and A. Sanna and L. Moscadelli and K. L. J. Rygl and A. Bartkiewicz and B. Hu and L. H. Quiroga-Nuñez and H. J. van Langevelde},
 doi = {10.3847/1538-4357/ab4a11},
 issn = {0004-637X},
 issue = {2},
 journal = {ApJ},
 month = {11},
 pages = {131},
 title = {Trigonometric Parallaxes of High-mass Star-forming Regions: Our View of the Milky Way},
 url = {https://iopscience.iop.org/article/10.3847/1538-4357/ab4a11},
 volume = {885},
 year = {2019}
}

@article{SK80,
 adsurl = {https://ui.adsabs.harvard.edu/abs/1980ApJ...242...74S},
 author = {{Simard-Normandin}, M. and {Kronberg}, P.~P.},
 doi = {10.1086/158445},
 journal = {\apj},
 month = {November},
 pages = {74-94},
 title = {{Rotation measures and the galactic magnetic field.}},
 volume = {242},
 year = {1980}
}

@article{Sun2008,
 author = {X. H. Sun and W. Reich and A. Waelkens and T. A. Enßlin},
 doi = {10.1051/0004-6361:20078671},
 issn = {0004-6361},
 issue = {2},
 journal = {A\&A},
 month = {1},
 pages = {573-592},
 title = {Radio observational constraints on Galactic 3D-emission models},
 url = {http://www.aanda.org/10.1051/0004-6361:20078671},
 volume = {477},
 year = {2008}
}

@article{Thomson80,
 adsurl = {https://ui.adsabs.harvard.edu/abs/1980MNRAS.191..863T},
 author = {{Thomson}, R.~C. and {Nelson}, A.~H.},
 doi = {10.1093/mnras/191.4.863},
 journal = {\mnras},
 month = {June},
 pages = {863-870},
 title = {{The interpretation of pulsar rotation measures and the magnetic field of the galaxy}},
 volume = {191},
 year = {1980}
}

@article{Unger2024,
 author = {Michael Unger and Glennys R. Farrar},
 doi = {10.3847/1538-4357/ad4a54},
 issn = {0004-637X},
 issue = {1},
 journal = {ApJ},
 month = {7},
 pages = {95},
 title = {The Coherent Magnetic Field of the Milky Way},
 volume = {970},
 year = {2024}
}

@article{Uyaniker2003,
 author = {B. Uyan{\i}ker and T. L. Landecker and A. D. Gray and R. Kothes},
 doi = {10.1086/346234},
 issn = {0004-637X},
 issue = {2},
 journal = {ApJ},
 month = {3},
 pages = {785-800},
 title = {Radio Polarization from the Galactic Plane in Cygnus},
 volume = {585},
 year = {2003}
}

@article{VanEck2011,
 author = {C. L. {Van Eck} and J. C. Brown and J. M. Stil and K. Rae and S. A. Mao and B. M. Gaensler and A. Shukurov and A. R. Taylor and M. Haverkorn and P. P. Kronberg and N. M. McClure-Griffiths},
 doi = {10.1088/0004-637X/728/2/97},
 issn = {0004-637X},
 issue = {2},
 journal = {ApJ},
 month = {2},
 pages = {97},
 title = {MODELING THE MAGNETIC FIELD IN THE GALACTIC DISK USING NEW ROTATION MEASURE OBSERVATIONS FROM THE VERY LARGE ARRAY},
 url = {https://iopscience.iop.org/article/10.1088/0004-637X/728/2/97},
 volume = {728},
 year = {2011}
}

@article{VanEck2021,
 author = {C. L. {Van Eck} and J. C. Brown and A. Ordog and R. Kothes and T. L. Landecker and B. Cooper and K. M. Rae and D. A. Del Rizzo and A. D. Gray and R. Ransom and R. I Reid and B. Uyaniker},
 doi = {10.3847/1538-4365/abe389},
 issn = {0067-0049},
 issue = {2},
 journal = {ApJS},
 month = {4},
 pages = {48},
 title = {Revisiting Rotation Measures from the Canadian Galactic Plane Survey: the Magnetic Field in the Disk of the Outer Galaxy},
 url = {https://iopscience.iop.org/article/10.3847/1538-4365/abe389},
 volume = {253},
 year = {2021}
}

@article{Virtanen2020,
 author = {Pauli Virtanen and Ralf Gommers and Travis E. Oliphant and Matt Haberland and Tyler Reddy and David Cournapeau and Evgeni Burovski and Pearu Peterson and Warren Weckesser and Jonathan Bright and Stéfan J. van der Walt and Matthew Brett and Joshua Wilson and K. Jarrod Millman and Nikolay Mayorov and Andrew R. J. Nelson and Eric Jones and Robert Kern and Eric Larson and C J Carey and İlhan Polat and Yu Feng and Eric W. Moore and Jake VanderPlas and Denis Laxalde and Josef Perktold and Robert Cimrman and Ian Henriksen and E. A. Quintero and Charles R. Harris and Anne M. Archibald and Antônio H. Ribeiro and Fabian Pedregosa and Paul van Mulbregt and Aditya Vijaykumar and Alessandro Pietro Bardelli and Alex Rothberg and Andreas Hilboll and Andreas Kloeckner and Anthony Scopatz and Antony Lee and Ariel Rokem and C. Nathan Woods and Chad Fulton and Charles Masson and Christian Häggström and Clark Fitzgerald and David A. Nicholson and David R. Hagen and Dmitrii V. Pasechnik and Emanuele Olivetti and Eric Martin and Eric Wieser and Fabrice Silva and Felix Lenders and Florian Wilhelm and G. Young and Gavin A. Price and Gert-Ludwig Ingold and Gregory E. Allen and Gregory R. Lee and Hervé Audren and Irvin Probst and Jörg P. Dietrich and Jacob Silterra and James T Webber and Janko Slavič and Joel Nothman and Johannes Buchner and Johannes Kulick and Johannes L. Schönberger and José Vinícius de Miranda Cardoso and Joscha Reimer and Joseph Harrington and Juan Luis Cano Rodríguez and Juan Nunez-Iglesias and Justin Kuczynski and Kevin Tritz and Martin Thoma and Matthew Newville and Matthias Kümmerer and Maximilian Bolingbroke and Michael Tartre and Mikhail Pak and Nathaniel J. Smith and Nikolai Nowaczyk and Nikolay Shebanov and Oleksandr Pavlyk and Per A. Brodtkorb and Perry Lee and Robert T. McGibbon and Roman Feldbauer and Sam Lewis and Sam Tygier and Scott Sievert and Sebastiano Vigna and Stefan Peterson and Surhud More and Tadeusz Pudlik and Takuya Oshima and Thomas J. Pingel and Thomas P. Robitaille and Thomas Spura and Thouis R. Jones and Tim Cera and Tim Leslie and Tiziano Zito and Tom Krauss and Utkarsh Upadhyay and Yaroslav O. Halchenko and Yoshiki Vázquez-Baeza},
 doi = {10.1038/s41592-019-0686-2},
 issn = {1548-7091},
 issue = {3},
 journal = {Nature Methods},
 month = {3},
 pages = {261-272},
 title = {SciPy 1.0: fundamental algorithms for scientific computing in Python},
 url = {https://www.nature.com/articles/s41592-019-0686-2},
 volume = {17},
 year = {2020}
}

@ARTICLE{Wolleben2010a,
       author = {{Wolleben}, M. and {Fletcher}, A. and {Landecker}, T.~L. and {Carretti}, E. and {Dickey}, J.~M. and {Gaensler}, B.~M. and {Haverkorn}, M. and {McClure-Griffiths}, N. and {Reich}, W. and {Taylor}, A.~R.},
        title = "{Antisymmetry in the Faraday Rotation Sky Caused by a Nearby Magnetized Bubble}",
      journal = {\apjl},
         year = 2010,
        month = nov,
       volume = {724},
       number = {1},
        pages = {L48-L52},
          doi = {10.1088/2041-8205/724/1/L48},
archivePrefix = {arXiv},
       eprint = {1011.0341},
 primaryClass = {astro-ph.GA},
       adsurl = {https://ui.adsabs.harvard.edu/abs/2010ApJ...724L..48W}
}

@article{Yao2017,
 author = {J. M. Yao and R. N. Manchester and N. Wang},
 doi = {10.3847/1538-4357/835/1/29},
 issn = {0004-637X},
 issue = {1},
 journal = {ApJ},
 month = {1},
 pages = {29},
 title = {A NEW ELECTRON-DENSITY MODEL FOR ESTIMATION OF PULSAR AND FRB DISTANCES},
 url = {https://iopscience.iop.org/article/10.3847/1538-4357/835/1/29},
 volume = {835},
 year = {2017}
}

@article{ Sun2025,
	author = {Sun, X. and Haverkorn, M. and Carretti, E. and Landecker, T. and Gaensler, B. M. and Poppi, S. and Staveley-Smith, L. and Gao, X. and Han, J.},
	title = {The Southern Twenty-centimetre All-sky Polarization Survey (STAPS): Survey description and maps},
	DOI= "10.1051/0004-6361/202453326",
	url= "https://doi.org/10.1051/0004-6361/202453326",
	journal = {A\&A},
	year = 2025,
	volume = 694,
	pages = "A169",
}

@article{Gorski_2005,
   title={HEALPix: A Framework for High‐Resolution Discretization and Fast Analysis of Data Distributed on the Sphere},
   volume={622},
   ISSN={1538-4357},
   url={http://dx.doi.org/10.1086/427976},
   DOI={10.1086/427976},
   number={2},
   journal={\apj},
   publisher={American Astronomical Society},
   author={Gorski, K. M. and Hivon, E. and Banday, A. J. and Wandelt, B. D. and Hansen, F. K. and Reinecke, M. and Bartelmann, M.},
   year={2005},
   month=apr, pages={759–771} }

@article{heald2009,
  title={The Westerbork SINGS survey-II Polarization, Faraday rotation, and magnetic fields},
  author={Heald, G and Braun, R and Edmonds, R},
  journal={A\&A},
  volume={503},
  number={2},
  pages={409--435},
  year={2009},
  publisher={EDP Sciences}
}

@software{Purcell2020,
       author = {Purcell, C.~R. and {Van Eck}, C.~L. and West, J. and Sun, X.~H. and Gaensler, B.~M.},
        title = "{RM-Tools: Rotation measure (RM) synthesis and Stokes QU-fitting}",
 howpublished = {Astrophysics Source Code Library, record ascl:2005.003},
         year = 2020,
        month = may,
          eid = {ascl:2005.003},
       adsurl = {https://ui.adsabs.harvard.edu/abs/2020ascl.soft05003P}
}

@article{Manchester_2005,
   title={The Australia Telescope National Facility Pulsar Catalogue},
   volume={129},
   ISSN={1538-3881},
   url={http://dx.doi.org/10.1086/428488},
   DOI={10.1086/428488},
   number={4},
   journal={AJ},
   publisher={American Astronomical Society},
   author={Manchester, R. N. and Hobbs, G. B. and Teoh, A. and Hobbs, M.},
   year={2005},
   month=apr, pages={1993–2006} }

@ARTICLE{Meyerdierks1991,
       author = {Meyerdierks, H. and Heithausen, A. and Reif, K.},
        title = "{The North Celestial Pole Loop}",
      journal = {\aap},
         year = 1991,
        month = may,
       volume = {245},
       number = {1},
        pages = {247-256},
       adsurl = {https://ui.adsabs.harvard.edu/abs/1991A\&A...245..247M}
}

@ARTICLE{Han1997,
       author = {Han, J.~L. and Manchester, R.~N. and Berkhuijsen, E.~M. and Beck, R.},
        title = "{Antisymmetric rotation measures in our Galaxy: evidence for an A0 dynamo.}",
      journal = {\aap},
         year = 1997,
        month = jun,
       volume = {322},
        pages = {98-102},
       adsurl = {https://ui.adsabs.harvard.edu/abs/1997A\&A...322...98H}
}

@ARTICLE{Han1994,
       author = {Han, J.~L. and Qiao, G.~J.},
        title = "{The magnetic field in the disk of our Galaxy}",
      journal = {\aap},
         year = 1994,
        month = aug,
       volume = {288},
        pages = {759-772},
       adsurl = {https://ui.adsabs.harvard.edu/abs/1994A\&A...288..759H}
}

@article{Wolleben2021,
doi = {10.3847/1538-3881/abf7c1},
url = {https://dx.doi.org/10.3847/1538-3881/abf7c1},
year = {2021},
month = {jun},
publisher = {The American Astronomical Society},
volume = {162},
number = {1},
pages = {35},
author = {{Wolleben}, M. and Landecker, T. L. and Douglas, K. A. and Gray, A. D. and Ordog, A. and Dickey, J. M. and Hill, A. S. and Carretti, E. and Brown, J. C. and Gaensler, B. M. and Han, J. L. and Haverkorn, M. and Kothes, R. and Leahy, J. P. and McClure-Griffiths, N. and McConnell, D. and Reich, W. and Taylor, A. R. and Thomson, A. J. M. and West, J. L.},
title = {The Global Magneto-ionic Medium Survey: A Faraday Depth Survey of the Northern Sky Covering 1280–1750 MHz},
journal = {AJ},
}

@ARTICLE{Wolleben2019,
       author = {{Wolleben}, M. and Landecker, T.~L. and Carretti, E. and Dickey, J.~M. and {Fletcher}, A. and {McClure-Griffiths}, N.~M. and {McConnell}, D. and {Thomson}, A.~J.~M. and {Hill}, A.~S. and {Gaensler}, B.~M. and {Han}, J. -L. and {Haverkorn}, M. and {Leahy}, J.~P. and {Reich}, W. and {Taylor}, A.~R.},
        title = "{The Global Magneto-Ionic Medium Survey: Polarimetry of the Southern Sky from 300 to 480 MHz}",
      journal = {\aj},
         year = 2019,
        month = jul,
       volume = {158},
       number = {1},
          eid = {44},
        pages = {44},
          doi = {10.3847/1538-3881/ab22b0},
archivePrefix = {arXiv},
       eprint = {1905.12685},
 primaryClass = {astro-ph.GA},
       adsurl = {https://ui.adsabs.harvard.edu/abs/2019AJ....158...44W}
}

@ARTICLE{Taylor2009,
       author = {{Taylor}, A.~R. and {Stil}, J.~M. and {Sunstrum}, C.},
        title = "{A Rotation Measure Image of the Sky}",
      journal = {\apj},
         year = 2009,
        month = sep,
       volume = {702},
       number = {2},
        pages = {1230-1236},
          doi = {10.1088/0004-637X/702/2/1230},
       adsurl = {https://ui.adsabs.harvard.edu/abs/2009ApJ...702.1230T}
}

@ARTICLE{GaiaDR1_2016,
       author = {{Gaia Collaboration} and {Brown}, A.~G.~A. and {Vallenari}, A. and {Prusti}, T. and {de Bruijne}, J.~H.~J. and {Mignard}, F. and {Drimmel}, R. and {Babusiaux}, C. and {Bailer-Jones}, C.~A.~L. and {Bastian}, U. and {Biermann}, M. and {Evans}, D.~W. and {Eyer}, L. and {Jansen}, F. and {Jordi}, C. and {Katz}, D. and {Klioner}, S.~A. and {Lammers}, U. and {Lindegren}, L. and {Luri}, X. and {O'Mullane}, W. and {Panem}, C. and {Pourbaix}, D. and {Randich}, S. and {Sartoretti}, P. and {Siddiqui}, H.~I. and {Soubiran}, C. and {Valette}, V. and {van Leeuwen}, F. and {Walton}, N.~A. and {Aerts}, C. and {Arenou}, F. and {Cropper}, M. and {H{\o}g}, E. and {Lattanzi}, M.~G. and {Grebel}, E.~K. and {Holland}, A.~D. and {Huc}, C. and {Passot}, X. and {Perryman}, M. and {Bramante}, L. and {Cacciari}, C. and {Casta{\~n}eda}, J. and {Chaoul}, L. and {Cheek}, N. and {De Angeli}, F. and {Fabricius}, C. and {Guerra}, R. and {Hern{\'a}ndez}, J. and {Jean-Antoine-Piccolo}, A. and {Masana}, E. and {Messineo}, R. and {Mowlavi}, N. and {Nienartowicz}, K. and {Ord{\'o}{\~n}ez-Blanco}, D. and {Panuzzo}, P. and {Portell}, J. and {Richards}, P.~J. and {Riello}, M. and {Seabroke}, G.~M. and {Tanga}, P. and {Th{\'e}venin}, F. and {Torra}, J. and {Els}, S.~G. and {Gracia-Abril}, G. and {Comoretto}, G. and {Garcia-Reinaldos}, M. and {Lock}, T. and {Mercier}, E. and {Altmann}, M. and {Andrae}, R. and {Astraatmadja}, T.~L. and {Bellas-Velidis}, I. and {Benson}, K. and {Berthier}, J. and {Blomme}, R. and {Busso}, G. and {Carry}, B. and {Cellino}, A. and {Clementini}, G. and {Cowell}, S. and {Creevey}, O. and {Cuypers}, J. and {Davidson}, M. and {De Ridder}, J. and {de Torres}, A. and {Delchambre}, L. and {Dell'Oro}, A. and {Ducourant}, C. and {Fr{\'e}mat}, Y. and {Garc{\'\i}a-Torres}, M. and {Gosset}, E. and {Halbwachs}, J. -L. and {Hambly}, N.~C. and {Harrison}, D.~L. and {Hauser}, M. and {Hestroffer}, D. and {Hodgkin}, S.~T. and {Huckle}, H.~E. and {Hutton}, A. and {Jasniewicz}, G. and {Jordan}, S. and {Kontizas}, M. and {Korn}, A.~J. and {Lanzafame}, A.~C. and {Manteiga}, M. and {Moitinho}, A. and {Muinonen}, K. and {Osinde}, J. and {Pancino}, E. and {Pauwels}, T. and {Petit}, J. -M. and {Recio-Blanco}, A. and {Robin}, A.~C. and {Sarro}, L.~M. and {Siopis}, C. and {Smith}, M. and {Smith}, K.~W. and {Sozzetti}, A. and {Thuillot}, W. and {van Reeven}, W. and {Viala}, Y. and {Abbas}, U. and {Abreu Aramburu}, A. and {Accart}, S. and {Aguado}, J.~J. and {Allan}, P.~M. and {Allasia}, W. and {Altavilla}, G. and {{\'A}lvarez}, M.~A. and {Alves}, J. and {Anderson}, R.~I. and {Andrei}, A.~H. and {Anglada Varela}, E. and {Antiche}, E. and {Antoja}, T. and {Ant{\'o}n}, S. and {Arcay}, B. and {Bach}, N. and {Baker}, S.~G. and {Balaguer-N{\'u}{\~n}ez}, L. and {Barache}, C. and {Barata}, C. and {Barbier}, A. and {Barblan}, F. and {Barrado y Navascu{\'e}s}, D. and {Barros}, M. and {Barstow}, M.~A. and {Becciani}, U. and {Bellazzini}, M. and {Bello Garc{\'\i}a}, A. and {Belokurov}, V. and {Bendjoya}, P. and {Berihuete}, A. and {Bianchi}, L. and {Bienaym{\'e}}, O. and {Billebaud}, F. and {Blagorodnova}, N. and {Blanco-Cuaresma}, S. and {Boch}, T. and {Bombrun}, A. and {Borrachero}, R. and {Bouquillon}, S. and {Bourda}, G. and {Bouy}, H. and {Bragaglia}, A. and {Breddels}, M.~A. and {Brouillet}, N. and {Br{\"u}semeister}, T. and {Bucciarelli}, B. and {Burgess}, P. and {Burgon}, R. and {Burlacu}, A. and {Busonero}, D. and {Buzzi}, R. and {Caffau}, E. and {Cambras}, J. and {Campbell}, H. and {Cancelliere}, R. and {Cantat-Gaudin}, T. and {Carlucci}, T. and {Carrasco}, J.~M. and {Castellani}, M. and {Charlot}, P. and {Charnas}, J. and {Chiavassa}, A. and {Clotet}, M. and {Cocozza}, G. and {Collins}, R.~S. and {Costigan}, G. and {Crifo}, F. and {Cross}, N.~J.~G. and {Crosta}, M. and {Crowley}, C. and {Dafonte}, C. and {Damerdji}, Y. and {Dapergolas}, A. and {David}, P. and {David}, M. and {De Cat}, P.},
        title = "{Gaia Data Release 1. Summary of the astrometric, photometric, and survey properties}",
      journal = {\aap},
         year = 2016,
        month = nov,
       volume = {595},
          eid = {A2},
        pages = {A2},
          doi = {10.1051/0004-6361/201629512},
archivePrefix = {arXiv},
       eprint = {1609.04172},
 primaryClass = {astro-ph.IM},
       adsurl = {https://ui.adsabs.harvard.edu/abs/2016A\&A...595A...2G}
}

@article{Pshirkov_2011,
doi = {10.1088/0004-637X/738/2/192},
url = {https://dx.doi.org/10.1088/0004-637X/738/2/192},
year = {2011},
month = {aug},
publisher = {The American Astronomical Society},
volume = {738},
number = {2},
pages = {192},
author = {Pshirkov, M. S. and Tinyakov, P. G. and Kronberg, P. P. and Newton-McGee, K. J.},
title = {DERIVING THE GLOBAL STRUCTURE OF THE GALACTIC MAGNETIC FIELD FROM FARADAY ROTATION MEASURES OF EXTRAGALACTIC SOURCES},
journal = {\apj},
}

@ARTICLE{Maconi2025,
       author = {Maconi, E. and Reissl, S. and Soler, J.~D. and Girichidis, P. and Klessen, R.~S. and Bracco, A. and Hutschenreuter, S.},
        title = "{Modeling Local Bubble analogs: II. Synthetic Faraday rotation maps}",
      journal = {\aap},
         year = 2025,
        month = jun,
       volume = {698},
          eid = {A84},
        pages = {A84},
          doi = {10.1051/0004-6361/202451477},
archivePrefix = {arXiv},
       eprint = {2504.09701},
 primaryClass = {astro-ph.GA},
       adsurl = {https://ui.adsabs.harvard.edu/abs/2025A\&A...698A..84M}
}

@article{Halal2024,
author = {Halal, George and Clark, S. and Tahani, Mehrnoosh},
year = {2024},
month = {09},
pages = {54},
title = {Imprints of the Local Bubble and Dust Complexity on Polarized Dust Emission},
volume = {973},
journal = {\apj},
doi = {10.3847/1538-4357/ad61e0}
}

@article{Pakmor2017,
    author = {Pakmor, Rüdiger and Gómez, Facundo A. and Grand, Robert J. J. and Marinacci, Federico and Simpson, Christine M. and Springel, Volker and Campbell, David J. R. and Frenk, Carlos S. and Guillet, Thomas and Pfrommer, Christoph and White, Simon D. M.},
    title = {Magnetic field formation in the Milky Way like disc galaxies of the Auriga project},
    journal = {\mnras},
    volume = {469},
    number = {3},
    pages = {3185-3199},
    year = {2017},
    month = {05},
    issn = {0035-8711},
    doi = {10.1093/mnras/stx1074},
    url = {https://doi.org/10.1093/mnras/stx1074},
    eprint = {https://academic.oup.com/mnras/article-pdf/469/3/3185/49203504/mnras\_469\_3\_3185.pdf},
}

@ARTICLE{Han2017,
       author = {Han, J. L.},
        title = "{Observing Interstellar and Intergalactic Magnetic Fields}",
      journal = {\araa},
         year = 2017,
        month = aug,
       volume = {55},
       number = {1},
        pages = {111-157},
          doi = {10.1146/annurev-astro-091916-055221},
       adsurl = {https://ui.adsabs.harvard.edu/abs/2017ARA\&A..55..111H}
}

@article{Jansson_2012,
   title={A NEW MODEL OF THE GALACTIC MAGNETIC FIELD},
   volume={757},
   ISSN={1538-4357},
   url={http://dx.doi.org/10.1088/0004-637X/757/1/14},
   DOI={10.1088/0004-637x/757/1/14},
   number={1},
   journal={\apj},
   publisher={American Astronomical Society},
   author={Jansson, Ronnie and Farrar, Glennys R.},
   year={2012},
   month=aug, pages={14} }

@article{Zhang_2023,
   title={Parameters of 220 million stars from Gaia BP/RP spectra},
   volume={524},
   ISSN={1365-2966},
   url={http://dx.doi.org/10.1093/mnras/stad1941},
   DOI={10.1093/mnras/stad1941},
   number={2},
   journal={\mnras},
   publisher={Oxford University Press (OUP)},
   author={Zhang, Xiangyu and Green, Gregory M and Rix, Hans-Walter},
   year={2023},
   month=jun, pages={1855–1884} }

@ARTICLE{Erceg2024b,
       author = {{Erceg}, Ana and {Jeli{\'c}}, Vibor and {Haverkorn}, Marijke and {Gajovi{\'c}}, Lovorka and {Hardcastle}, Martin and {Shimwell}, Timothy W. and {Tasse}, Cyril},
        title = "{Faraday tomography of LoTSS-DR2 data. III. Revealing the Local Bubble and the complex of local interstellar clouds in the high-latitude inner Galaxy}",
      journal = {\aap},
         year = 2024,
        month = aug,
       volume = {688},
          eid = {A200},
        pages = {A200},
          doi = {10.1051/0004-6361/202450082},
archivePrefix = {arXiv},
       eprint = {2406.14679},
 primaryClass = {astro-ph.GA},
       adsurl = {https://ui.adsabs.harvard.edu/abs/2024A&A...688A.200E}
}

@ARTICLE{VanEck2017,
       author = {{Van Eck}, C.~L. and {Haverkorn}, M. and {Alves}, M.~I.~R. and {Beck}, R. and {de Bruyn}, A.~G. and {En{\ss}lin}, T. and {Farnes}, J.~S. and {Ferri{\`e}re}, K. and {Heald}, G. and {Horellou}, C. and {Horneffer}, A. and {Iacobelli}, M. and {Jeli{\'c}}, V. and {Mart{\'\i}-Vidal}, I. and {Mulcahy}, D.~D. and {Reich}, W. and {R{\"o}ttgering}, H.~J.~A. and {Scaife}, A.~M.~M. and {Schnitzeler}, D.~H.~F.~M. and {Sobey}, C. and {Sridhar}, S.~S.},
        title = "{Faraday tomography of the local interstellar medium with LOFAR: Galactic foregrounds towards IC 342}",
      journal = {\aap},
         year = 2017,
        month = jan,
       volume = {597},
          eid = {A98},
        pages = {A98},
          doi = {10.1051/0004-6361/201629707},
archivePrefix = {arXiv},
       eprint = {1612.00710},
 primaryClass = {astro-ph.GA},
       adsurl = {https://ui.adsabs.harvard.edu/abs/2017A\&A...597A..98V}
}

@article{Zaroubi2015,
    author = {Zaroubi, S. and Jelić, V. and de Bruyn, A. G. and Boulanger, F. and Bracco, A. and Kooistra, R. and Alves, M. I. R. and Brentjens, M. A. and Ferrière, K. and Ghosh, T. and Koopmans, L. V. E. and Levrier, F. and Miville-Deschênes, M.-A. and Montier, L. and Pandey, V. N. and Soler, J. D.},
    title = {Galactic interstellar filaments as probed by LOFAR and Planck},
    journal = {MNRASL},
    volume = {454},
    number = {1},
    pages = {L46-L50},
    year = {2015},
    month = {09},
    issn = {1745-3925},
    doi = {10.1093/mnrasl/slv123},
    url = {https://doi.org/10.1093/mnrasl/slv123},
    eprint = {https://academic.oup.com/mnrasl/article-pdf/454/1/L46/54663059/mnrasl_454_1_l46.pdf},
}

@article{ Soler2019a,
	author = {Soler, J. D. and {Beuther, H.} and {Rugel, M.} and {Wang, Y.} and {Clark, P. C.} and {Glover, S. C. O.} and {Goldsmith, P. F.} and {Heyer, M.} and {Anderson, L. D.} and {Goodman, A.} and {Henning, Th.} and {Kainulainen, J.} and {Klessen, R. S.} and {Longmore, S. N.} and {McClure-Griffiths, N. M.} and {Menten, K. M.} and {Mottram, J. C.} and {Ott, J.} and {Ragan, S. E.} and {Smith, R. J.} and {Urquhart, J. S.} and {Bigiel, F.} and {Hennebelle, P.} and {Roy, N.} and {Schilke, P.}},
	title = {Histogram of oriented gradients: a technique for the study of molecular cloud formation},
	DOI= "10.1051/0004-6361/201834300",
	url= "https://doi.org/10.1051/0004-6361/201834300",
	journal = {A\&A},
	year = 2019,
	volume = 622,
	pages = "A166",
}

@ARTICLE{Wolleben2010b,
       author = {{Wolleben}, M. and {Landecker}, T.~L. and {Hovey}, G.~J. and {Messing}, R. and {Davison}, O.~S. and {House}, N.~L. and {Somaratne}, K.~H.~M.~S. and {Tashev}, I.},
        title = "{Rotation Measure Synthesis of Galactic Polarized Emission with the DRAO 26-m Telescope}",
      journal = {\aj},
         year = 2010,
        month = apr,
       volume = {139},
       number = {4},
        pages = {1681-1690},
          doi = {10.1088/0004-6256/139/4/1681},
archivePrefix = {arXiv},
       eprint = {1002.2312},
 primaryClass = {astro-ph.IM},
       adsurl = {https://ui.adsabs.harvard.edu/abs/2010AJ....139.1681W}
}

@ARTICLE{Adak2020,
       author = {{Adak}, Debabrata and {Ghosh}, Tuhin and {Boulanger}, Francois and {Haud}, Urmas and {Kalberla}, Peter and {Martin}, Peter G. and {Bracco}, Andrea and {Souradeep}, Tarun},
        title = "{Dust polarization modelling at large scale over the northern Galactic cap using EBHIS and Planck data}",
      journal = {\aap},
         year = 2020,
        month = aug,
       volume = {640},
          eid = {A100},
        pages = {A100},
          doi = {10.1051/0004-6361/201936124},
archivePrefix = {arXiv},
       eprint = {1906.07445},
 primaryClass = {astro-ph.GA},
       adsurl = {https://ui.adsabs.harvard.edu/abs/2020A\&A...640A.100A}
}

@article{Brown_2007,
   title={Rotation Measures of Extragalactic Sources behind the Southern Galactic Plane: New Insights into the Large‐Scale Magnetic Field of the Inner Milky Way},
   volume={663},
   ISSN={1538-4357},
   url={http://dx.doi.org/10.1086/518499},
   DOI={10.1086/518499},
   number={1},
   journal={\apj},
   publisher={American Astronomical Society},
   author={Brown, J. C. and Haverkorn, M. and Gaensler, B. M. and Taylor, A. R. and Bizunok, N. S. and McClure‐Griffiths, N. M. and Dickey, J. M. and Green, A. J.},
   year={2007},
   month=jul, pages={258–266} }

@article{Soler_2025,
   title={Kinetic tomography of the Galactic plane within 1.25 kiloparsecs from the Sun: The interstellar flows revealed by HI and CO line emission and 3D dust},
   volume={695},
   ISSN={1432-0746},
   url={http://dx.doi.org/10.1051/0004-6361/202453022},
   DOI={10.1051/0004-6361/202453022},
   journal={A\&A},
   publisher={EDP Sciences},
   author={Soler, J. D. and Molinari, S. and Glover, S. C. O. and Smith, R. J. and Klessen, R. S. and Benjamin, R. A. and Hennebelle, P. and Peek, J. E. G. and Beuther, H. and Edenhofer, G. and Zari, E. and Swiggum, C. and Zucker, C.},
   year={2025},
   month=mar, pages={A222} }

@ARTICLE{Jow2018,
       author = {{Jow}, Dylan L. and {Hill}, Ryley and {Scott}, Douglas and {Soler}, J.~D. and {Martin}, P.~G. and {Devlin}, M.~J. and {Fissel}, L.~M. and {Poidevin}, F.},
        title = "{An application of an optimal statistic for characterizing relative orientations}",
      journal = {\mnras},
         year = 2018,
        month = feb,
       volume = {474},
       number = {1},
        pages = {1018-1027},
          doi = {10.1093/mnras/stx2736},
archivePrefix = {arXiv},
       eprint = {1708.04063},
 primaryClass = {astro-ph.IM},
       adsurl = {https://ui.adsabs.harvard.edu/abs/2018MNRAS.474.1018J}
}

@article{Henriksen2018,
    author = {Henriksen, R N and Woodfinden, A and Irwin, J A},
    title = {Exact axially symmetric galactic dynamos},
    journal = {\mnras},
    volume = {476},
    number = {1},
    pages = {635-645},
    year = {2018},
    month = {02},
    issn = {0035-8711},
    doi = {10.1093/mnras/sty256},
    url = {https://doi.org/10.1093/mnras/sty256},
    eprint = {https://academic.oup.com/mnras/article-pdf/476/1/635/24215952/sty256.pdf},
}

@ARTICLE{Henriksen2017,
       author = {{Henriksen}, R.~N.},
        title = "{Magnetic spiral arms in galaxy haloes}",
      journal = {\mnras},
         year = 2017,
        month = aug,
       volume = {469},
       number = {4},
        pages = {4806-4830},
          doi = {10.1093/mnras/stx1169},
archivePrefix = {arXiv},
       eprint = {1704.06958},
 primaryClass = {astro-ph.GA},
       adsurl = {https://ui.adsabs.harvard.edu/abs/2017MNRAS.469.4806H}
}

@article{Dobbs2016,
    author = {Dobbs, C. L. and Price, D. J. and Pettitt, A. R. and Bate, M. R. and Tricco, T. S.},
    title = {Magnetic field evolution and reversals in spiral galaxies},
    journal = {\mnras},
    volume = {461},
    number = {4},
    pages = {4482-4495},
    year = {2016},
    month = {07},
    issn = {0035-8711},
    doi = {10.1093/mnras/stw1625},
    url = {https://doi.org/10.1093/mnras/stw1625},
    eprint = {https://academic.oup.com/mnras/article-pdf/461/4/4482/8113066/stw1625.pdf},
}

@article{Grand2017,
    author = {Grand, Robert J. J. and Gómez, Facundo A. and Marinacci, Federico and Pakmor, Rüdiger and Springel, Volker and Campbell, David J. R. and Frenk, Carlos S. and Jenkins, Adrian and White, Simon D. M.},
    title = {The Auriga Project: the properties and formation mechanisms of disc galaxies across cosmic time},
    journal = {\mnras},
    volume = {467},
    number = {1},
    pages = {179-207},
    year = {2017},
    month = {01},
    issn = {0035-8711},
    doi = {10.1093/mnras/stx071},
    url = {https://doi.org/10.1093/mnras/stx071},
    eprint = {https://academic.oup.com/mnras/article-pdf/467/1/179/10327352/stx071.pdf},
}

@article{Pakmor2018,
    author = {Pakmor, Rüdiger and Guillet, Thomas and Pfrommer, Christoph and Gómez, Facundo A and Grand, Robert J J and Marinacci, Federico and Simpson, Christine M and Springel, Volker},
    title = {Faraday rotation maps of disc galaxies},
    journal = {\mnras},
    volume = {481},
    number = {4},
    pages = {4410-4418},
    year = {2018},
    month = {09},
    issn = {0035-8711},
    doi = {10.1093/mnras/sty2601},
    url = {https://doi.org/10.1093/mnras/sty2601},
    eprint = {https://academic.oup.com/mnras/article-pdf/481/4/4410/25900081/sty2601.pdf},
}

@ARTICLE{Han2018,
       author = {{Han}, J.~L. and {Manchester}, R.~N. and {van Straten}, W. and {Demorest}, P.},
        title = "{Pulsar Rotation Measures and Large-scale Magnetic Field Reversals in the Galactic Disk}",
      journal = {\apjs},
         year = 2018,
        month = jan,
       volume = {234},
       number = {1},
          eid = {11},
        pages = {11},
          doi = {10.3847/1538-4365/aa9c45},
archivePrefix = {arXiv},
       eprint = {1712.01997},
 primaryClass = {astro-ph.GA},
       adsurl = {https://ui.adsabs.harvard.edu/abs/2018ApJS..234...11H}
}

@ARTICLE{ordog2025DRAGONS,
       author = {{Ordog}, Anna and {Booth}, Rebecca A. and {Landecker}, T.~L. and {Carretti}, Ettore and {Hill}, Alex S. and {Brown}, Jo-Anne C. and {Davydov}, Artem and {Moutinho Caffarello}, Leonardo and {Galler}, Luca B. and {Flygare}, Jonas and {West}, Jennifer L. and {Willis}, A.~G. and {Tahani}, Mehrnoosh and {Hovey}, G.~J. and {Lagoy}, Dustin and {Harrison}, Stephen and {Smith}, Mike and {Baard}, Charl and {Messing}, Rob H. and {Del Rizzo}, D.~A. and {Robert}, Benoit and {Robishaw}, Timothy and {Dickey}, John M. and {Morgan}, George and {Kennedy}, Ian R. and {Haverkorn}, Marijke and {Bracco}, Andrea and {Conway}, John},
        title = "{GMIMS-DRAGONS: A Faraday Depth Survey of the Northern Sky Covering 350 to 1030 MHz}",
      journal = {\apjs},
         year = 2025,
        month = oct,
          eid = {arXiv:2510.09759},
         volume = {in press},
archivePrefix = {arXiv},
       eprint = {2510.09759},
 primaryClass = {astro-ph.IM},
       adsurl = {https://ui.adsabs.harvard.edu/abs/2025arXiv251009759O}
}

@ARTICLE{Gaensler2001,
       author = {{Gaensler}, B.~M. and {Dickey}, John M. and {McClure-Griffiths}, N.~M. and {Green}, A.~J. and {Wieringa}, M.~H. and {Haynes}, R.~F.},
        title = "{Radio Polarization from the Inner Galaxy at Arcminute Resolution}",
      journal = {\apj},
         year = 2001,
        month = mar,
       volume = {549},
       number = {2},
        pages = {959-978},
          doi = {10.1086/319468},
archivePrefix = {arXiv},
       eprint = {astro-ph/0010518},
 primaryClass = {astro-ph},
       adsurl = {https://ui.adsabs.harvard.edu/abs/2001ApJ...549..959G}
}

@ARTICLE{Hill2018,
       author = {{Hill}, Alex S.},
        title = "{Is There a Polarization Horizon?}",
      journal = {Galaxies},
         year = 2018,
        month = nov,
       volume = {6},
       number = {4},
          eid = {129},
        pages = {129},
          doi = {10.3390/galaxies6040129},
archivePrefix = {arXiv},
       eprint = {1810.12008},
 primaryClass = {astro-ph.GA},
       adsurl = {https://ui.adsabs.harvard.edu/abs/2018Galax...6..129H}
}

@ARTICLE{Chamandy2013,
       author = {{Chamandy}, Luke and {Subramanian}, Kandaswamy and {Shukurov}, Anvar},
        title = "{Galactic spiral patterns and dynamo action - I. A new twist on magnetic arms}",
      journal = {\mnras},
         year = 2013,
        month = feb,
       volume = {428},
       number = {4},
        pages = {3569-3589},
          doi = {10.1093/mnras/sts297},
archivePrefix = {arXiv},
       eprint = {1207.6239},
 primaryClass = {astro-ph.GA},
       adsurl = {https://ui.adsabs.harvard.edu/abs/2013MNRAS.428.3569C}
}

@ARTICLE{Brown2001,
       author = {{Brown}, J.~C. and {Taylor}, A.~R.},
        title = "{The Structure of the Magnetic Field in the Outer Galaxy from Rotation Measure Observations through the Disk}",
      journal = {\apjl},
         year = 2001,
        month = dec,
       volume = {563},
       number = {1},
        pages = {L31-L34},
          doi = {10.1086/338358},
       adsurl = {https://ui.adsabs.harvard.edu/abs/2001ApJ...563L..31B}
}

@article{Zonca2019,
  doi = {10.21105/joss.01298},
  url = {https://doi.org/10.21105/joss.01298},
  year = {2019},
  month = mar,
  publisher = {The Open Journal},
  volume = {4},
  number = {35},
  pages = {1298},
  author = {Andrea Zonca and Leo Singer and Daniel Lenz and Martin Reinecke and Cyrille Rosset and Eric Hivon and Krzysztof Gorski},
  title = {healpy: equal area pixelization and spherical harmonics transforms for data on the sphere in Python},
  journal = {Journal of Open Source Software}
}

@article{Hunter_2007,
 adsurl = {https://ui.adsabs.harvard.edu/abs/2007CSE.....9...90H},
 author = {{Hunter}, John D.},
 doi = {10.1109/MCSE.2007.55},
 journal = {Computing in Science and Engineering},
 month = {May},
 number = {3},
 pages = {90-95},
 title = {{Matplotlib: A 2D Graphics Environment}},
 volume = {9},
 year = {2007}
}

@article{Astropy_2022,
 adsurl = {https://ui.adsabs.harvard.edu/abs/2022ApJ...935..167A},
 archiveprefix = {arXiv},
 author = {{Astropy Collaboration} and {Price-Whelan}, Adrian M. and {Lim}, Pey Lian and {Earl}, Nicholas and {Starkman}, Nathaniel and {Bradley}, Larry and {Shupe}, David L. and {Patil}, Aarya A. and {Corrales}, Lia and {Brasseur}, C.~E. and {N{\"o}the}, Maximilian and {Donath}, Axel and {Tollerud}, Erik and {Morris}, Brett M. and {Ginsburg}, Adam and {Vaher}, Eero and {Weaver}, Benjamin A. and {Tocknell}, James and {Jamieson}, William and {van Kerkwijk}, Marten H. and {Robitaille}, Thomas P. and {Merry}, Bruce and {Bachetti}, Matteo and {G{\"u}nther}, H. Moritz and {Aldcroft}, Thomas L. and {Alvarado-Montes}, Jaime A. and {Archibald}, Anne M. and {B{\'o}di}, Attila and {Bapat}, Shreyas and {Barentsen}, Geert and {Baz{\'a}n}, Juanjo and {Biswas}, Manish and {Boquien}, M{\'e}d{\'e}ric and {Burke}, D.~J. and {Cara}, Daria and {Cara}, Mihai and {Conroy}, Kyle E. and {Conseil}, Simon and {Craig}, Matthew W. and {Cross}, Robert M. and {Cruz}, Kelle L. and {D'Eugenio}, Francesco and {Dencheva}, Nadia and {Devillepoix}, Hadrien A.~R. and {Dietrich}, J{\"o}rg P. and {Eigenbrot}, Arthur Davis and {Erben}, Thomas and {Ferreira}, Leonardo and {Foreman-Mackey}, Daniel and {Fox}, Ryan and {Freij}, Nabil and {Garg}, Suyog and {Geda}, Robel and {Glattly}, Lauren and {Gondhalekar}, Yash and {Gordon}, Karl D. and {Grant}, David and {Greenfield}, Perry and {Groener}, Austen M. and {Guest}, Steve and {Gurovich}, Sebastian and {Handberg}, Rasmus and {Hart}, Akeem and {Hatfield-Dodds}, Zac and {Homeier}, Derek and {Hosseinzadeh}, Griffin and {Jenness}, Tim and {Jones}, Craig K. and {Joseph}, Prajwel and {Kalmbach}, J. Bryce and {Karamehmetoglu}, Emir and {Ka{\l}uszy{\'n}ski}, Miko{\l}aj and {Kelley}, Michael S.~P. and {Kern}, Nicholas and {Kerzendorf}, Wolfgang E. and {Koch}, Eric W. and {Kulumani}, Shankar and {Lee}, Antony and {Ly}, Chun and {Ma}, Zhiyuan and {MacBride}, Conor and {Maljaars}, Jakob M. and {Muna}, Demitri and {Murphy}, N.~A. and {Norman}, Henrik and {O'Steen}, Richard and {Oman}, Kyle A. and {Pacifici}, Camilla and {Pascual}, Sergio and {Pascual-Granado}, J. and {Patil}, Rohit R. and {Perren}, Gabriel I. and {Pickering}, Timothy E. and {Rastogi}, Tanuj and {Roulston}, Benjamin R. and {Ryan}, Daniel F. and {Rykoff}, Eli S. and {Sabater}, Jose and {Sakurikar}, Parikshit and {Salgado}, Jes{\'u}s and {Sanghi}, Aniket and {Saunders}, Nicholas and {Savchenko}, Volodymyr and {Schwardt}, Ludwig and {Seifert-Eckert}, Michael and {Shih}, Albert Y. and {Jain}, Anany Shrey and {Shukla}, Gyanendra and {Sick}, Jonathan and {Simpson}, Chris and {Singanamalla}, Sudheesh and {Singer}, Leo P. and {Singhal}, Jaladh and {Sinha}, Manodeep and {Sip{\H{o}}cz}, Brigitta M. and {Spitler}, Lee R. and {Stansby}, David and {Streicher}, Ole and {{\v{S}}umak}, Jani and {Swinbank}, John D. and {Taranu}, Dan S. and {Tewary}, Nikita and {Tremblay}, Grant R. and {de Val-Borro}, Miguel and {Van Kooten}, Samuel J. and {Vasovi{\'c}}, Zlatan and {Verma}, Shresth and {de Miranda Cardoso}, Jos{\'e} Vin{\'\i}cius and {Williams}, Peter K.~G. and {Wilson}, Tom J. and {Winkel}, Benjamin and {Wood-Vasey}, W.~M. and {Xue}, Rui and {Yoachim}, Peter and {Zhang}, Chen and {Zonca}, Andrea and {Astropy Project Contributors}},
 doi = {10.3847/1538-4357/ac7c74},
 eid = {167},
 eprint = {2206.14220},
 journal = {\apj},
 month = {August},
 number = {2},
 pages = {167},
 primaryclass = {astro-ph.IM},
 title = {{The Astropy Project: Sustaining and Growing a Community-oriented Open-source Project and the Latest Major Release (v5.0) of the Core Package}},
 volume = {935},
 year = {2022}
}

@Article{         harris2020array,
 title         = {Array programming with {NumPy}},
 author        = {Charles R. Harris and K. Jarrod Millman and St{\'{e}}fan J.
                 van der Walt and Ralf Gommers and Pauli Virtanen and David
                 Cournapeau and Eric Wieser and Julian Taylor and Sebastian
                 Berg and Nathaniel J. Smith and Robert Kern and Matti Picus
                 and Stephan Hoyer and Marten H. van Kerkwijk and Matthew
                 Brett and Allan Haldane and Jaime Fern{\'{a}}ndez del
                 R{\'{i}}o and Mark Wiebe and Pearu Peterson and Pierre
                 G{\'{e}}rard-Marchant and Kevin Sheppard and Tyler Reddy and
                 Warren Weckesser and Hameer Abbasi and Christoph Gohlke and
                 Travis E. Oliphant},
 year          = {2020},
 month         = sep,
 journal       = {Nature},
 volume        = {585},
 number        = {7825},
 pages         = {357--362},
 doi           = {10.1038/s41586-020-2649-2},
 publisher     = {Springer Science and Business Media {LLC}},
 url           = {https://doi.org/10.1038/s41586-020-2649-2}
}
\bibliographystyle{aasjournal}

\end{document}